\documentclass{jfm}
\usepackage{graphicx}
\usepackage{epstopdf, epsfig}
\usepackage{hyperref}
\hypersetup{
	colorlinks = true,
	urlcolor   = blue,
	citecolor  = black,
}

\usepackage{float}
\usepackage{amsmath}
\usepackage{amssymb}
\usepackage{esint}
\usepackage{bm}
\usepackage{cancel}
\usepackage{subcaption}
\usepackage{graphicx}  
\usepackage{mwe}
\usepackage{tikz}
\usepackage{xcolor}
\usepackage{stackengine}
\usepackage{wasysym}
\usepackage{comment}
\usepackage{relsize}

\definecolor{color_1}{rgb}{1, 0, 0}
\definecolor{color_2}{rgb}{0, 1, 0}
\definecolor{color_3}{rgb}{0, 0, 1}
\definecolor{color_4}{rgb}{0, 1, 1}
\definecolor{color_5}{rgb}{1, 0, 1}
\definecolor{color_6}{rgb}{0.8500, 0.3250, 0.0980}
\definecolor{color_7}{rgb}{0.4940, 0.1840, 0.5560}
\definecolor{color_8}{rgb}{0.3010, 0.7450, 0.9330}

\usepackage{tikz}
\DeclareRobustCommand\full  {\tikz[baseline=-0.6ex]\draw[thick] (0,0)--(0.25,0);}

\usepackage[textsize=scriptsize]{todonotes}

\usepackage[normalem]{ulem}

\shorttitle{Pairwise interaction of spherical particles aligned in oscillatory flow}
\shortauthor{F. Kleischmann, P. Luzzatto-Fegiz, E. Meiburg 
and B. Vowinckel}

\title{Pairwise interaction of spherical particles aligned in high-frequency oscillatory flow} 

\author{F. Kleischmann\aff{1,2,3}
  \corresp{\email{fabian.kleischmann@tu-dresden.de}},
  P. Luzzatto-Fegiz\aff{3},
  E. Meiburg\aff{3},\\
 \and B. Vowinckel\aff{1,2}}

\affiliation{\aff{1}Leichtweiß-Institute for Hydraulic Engineering and Water Resources, Technische Universität Braunschweig, 38106 Braunschweig, Germany
\aff{2}Institute of Urban and Industrial Water Management, Technische Universität Dresden, \\ 01062 Dresden, Germany
\aff{3}Department of Mechanical Engineering, UC Santa Barbara, Santa Barbara, CA 93106, USA}

\begin{document}

\maketitle

\begin{abstract}
We present a systematic simulation campaign to investigate the pairwise interaction of two mobile, monodisperse particles submerged in a viscous fluid and subjected to monochromatic oscillating flows. 
To this end, we employ the immersed boundary method to geometrically resolve the flow around the two particles in a non-inertial reference frame. 
We neglect gravity to focus on fluid-particle interactions associated with particle inertia and consider particles of three different density ratios aligned along the axis of oscillation.
We systematically vary the initial particle distance and the frequency based on which the particles show either attractive or repulsive behavior by approaching or moving away from each other, respectively.
This behavior is consistently confirmed for the three density ratios investigated, although particle inertia dictates the overall magnitude of the particle dynamics. 
Based on this, threshold conditions for the transition from attraction to repulsion are introduced that obey the same power law for all density ratios investigated. 
We furthermore analyze the flow patterns by suitable averaging and decomposition of the flow fields and find competing effects of the vorticity induced by the fluid-particle interactions. 
Based on these flow patterns, we derive a circulation-based criterion that provides a quantitative measure to categorize the different cases. 
It is shown that such a criterion provides a consistent measure to distinguish the attractive and repulsive arrangements.

\end{abstract}


%%%%%%%%%%%%%%%%%%%%%%%%%%%%%%%%%
%%%%%%%%%%%%%%%%%%%%%%%%%%%%%%%%%
\section{Introduction}\label{sec:Introduction}
%%%%%%%%%%%%%%%%%%%%%%%%%%%%%%%%%
%%%%%%%%%%%%%%%%%%%%%%%%%%%%%%%%%

The behavior of particles in oscillating flows is a widespread phenomenon in many engineering systems and has attracted much attention in recent decades. 
In industrial applications, oscillating flows are used to improve agglomeration and removal of particles in water treatment facilities \citep{2019_Halfi_etal, 2020_Halfi_etal} and in diesel engines \citep{2005_Katoshevski_etal,2016_Ruzal_etal,2019_Gupta_etal}.
Recently, research has also been conducted to investigate how oscillations can be used as a non-invasive biomedical application for drug delivery inside the ear \citep{2021_Sumner_etal, 2023_Harte_etal}.

Of particular interest for the present study are applications and investigations of fluid-particle interactions in oscillating conditions where gravity is negligible and at the same time the inertia of the particles is of decisive importance.
Such conditions exist in space, for instance, and have already been used for a variety of scientific studies.
For example, experiments were performed as part of the First International Microgravity Laboratory in $1992$ on board a space shuttle to investigate the growth of crystals in the absence of gravity to gain insights into their growth process \citep{1996a_Trolinger_etal,1996b_Trolinger_etal}.
Studies of the same kind were recently conducted on board the International Space Station to examine the flocculation behavior of a variety of cohesive particles \citep{2021_Kleischmann_etal}.
Such experimental setups are subjected to small oscillations, known as g-jitter, caused by the spacecraft's propulsion system and on-board machinery.
For obvious reasons, such experimental campaigns involve a great deal of effort as well as expense and are, therefore, rare.
Nevertheless, such special conditions can also be found in earth-bound environments.
In the field of microfluidics, the inertial properties of particles can be used to manipulate their dynamics in oscillatory fluid flows \citep{2018_Mutlu_etal, 2020_Mutlu_etal, 2019_Dietsche_etal}.
As the particles are usually only a few micrometers in size, the settling due to gravity can be largely neglected \citep{2023_Owen_etal}.

Regardless of the area of application or the specifications of the setup, the behavior of particles in oscillating flows is determined by the fluid-particle interaction of each individual particle.
For this purpose, \cite{2001_Coimbra_Rangel} proposed an analytical solution for particle motion in an oscillating background flow based on the work of \cite{1947_Tchen}.
The theoretical prediction has subsequently been confirmed by the experiments of \cite{2004_Coimbra_etal} and \cite{2005_LEsperance_etal, 2006_LEsperance_etal}.
In these experimental studies, the authors investigated the response of an individual spherical particle oscillating in a viscous liquid-filled container, neutralizing gravity by tethering the particle to prevent vertical movement based on its density.
Similar experiments were performed by \cite{2006b_Hassan_etal}, \cite{2007b_Hassan_Kawaji},  \cite{2008_Hassan_Kawaji}, and \cite{2010_Saadatmand_Kawaji}.
\cite{2006_Simic-Stefani_etal} extended the idea of studying the particle motion in oscillatory flow without the impact of gravity to experiments in microgravity on parabolic flights.
However, for parabolic flights or comparable methods, such as the drop tower, the state of weightlessness only lasts for a very short period of time and long-term studies are not possible.
Numerical simulations, on the other hand, easily achieve microgravity by setting gravitational acceleration to zero \citep[e.g.][]{2006_Simic-Stefani_etal,2010_Saadatmand_Kawaji,2022_Satish_etal}. 
Consistent findings across theory, experiments, and numerical simulations demonstrate that particle excursion increases with increasing deviation of the density ratio from unity. Hence, particle motion reflects the oscillation frequency with a delay related to particle inertia.

Another important aspect is the flow field generated by the fluid-particle interaction, in which either the fluid or the particle oscillates, characterized by the instantaneous flow field or the flow field averaged over one oscillation period, which is commonly referred to as steady streaming.
Due to the fact that the averaged flow field is usually non-zero, it causes a change in particle motion that deviates from the prevailing oscillatory motion \citep{1966_Riley}.
Analytical models for steady streaming results of flows around a sphere have been developed by \cite{1955_Lane} and \cite{1966_Riley, 1967_Riley}, in which inner and outer solutions were obtained by perturbation theory and associated streamlines were described.
To this end, vortices near the particle surface (within the oscillating boundary layer) were defined as primary vortices whereas secondary vortices emerge further away from the particle that in turn are driven by the rotation of the primary vortices.
In this context, \cite{1966_Riley} has elaborated on two distinct cases for small oscillation amplitudes. 
For larger inertial forces, the field of the steady streaming shows a pattern consisting of both the primary and secondary vortices. 
For smaller inertial forces, only one dominant structure of primary vortices exists, that expands over the whole domain \citep{1964_Rott}.
In case of a stationary sphere in an oscillating fluid, \cite{2023_Li_etal} recently provided a phase diagram that can be utilized to determine the presence of only primary or primary and secondary vortices by means of the non-dimensional frequency and amplitude of the oscillation.

The flow structures around a single spherical particle were extensively investigated by theoretical and numerical studies \citep{1994_Chang_Maxey, 1997_Alassar_Badr, 2002_Blackburn, 2008_Alassar, 2017_Fabre_etal, 2018_Jalal}.
Together with experimental measurements \citep{2007_Kotas_etal, 2008_Otto_etal}, the size and the rotational directions of the structures of the steady streaming were analyzed. 
According to \cite{2002_Voth_etal} and \cite{2009_Klotsa_etal}, detailed knowledge of the flow structures of an individual particle should provide information about the interaction behavior of two particles.
Hence, experimental and numerical studies explored the interaction of two particles under either vertical \citep{2002_Voth_etal} or horizontal oscillations \citep{2007_Klotsa_etal, 2009_Klotsa_etal, 2022_vanOverveld_etal, 2023_vanOverveld_etal}. 
The latter studies, in which the particles were positioned on the bottom plate of the fluid container perpendicular to the oscillation, focused on the equilibrium distance between the particles. 
A major finding was that the equilibrium is determined by a balance of short-range repulsive and long-range attractive forces induced by the flow structures.

Although these studies provide valuable insights into the particle behavior subjected to oscillating flows, they are limited to initial particle configurations that are perpendicular with respect to the oscillation direction.
Moreover, the particle behavior is affected by the presence of the bottom plate and the effect of gravity.
Despite the fact that \cite{2007_Klotsa_etal, 2009_Klotsa_etal} and \cite{2022_vanOverveld_etal,2023_vanOverveld_etal} ensured that the friction between the particles and the bottom plate is minimized, the flow field that is formed is nevertheless restricted by the bottom plate.
This restriction is reflected in particular by the fact that the flow structures
are limited below the particles, which introduces a distortion of the flow field that might alter the system.
To avoid this effect, \cite{2017_Fabre_etal} and \cite{2018_Jalal} numerically investigated two stationary particles in an oscillating domain.
The studies provide key results for interpreting the interaction forces that might lead to either attraction or repulsion of the particles.
However, for freely moving particles the distance changes over time and, consequently, this may also affect the particle interaction.

Despite the large number of studies on the interaction of two particles in an oscillating flow, no investigations have been conducted, in which the particles can move freely, independently of gravity and without spatial constraints. The present study addresses this research gap.
We perform particle-resolved direct numerical simulations of two mobile particles in an oscillating box filled with a viscous fluid.
The setup enables us to analyze the behavior of the freely moving particles as well as the resulting flow structures, with respect to the change of the distance between the particles as a function of time.
In order to focus only on the hydrodynamic effects of the fluid-particle interactions, we neglect gravity and ensure that the particles are not affected by short-range wall effects from the outer boundaries of the domain.
Based on this, we aim to identify the impact of the particle arrangement, in terms of mutual distance and orientation angle, the oscillation frequency and the particle inertia, characterized by different particle densities and tie these observations to the respective flow characteristics.

The article is structured as follows. 
First, we introduce the numerical model along with the description of the computational approach 
and provide a comprehensive validation in \S \ref{sec:computational_model}. 
We then analyze the impact of the non dimensional frequency $S$, the initial inter-particle distance $\zeta_i$ and orientation of the particle alignment $\theta_i$, as well as the density ratio $\rho_s$ on the mutual particle behavior.
This is followed by a detailed analyses of the resulting flow fields in \S \ref{sec:flow_field_analysis} and finally, a summary of the conclusions and a brief outlook in \S \ref{sec:conclusions}.

%%%%%%%%%%%%%%%%%%%%%%%%%%%%%%%%%
%%%%%%%%%%%%%%%%%%%%%%%%%%%%%%%%%
\section{Computational model}\label{sec:computational_model}
%%%%%%%%%%%%%%%%%%%%%%%%%%%%%%%%%
%%%%%%%%%%%%%%%%%%%%%%%%%%%%%%%%%

%%%%%%%%%%%%%%%%%%%%%%%%%%%%%%%%%
\subsection{Numerical setup}\label{sec:numerical_setup}
%%%%%%%%%%%%%%%%%%%%%%%%%%%%%%%%%

\begin{figure}
  \centerline{\includegraphics[width=0.4\textwidth]{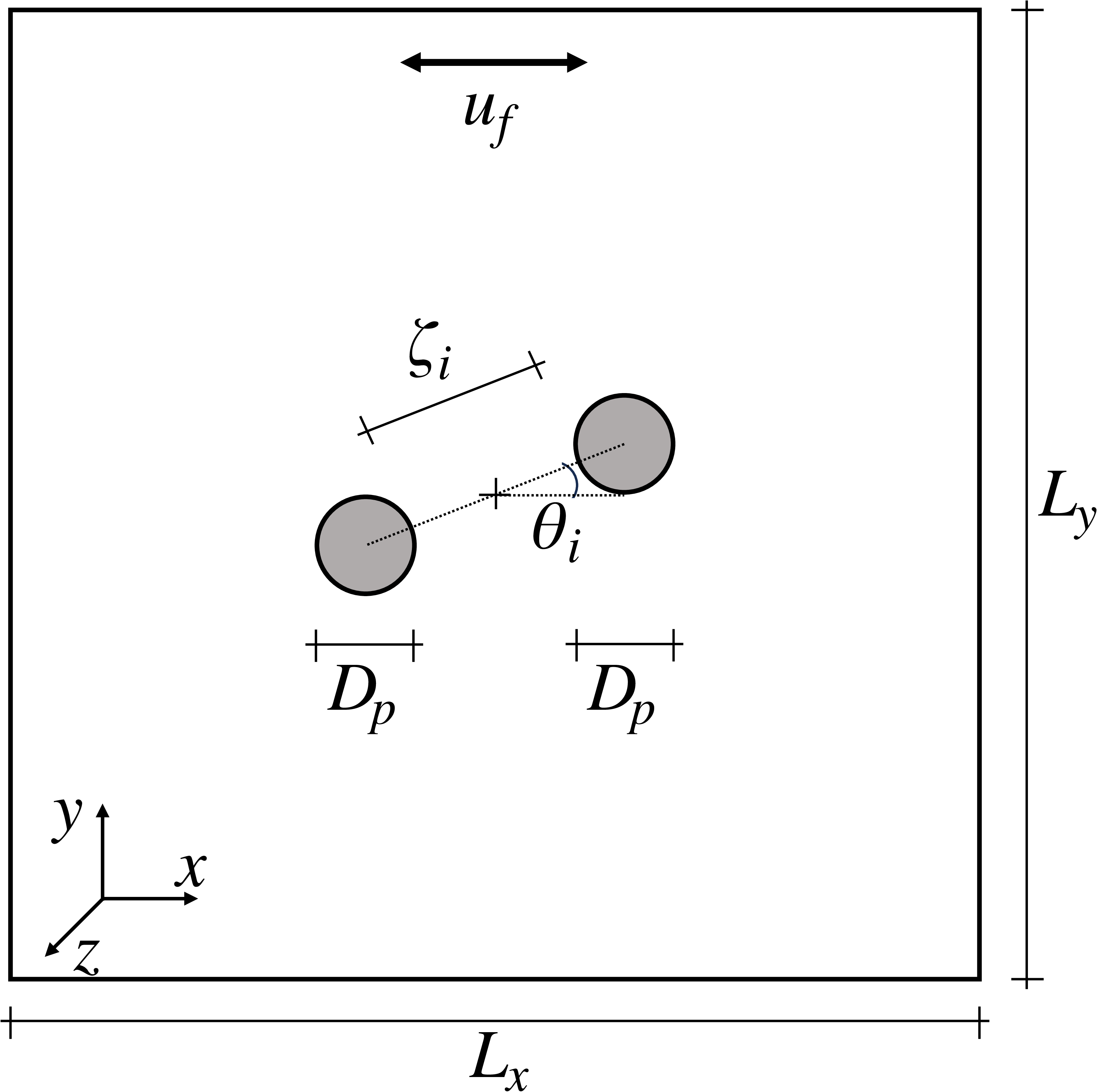}}% Images in 100% size
  \caption{Two-dimensional sketch of the cubic numerical domain with size $L_{x,y,z} = L = 10~D_p$. Two mobile spherical particles with identical diameters $D_p$ are initially placed so that the center of the initial particle \mbox{distance $\zeta_i$} coincides with the center of the domain. The arrangement is aligned with an initial angle $\theta_i$ with respect to the direction of oscillation. The domain oscillates with velocity $u_f$ in $x$-direction, as indicated by the arrows.}
\label{fig:Domain_Setup}
\textbf{}\end{figure}

The scenario under consideration is a cubic container with dimensions of $L_{x,y,z} = L = 10 D_p$, where $D_p$ is the particle diameter.
The container is filled with a viscous, incompressible fluid characterized by its density $\rho_f$ and kinematic viscosity $\nu_f$. 
As illustrated in figure \ref{fig:Domain_Setup}, two monodisperse particles are submerged in this container and placed with an initial particle distance~$\zeta_i$ as well as an initial orientation $\theta_i$ with respect to the horizontal direction of oscillation (black arrows).
Initially, the fluid is at rest and the center of $\zeta_i$ coincides with the center of the domain.
The origin of the Cartesian coordinate system is located in the back lower left corner and oriented according to the axes in figure \ref{fig:Domain_Setup}.
The particles are free to move in response to the container that oscillates in the $x$-direction with  $u_f =- A_f \Omega\sin{\Omega t}$.
Here, $A_f$ is the distance amplitude and $A_f \Omega$ the velocity amplitude, $t$ denotes time, $\Omega = 2 \pi f$ is the angular frequency and $f$ the frequency of the oscillatory motion of the domain.
The walls of the domain are characterized by stress-free boundary conditions, i.e. $d u_t / d n=0$ and $u_n=0$.
Here, $u_t$ and $u_n$ are the  tangential and the normal fluid velocity components relative to the wall, respectively, and $n$ is the normal direction on that same wall. 
A no-slip condition was applied on the particle surface.

%%%%%%%%%%%%%%%%%%%%%%%%%%%%%%%%%
\subsection{Governing equations}\label{sec:governingEqs}
%%%%%%%%%%%%%%%%%%%%%%%%%%%%%%%%%

We use Direct Numerical Simulations (DNS) to compute the dynamics inside the oscillating container and apply the Immersed Boundary Method (IBM) to account for the fluid-particle interaction \citep[][]{2005_Uhlmann,2012_Kempe_Froehlich,2017_Biegert_etal}. 
The simulations are performed in a non-inertial reference frame, accounting for the acceleration of the applied external oscillation. 
Since there are no free surfaces or density variations in the fluid, the effect of the frame acceleration is accounted for in the particle motion only.
Therefore, we can solve the Navier-Stokes equations and the continuity equation for an incompressible Newtonian fluid given by
\begin{equation} \label{eq:navier_stokes}
    \frac{\partial{\textbf{u}}}{\partial{t}}+\nabla\cdot(\textbf{u}\textbf{u}) = -\frac{1}{\rho_f}\:\nabla p + \nu_f \nabla^2 \textbf{u} + \textbf{f}_\text{IBM} \qquad , 
\end{equation}
\begin{equation}\label{eq:continuity}
    \nabla\cdot\textbf{u}=0 \qquad . 
\end{equation}
Here, $\textbf{u}=(u,v,w)^{T}$ represents the fluid velocity vector in Cartesian components, $p$ is the fluid pressure and $\textbf{f}_\text{IBM}$ the volume force based on the IBM. 
The $\textbf{f}_\text{IBM}$ connects the motion of the particle to the fluid phase and acts in the vicinity of the fluid-particle interface to enforce a no-slip condition on the particle surface.
Equations \eqref{eq:navier_stokes} and \eqref{eq:continuity} are integrated by a third-order low-storage Runge-Kutta scheme in time together with a second order finite difference method in space, respectively. 
The fast Fourier transform is applied to calculate the pressure correction for continuity by means of a direct solver.

In the context of the IBM, the motion of each individual spherical particle is calculated by solving the Newton-Euler equations, where Newton's equation for the translational particle velocity needs to be transformed into the non-inertial frame. 
In the non-inertial reference frame, which moves with $\textbf{u}_f = (u_f, 0, 0)^T$ relative to the inertial frame, we can calculate the non-inertial particle velocity as $\textbf{u}_p' = \textbf{u}_p - \textbf{u}_f$, where ${\textbf{u}_p=(u_p,v_p,w_p)^{T}}$ presents the translational velocity vector of the particle in an inertial frame. 
By re-arranging, we can state that 
\begin{equation} 
    \frac{\text{d}\textbf{u}_p}{\text{d}t} = \frac{\text{d}\textbf{u}_p'}{\text{d}t} + \frac{\text{d}\textbf{u}_f}{\text{d}t}    \hspace{0.5cm}.
    \label{eq:TransformationNonInertialFrame}
\end{equation}
If we take Newton's equation in an inertial frame
\begin{equation}\label{eq:NewtonEq_inertial}
    m_p \frac{\text{d}\textbf{u}_p}{\text{d}t}=  \textbf{F}_\text{h} + \textbf{F}_\text{c} + \rho_f V_p \frac{\text{d}\textbf{u}_f}{\text{d}t} \qquad , 
\end{equation}
and apply \eqref{eq:TransformationNonInertialFrame} to transform \eqref{eq:NewtonEq_inertial} to the non-inertial frame, we obtain the velocities in the non-inertial frame based on the equations of motion for translation
\begin{equation}\label{eq:NewtonEq_nonInertial}
    m_p \frac{\text{d}\textbf{u}_p'}{\text{d}t}=  \textbf{F}_\text{h} + \textbf{F}_\text{c} - \left(\rho_p - \rho_f \right) V_p \frac{\text{d}\textbf{u}_f}{\text{d}t} \qquad ,
\end{equation}
and  rotation
\begin{equation}\label{eq:EulerEq}
    I_p \frac{\text{d}\bm{\omega}_p}{\text{d}t}= \textbf{M}_h \: + \: \textbf{M}_c \qquad .
\end{equation}
Here, $m_p$ and $V_p$ are the particle mass and volume, respectively, $\rho_p$ the particle density, and $I_p = \pi \rho_p D_p ^ 5 / 60$ the moment of inertia with $\bm\omega_p = \left( \omega_{p,x}, \omega_{p,y}, \omega_{p,z} \right)^{T}$ the angular velocity vector. 
The last term of the right-hand side of \eqref{eq:NewtonEq_inertial} and \eqref{eq:NewtonEq_nonInertial} accounts for the pressure gradient of the background acceleration in the inertial and non-inertial frame, respectively. 
As an important consequence arising from the change of the observational frame, the background acceleration of the non-inertial frame is characterized by the motion of the particle mass minus the displaced fluid mass and, hence, represents the effect of particle inertia.
Note that \eqref{eq:EulerEq} remains unchanged for the translation from the inertial to the non-inertial reference frame, because we do not apply any rotational motion.
As noted earlier, we neglect gravity in \eqref{eq:NewtonEq_nonInertial} to limit the governing mechanism to the fluid-particle and particle-particle interactions. 
The first terms on the right hand side of \eqref{eq:NewtonEq_nonInertial} and \eqref{eq:EulerEq}, $\textbf{F}_\text{h}$ and $\textbf{M}_\text{h}$, denote the hydrodynamic forces and torques, respectively.
The fluid-particle interactions can be computed via   
\begin{equation}
    \begin{gathered}
        \textbf{F}_\text{h} = \oint\limits_{S_p}\, \bm\tau \cdot \textbf{n} \,dS \: , \qquad \textbf{M}_\text{h} = \oint\limits_{S_p}\textbf{r} \times (\bm{\tau} \cdot \textbf{n}) \,dS \: ,
    \end{gathered}
\end{equation}
where $\bm\tau = -p \textbf{I} \: + \: \mu_f \left[ \nabla \textbf{u} + \left(\nabla \textbf{u} \right)^T \right]$ represents  the hydrodynamic stress tensor that includes the identity tensor $\textbf{I}$ and the dynamic viscosity $\mu_f$ of the fluid. 
Furthermore, \mbox{$\textbf{n}$ is} the normal vector pointing outward from the particle surface $S_p$, and \textbf{r} is the position vector pointing from the particle center of mass to a point on $S_p$. 
The forces and torques due to collision are represented by $\textbf{F}_\text{c}$ and $\textbf{M}_\text{c}$, respectively.
Here, we account for normal and tangential contact as well as unresolved lubrication forces that emerge when fluid is squeezed out of particle gaps $\zeta_t \leq 2h$, with $\zeta_{t}$ the distance between the particles at any time and $h$ the grid cell size. 
Full details on the computation of short-range hydrodynamic as well as collision forces and torques can be found in \cite{2017_Biegert_etal} and \cite{2019_Vowinckel_etal}, but we note that those effects are not relevant for this study as no collisions occur in the cases examined.  

For our numerical study, we introduce characteristic scales to non-dimensionalize the governing equations where  $\ell$  and $\textbf{u}$ are the relevant  length and velocity scales, respectively, that appear in equations \eqref{eq:navier_stokes} -- \eqref{eq:EulerEq}:
\begin{equation} \label{eq:nonDim_scales}
    \begin{split}
        \begin{gathered}
            \vspace{0.1cm}
            \ell = D_{p} \tilde{\ell}, \quad %
            \textbf{u} = D_p \Omega  \tilde{\textbf{u}}, \quad %
            t = \frac{\tilde{t}}{\Omega}, \\
            \vspace{0.1cm}
            \rho = \rho_{f} \tilde{\rho}, \quad %
            p = \rho_f D_p^2 \Omega^2 \Tilde{p}, \quad %
            f_\textit{IBM} = D_p \Omega^2 \Tilde{f}_\textit{IBM}, \\
            \vspace{0.1cm}
            m = m_{f} \tilde{m} = \rho_f V_{p} \tilde{m}, \quad %
            V = D_p^3 \: \tilde{V}, \quad %
            F = m_f D_p \Omega^2 \tilde{F}, \\
            \vspace{0.1cm}
            I = m_f \: D_p^2 \tilde{I}, \quad %
            \omega = \Omega \:\tilde{\omega}, \quad %
            M = m_f D_p^2 \Omega^2 \Tilde{M},
       \end{gathered}
    \end{split}
\end{equation}
Here, $m_f$ is the mass of fluid of an equivalent particle volume and $F$ and $M$ represent forces or torques on the particles, respectively. 
The tilde symbol indicates the dimensionless variables. 
Applying \eqref{eq:nonDim_scales} to \eqref{eq:navier_stokes}--\eqref{eq:EulerEq} yields the dimensionless Navier-Stokes, continuity, and Newton-Euler equations with the Reynolds number $\Rey = u_{f,max}  D_p / \nu_f$, where $u_{f,max}=A_f \Omega$ is the velocity amplitude and $\rho_s = \rho_p / \rho_f$ the density ratio  

\begin{equation} \label{eq:nondim_navier_stokes}
    \frac{\partial{\tilde{\textbf{u}}}}{\partial{\tilde{t}}} + \tilde{\nabla} \cdot (\tilde{\textbf{u}}\tilde{\textbf{u}}) = -\tilde{\nabla} \tilde{p} + \frac{1}{\Rey} \tilde{\nabla}^2 \tilde{\textbf{u}} + \tilde{\textbf{f}}_\text{IBM} \quad ,
\end{equation}
\begin{equation}\label{eq:nondim_continuity}
    \tilde{\nabla}\cdot\tilde{\textbf{u}}=0 \quad ,
\end{equation}
\begin{equation}\label{eq:nondim_NewtonEq_nonInertial}
    \tilde{m}_p \frac{\text{d}\tilde{\textbf{u}}_p}{\text{d}\tilde{t}}=  \tilde{\textbf{F}}_\text{h} + \tilde{\textbf{F}}_\text{c} - \left(\rho_s - 1 \right) \tilde{V}_p \frac{\text{d}\tilde{\textbf{u}}_f}{\text{d}\tilde{t}} \quad ,
\end{equation}
\begin{equation}\label{eq:nondim_EulerEq}
    \tilde{I}_p \frac{\text{d}\tilde{\bm{\omega}}_p}{\text{d}\tilde{t}}= \tilde{\textbf{M}}_h \: + \: \tilde{\textbf{M}}_c \qquad .
\end{equation}

In addition, we introduce a streaming Reynolds number $\Rey_s$ which is defined by the product of the amplitude ratio $\epsilon = A_f / D_p$ and $\Rey$ to obtain $\Rey_s = \epsilon \Rey = A_f^2\Omega / \nu$.
Following \cite{2001_Coimbra_Rangel},  \cite{2004_Coimbra_etal} and \cite{2005_LEsperance_etal}, we also define a Strouhal number $Sl=\Omega D_p / (9 u_{f,max})$ for spherical objects to obtain the non-dimensional frequency $S=Sl \Rey /4=D_p^2 \Omega / (36 \nu_f)$. 
Note that $1/4$ is a geometric factor that allows for a straightforward comparison to the work of these authors who used the particle radius as a characteristic length scale, while we chose $D_p$ for the present study.
As an important consequence, the amplitude $A_f$ that controls $u_{f,max}$ cancels out for the definition of the non-dimensional frequency.
Note that $S$ represents the equivalent to the frequency parameter $M^2 = D_p^2 \Omega / (4 \nu_f) = 9 S$ that is often used in  literature  \citep[e.g.][]{1966_Riley, 1967_Riley}.
As stated in the Introduction, \cite{1966_Riley} distinguished two cases for a single particle to characterize the  flow pattern of the steady streaming under certain flow parameters:
(i) for $\Rey \ll 1$ and $S \ll 1/9$ the area over which the primary vortices propagate is much larger than $D_p/2$ and (ii) for $S \gg 1/9$ the primary vortices are confined to the particle surface and adjacent secondary vortices propagate into the far field.
Although the focus of this study is on the oscillating behavior of two particles, this classification is still useful for describing the obtained flow fields.

Moreover, we define a Stokes number $St=\tau_p / \tau_f$ using the characteristic particle time scale $\tau_p = | \rho_s - 1 | D_p^2 / (18 \nu_f)$ and the characteristic flow time scale $\tau_f = 1 / \Omega$, which yields $St = | \rho_s - 1 | 2 S$.
Since $St$ is determined by the linear combination of the density ratio $\rho_s$ and the dimensionless frequency $S$, we focus in the following on the influence of these two non-dimensional parameters on the behavior of the particles by varying the initial gap size $\zeta_i$.

In what follows, all equations, results and quantities are presented in dimensionless form unless stated otherwise. 
Therefore, we drop the tilde symbol for the remainder of the article.
In addition, to guarantee sufficient spatial resolution for the parameter space presented here, we discretize the domain by a uniform rectangular grid of a grid cell size $h = L / 200$ resulting in $D_p / h = 20$, if not stated otherwise. 
The timestep is chosen as $T / 1000$, where $T$ is the oscillation period, to also guarantee a high temporal resolution of the fluid motion.
We verified by means of test simulations that the particle motions and their mutual interaction are not affected by numerical parameters such as the timestep, spatial discretization and domain size.
Note that preliminary tests have shown that doubling the domain size only has an impact on the order of $\mathcal{O}(10^{-3})$ on the particle motion and is therefore negligible.

%%%%%%%%%%%%%%%%%%%%%%%%%%%%%%%%%
\subsection{Comparison to benchmark data} \label{sec:validation}
%%%%%%%%%%%%%%%%%%%%%%%%%%%%%%%%%

The numerical code used in the present study has already been validated and used for detailed studies of fluid-particle and particle-particle interactions, such as the rheological behavior of sediment beds under shear and  particles settling under gravity in quiescent fluids \citep{2017_Biegert_etal,2019_Vowinckel_etal,2021_Vowinckel_etal}. 
As a most relevant case for our scenario, \cite{2017_Biegert_etal} have obtained excellent agreement for a sphere settling in a large container \citep{2000_Mordant_Pinton} and for a settling sphere approaching a wall \citep{2002_tenCate_etal}. 
Since the setup under investigation considers two particles in an oscillating domain filled with fluid, we extend our validation to setups with oscillating motion patterns.
In this regard, we compare our computational results to analytical, experimental and numerical data of single- and two-particle setups in which either the fluid or the particles oscillate.

\begin{figure}
    \def\stackalignment{l}
    \centering
    \captionsetup[subfigure]{labelformat=empty}
    % Top
    \begin{subfigure}[b]{\textwidth}
         \centering
         \topinset{{\scriptsize (a)}}{\includegraphics[trim=12.5cm 2.7cm 14cm 1.9cm, clip,width=0.49\textwidth]{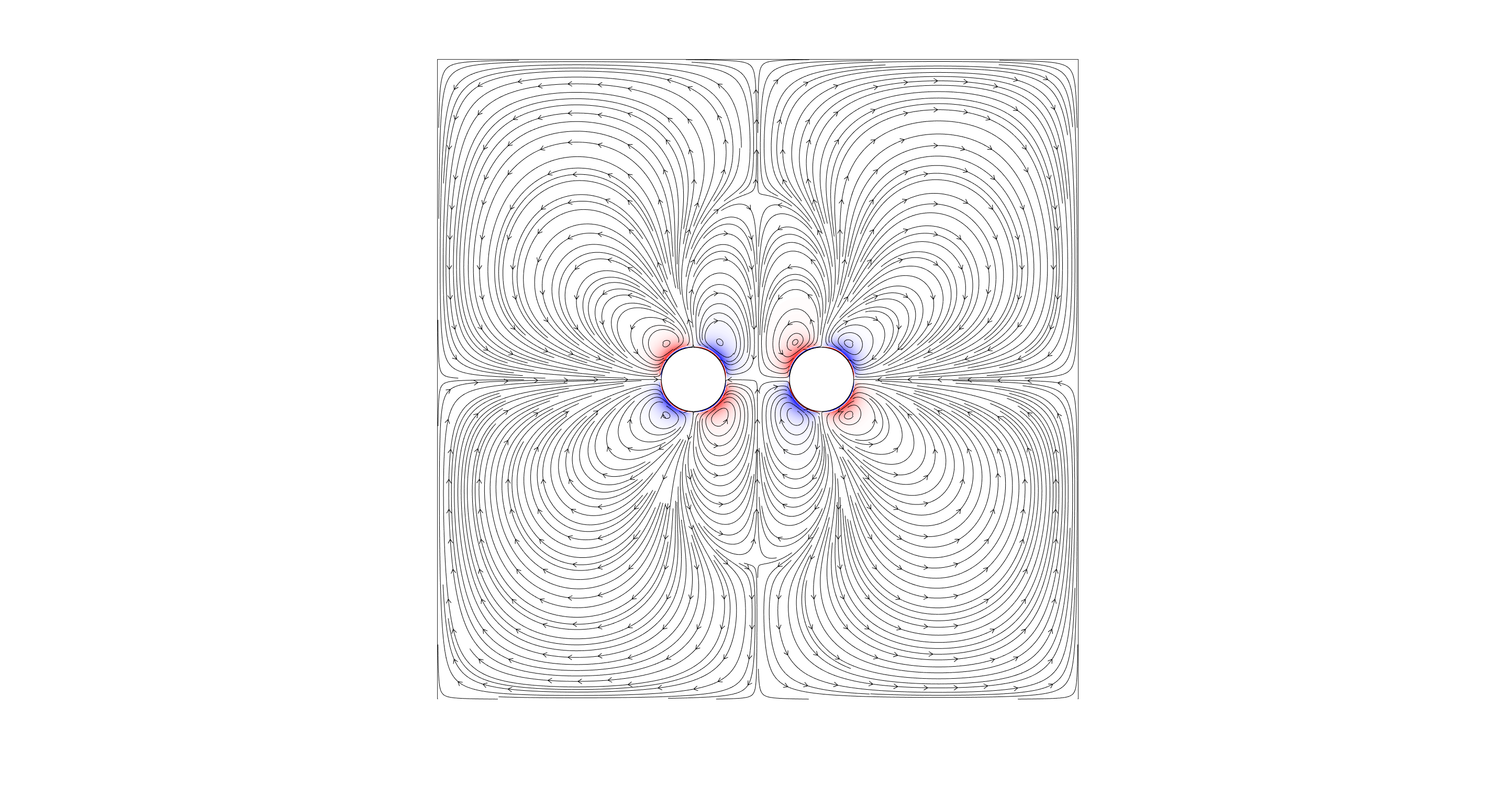}}{0.2cm}{0.05cm}
         \topinset{{\scriptsize (b)}}{\includegraphics[trim=6.8cm 3.9cm 7.4cm 2.6cm, clip,width=0.49\textwidth]{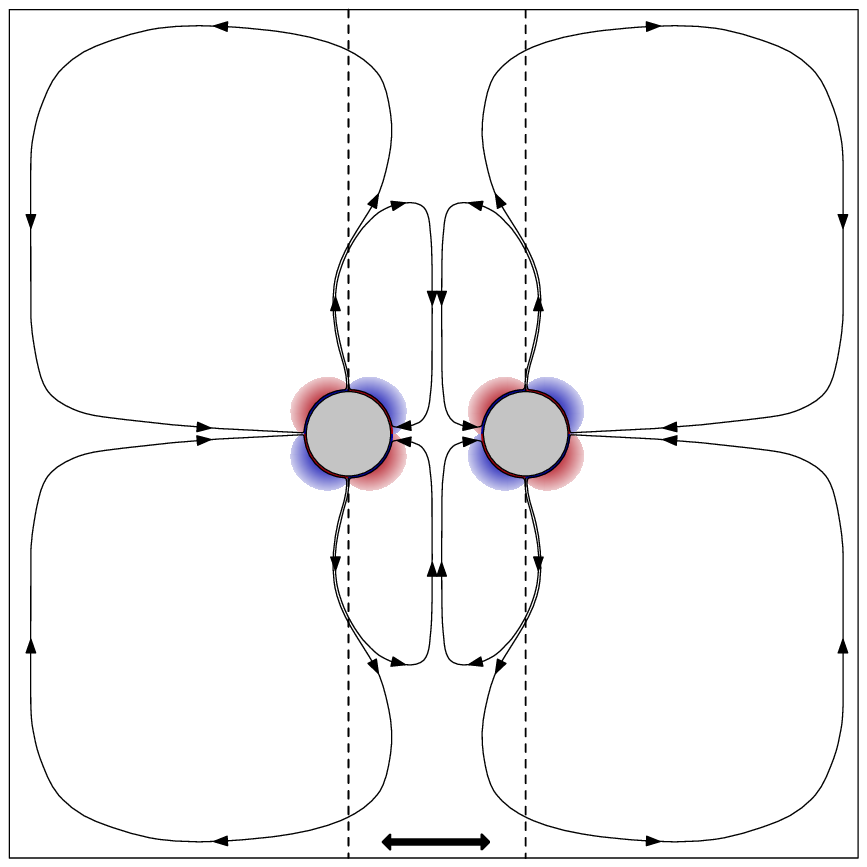}}{0.2cm}{0.05cm}
         \put(-5.3cm,5.75cm){{\smaller peripheral}}
         \put(-3.49cm,5.75cm){{\smaller central}}
         \put(-2.05cm,5.75cm){{\smaller peripheral}}
    \end{subfigure}
    \begin{subfigure}[t]{\textwidth}
        \centering
         \topinset{}{\includegraphics[width=0.65\textwidth]{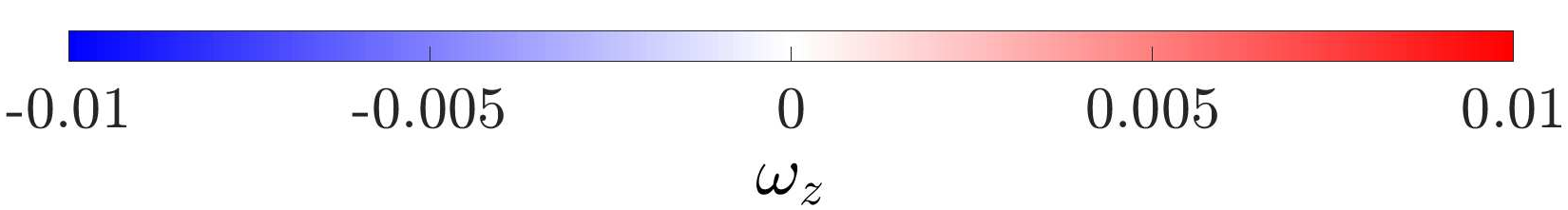}}{0.2cm}{0.05cm}
    \end{subfigure}
    \caption{Flow characteristics illustrated by streamlines and vorticity contours for steady streaming of two stationary and axially arranged particles in oscillatory flow. 
    (a) represents the numerical results of the setup with $S = 0.89$, $\Rey = 0.32$ and $\Rey_s = 0.0032$ and (b) a simplified sketch to highlight the quadrupole structures surrounding each particle as well as the direction of the imposed oscillation by the black arrows. 
    The vorticity coloring ranges from $-0.01$ (blue) for clockwise rotations to $0.01$ (red) for counterclockwise rotations.}
    \label{fig:FlowCharacteristics_TwoParticles}
\end{figure}

In a first step, we validate the excursions of the particle trajectories as well as the qualitative and quantitative streamline patterns for a setup with a single particle with benchmark data.
The results and thorough descriptions of the validation conditions are provided in \mbox{appendix \ref{app:appendixA}}.
In a next step, we examine the characteristics of the flow field generated by two stationary particles of equal diameter arranged in alignment with an unidirectional, monochromatic oscillatory fluid flow.
The particles are horizontally aligned, i.e. $\theta_i = 0^{\circ}$, and placed with a surface separation distance of $\zeta_i = \zeta_t = 1.00$ so that the center of the particle pair coincides with the center of the numerical domain.
The dimensions of the numerical domain are the same as described in \S \ref{sec:numerical_setup} with a cubic box of size \mbox{$L_{x,y,z} = L = 10$}.
The properties of the applied oscillation are $S = 0.89$, $\Rey = 0.32$ and $\Rey_s = 0.0032$. 
The flow characteristics of the steady streaming, which has been computed after $100 T$, are presented by the streamlines and vorticity contours in figure~\ref{fig:FlowCharacteristics_TwoParticles}(a).
The vorticity contours are based on the calculation of the polar component of the vorticity $\omega_z\left(x,y \right)$ and the coloring is according the presented color bar, which ranges from $-0.01$ (blue) for clockwise rotations to $0.01$ (red) for counterclockwise rotations.
The sketch in figure~\ref{fig:FlowCharacteristics_TwoParticles}(b) represents the fluid-particle interaction in a simplified form, highlighting that each particle is surrounded by four vortex structures, as indicated by the streamlines as well as the vorticity contours.
These flow patterns are usually referred to as quadrupoles \citep{1998_Longuet-Higgins, 2007_Tho_etal}.
As emphasized by \cite{2009_Klotsa_PhD}, the appearance of these quadrupole structures is due to the fact that the presented illustrations are planes through three-dimensional toroidal vortex structures surrounding each particle.
These rotate in such a way that the flow is directed toward the respective particle along the axis of oscillation and away from it perpendicular to the oscillation, as indicated in figure \ref{fig:FlowCharacteristics_TwoParticles}.
Similar flow properties of such planes have already been reported by \cite{2017_Fabre_etal} and \cite{2018_Jalal} for two stationary spheres and by \cite{2007_Tho_etal} for two bubbles attached to a wall in oscillating fluid flow.

With reference to the case distinctions for a single-particle setup according to \cite{1966_Riley} (cf. \S \ref{sec:governingEqs}), we note that the present case would represent an intermediate regime, as it could not be assigned to either the first (small $Re$ and small $S$) or the second class (large $S$).
The pattern of the steady streaming, on the other hand, would comply with the conditions for class one, as it apparently consists only of the primary vortices.
While the representation of the vorticity contours are only apparent in the vicinity of the respective particles, indicating that the highest vorticity is located close to particle surfaces, the vortex structures presented by the streamlines extend over the entire domain revealing the characteristic quadrupole pattern.
The extension as well as the shape of the vortex structures is limited by the walls of the numerical domain.
However, since the fluid velocities farther away from the particles rapidly decay to low values, their effects on the particle dynamics become negligible if the particles are relatively close to each other.
Nevertheless, depending on the properties of the oscillation and the particles, the characteristics of the vorticity and vortex structure change.
A more detailed analysis of the impact of these properties on the flow characteristics as well as an explanation of the division into central and peripheral sections is given in \S \ref{sec:flow_field_analysis} below.

In summary, we observe that our simulation approach is capable of reproducing the relevant flow features for two fixed spheres in oscillating fluid motion.
Considering the additional validation of the setups with a single sphere (appendix \ref{app:appendixA}), we conclude that the computational approach is suitable for detailed analyses of the interaction of two particles in oscillatory fluid flow.

%%%%%%%%%%%%%%%%%%%%%%%%%%%%%%%%%
%%%%%%%%%%%%%%%%%%%%%%%%%%%%%%%%%
\section{Particle dynamics}\label{sec:inter-particlebehavior}
%%%%%%%%%%%%%%%%%%%%%%%%%%%%%%%%%
%%%%%%%%%%%%%%%%%%%%%%%%%%%%%%%%%

%%%%%%%%%%%%%%%%%%%%%%%%%%%%%%%%%
\subsection{Initial Motion}\label{sec:initialMotion}
%%%%%%%%%%%%%%%%%%%%%%%%%%%%%%%%%

When the fluid container starts moving, the first oscillation is to a given direction, in this case to the right as determined by the way the oscillation is performed.
For an example of two particles with a density greater than that of the liquid, i.e. $\rho_s > 1$, the inertia of the particles is larger than that of the surrounding fluid.
Since we solve for particle motion in a non-inertial frame, both particles shift to the left for this case.
At the start of the simulation, the fluid is at rest and the particles do not experience a counterflow at the beginning of their movement. This condition results in a relatively large excursion.
After the reverse of the oscillation of the domain in the opposite direction (now to the left), the particles denser than the fluid tend to move again to the opposite direction (to the right).
However, after this initial oscillation period they encounter the previously created flow field induced by the fluid-particle interaction, which still points to the preceding direction of movement. 
As a consequence, the particle motion and, hence, the excursion are slowed down.

\begin{figure}
    \def\stackalignment{l}
    \centering
    \captionsetup[subfigure]{labelformat=empty}
    % Top
    \begin{subfigure}[b]{\textwidth}
         \centering
         \topinset{{\scriptsize (a)}}{\includegraphics[trim=2cm 0.3cm 4cm 1.3cm, clip,width=0.8\textwidth]{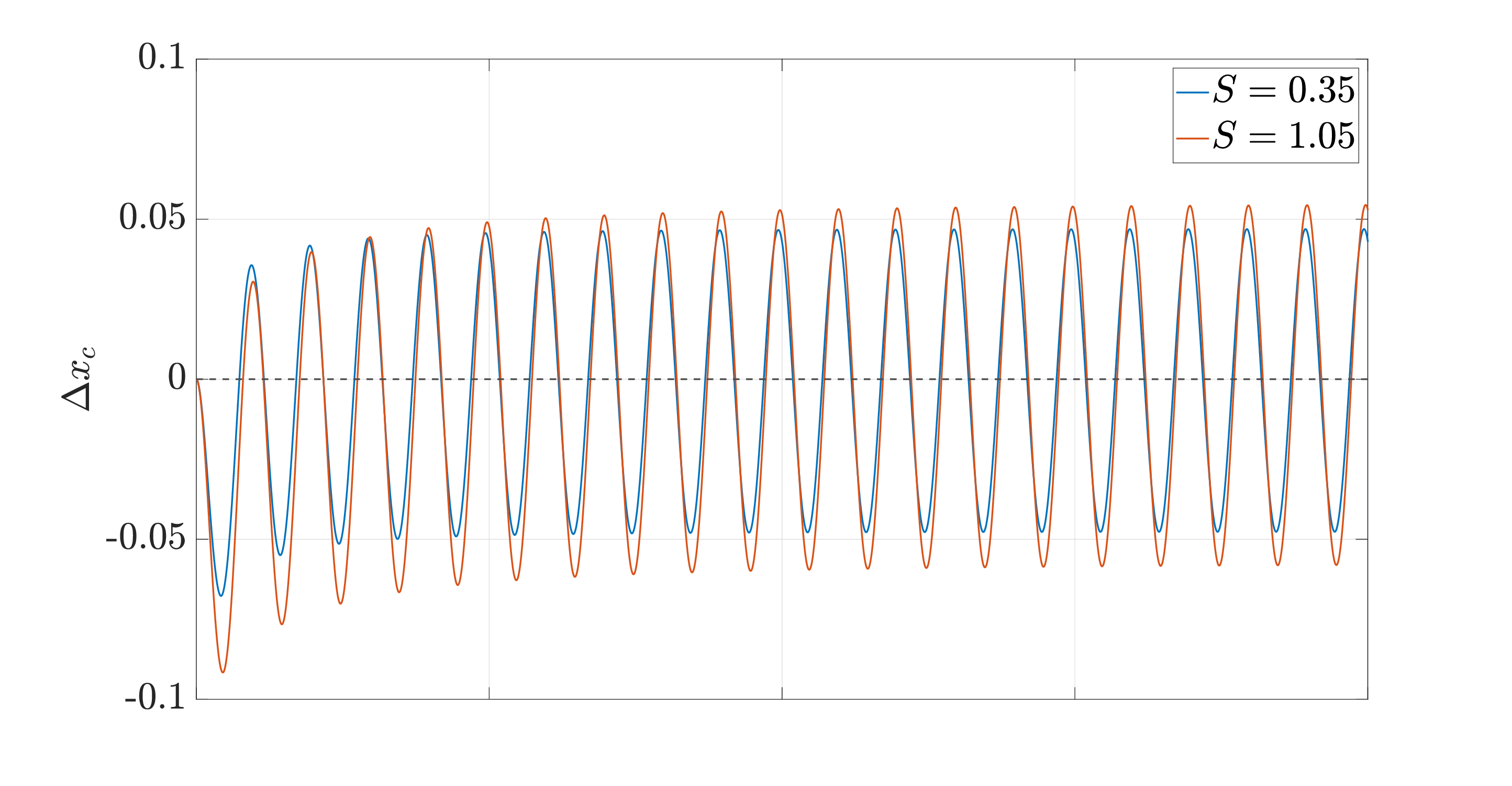}}{0.3cm}{0cm}
         \topinset{{\scriptsize (b)}}{\includegraphics[trim=2cm 0.3cm 4cm 1.3cm, clip,width=0.8\textwidth]{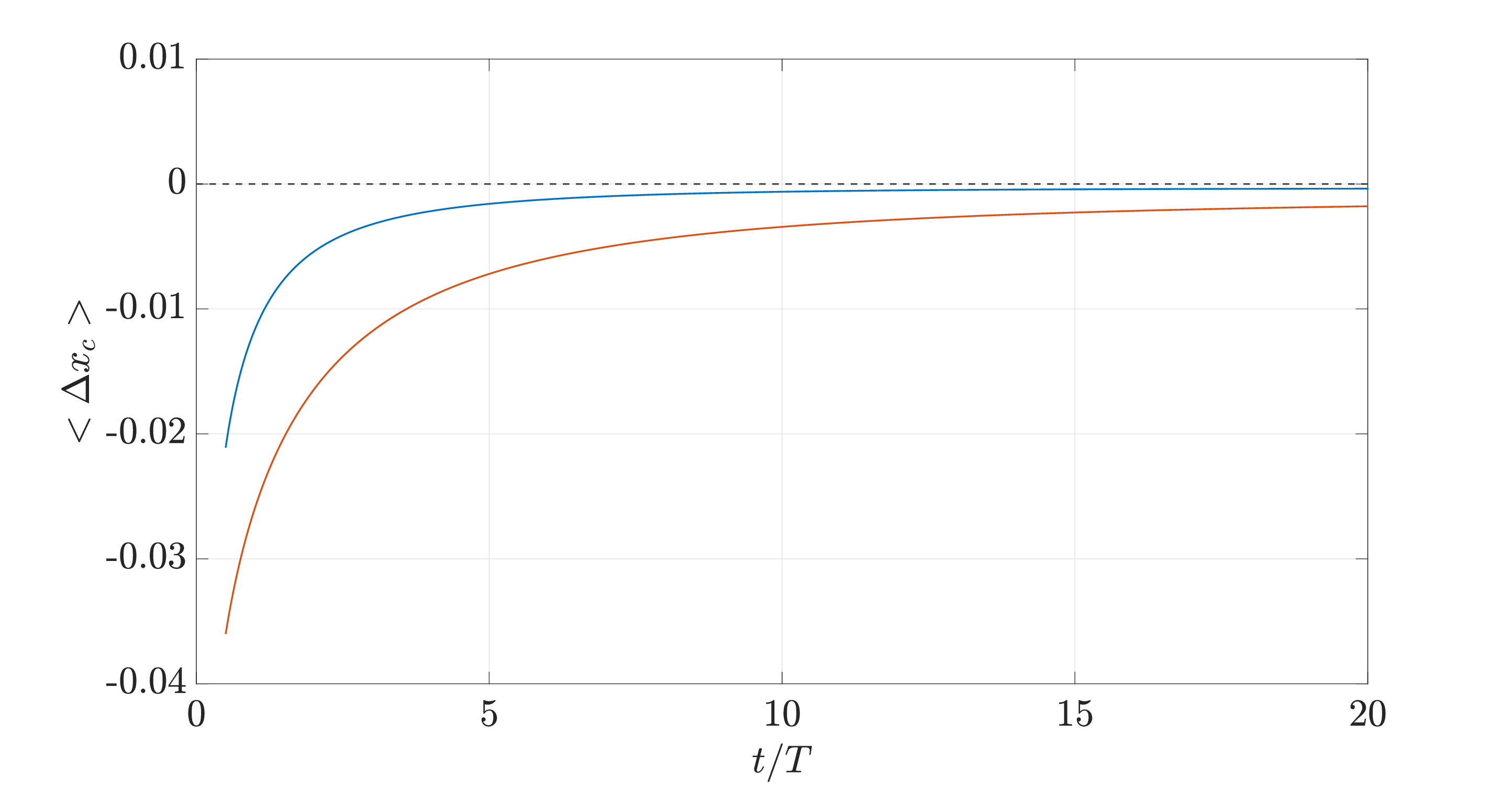}}{0.3cm}{0cm}
    \end{subfigure}
    \caption{Location of the midpoint between the particles relative to its initial location over time for (a) the instantaneous location $\Delta x_c$ and (b) the moving average $<\Delta x_c>$ with an averaging window equal to the oscillation period $T$.}
\label{fig:initial_motion} 
\end{figure}

As this initial displacement has now established an offset, in which the distance of both particles to the left wall of the domain is slightly smaller than to the right, the system thenceforth recovers by having the two particles drifting back towards the center of the domain. 
The particle pair approaches the center of the domain asymptotically so that for a horizontally aligned arrangement ($\theta_i = 0^{\circ}$) with $\rho_s = 4.68$ and $\zeta_i = 0.75$, the artifact from the initial conditions is compensated after approximately $20$ oscillation periods. 
This can be seen for the examples of $S = 0.35$ and $1.05$ in figure \ref{fig:initial_motion}, where figure~\ref{fig:initial_motion}(a) shows the instantaneous location of the center between the particles in relation to its initial position over time and figure \ref{fig:initial_motion}(b) illustrates the evolution of the inter-particle center over time as moving average with an averaging window of one oscillation period.

%%%%%%%%%%%%%%%%%%%%%%%%%%%%%%%%%
\subsection{Inter-particle distance over time}\label{sec:inter-particleDistance}
%%%%%%%%%%%%%%%%%%%%%%%%%%%%%%%%%

Since we could see in \S \ref{sec:initialMotion} that the influence of the initial condition of fluid at rest becomes negligible after a few oscillation periods, we can now analyze the near-field and far-field hydrodynamic interactions of two particles aligned with $\theta_i = 0^{\circ}$ for various initial particle distances $\zeta_i$ and oscillation frequencies $S$ over longer simulation times.
Moreover, we investigate the particle dynamics as a function of particle inertia expressed by the density ratio $\rho_s$. 

Since we neglect gravity and consider two particles with an initial orientation angle $\theta_i = 0^{\circ}$, the  particle coordinates remain constant in $y$- and $z$-direction at $L_y/2$ and $L_z/2$, respectively.
The initial positions in $x$-direction are modified to result in $\zeta_i = \{0.10, 0.25, 0.50, 0.75, 1.00, 1.50, 2.00, 3.00\}$. 
For all these runs, we use a spatial discretization of $D_p / h = 20$, as given in \S\ref{sec:numerical_setup}, except for $\zeta_i = 0.10$ where we employ an even higher resolution with $D_p / h = 40$ to prevent the lubrication model of \cite{2017_Biegert_etal} to become active for $\zeta_{t}<2h$.
These choices also allowed for a proper resolution of the particle gap even for the arrangement with a very small gap width.
For the following analyses, oscillations in a range of $S = [0.07, 2.09]$ with a constant amplitude of $\epsilon = 0.1$ were applied for a total number of $100$ oscillation periods.

\begin{figure}
    \centerline{\includegraphics[trim=2.0cm 0.4cm 3.1cm 1.3cm, clip,width=0.8\textwidth]{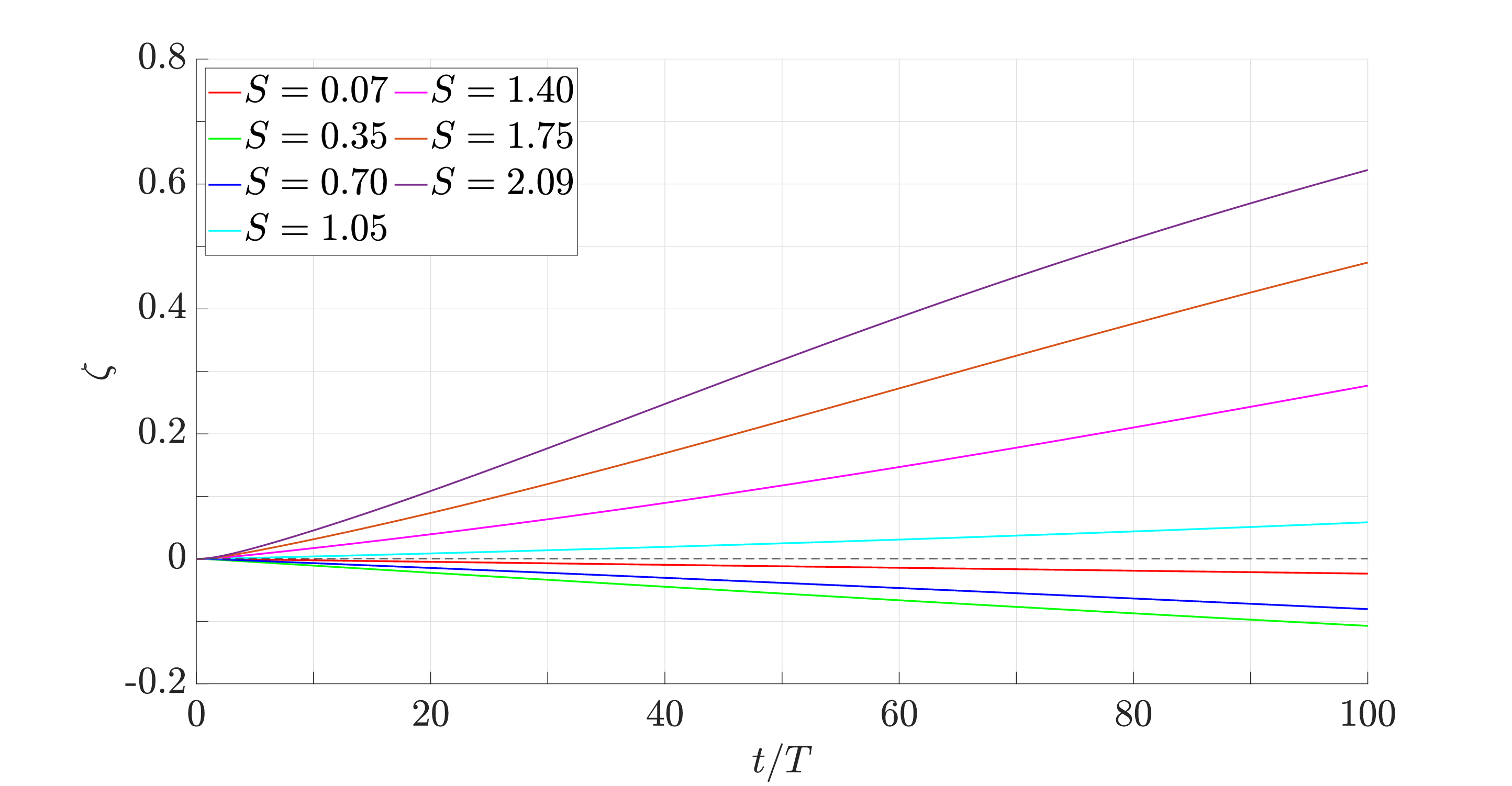}}% Images in 100% size
    \caption{Evolution of instantaneous $\zeta$ over $100$ oscillation periods for $\rho_s = 4.68$, $\zeta_i = 0.25$ and a variety of $S$.}
    \label{fig:zeta_varyingS}
\end{figure}

Figure \ref{fig:zeta_varyingS} illustrates the instantaneous evolution of $\zeta$ over time for various choices of $S$ for an representative setup with initial particle distance $\zeta_i = 0.25$ and $\rho_s = 4.68$.
The horizontal dashed line represents the respective state of equilibrium, in which $\zeta$ does not change with time. 
As can be seen from the results, different $S$ lead to different particle dynamics in terms of the change of the particle distance $\zeta$ despite identical initial conditions. 
In general, results with increasing values of $\zeta > 0$ over time show that the particles drift apart in a repulsive behavior.
Vice-versa, values of  $\zeta < 0$ indicate that the particles are approaching each other, representing an attractive motion.

\begin{figure}
    \def\stackalignment{l}
    \centering
    \captionsetup[subfigure]{labelformat=empty}
    % Top
    \begin{subfigure}[b]{\textwidth}
         \centering
         \topinset{{\scriptsize (a)}}{\includegraphics[trim=2cm 0.1cm 3.1cm 1.3cm, clip,width=0.8\textwidth]{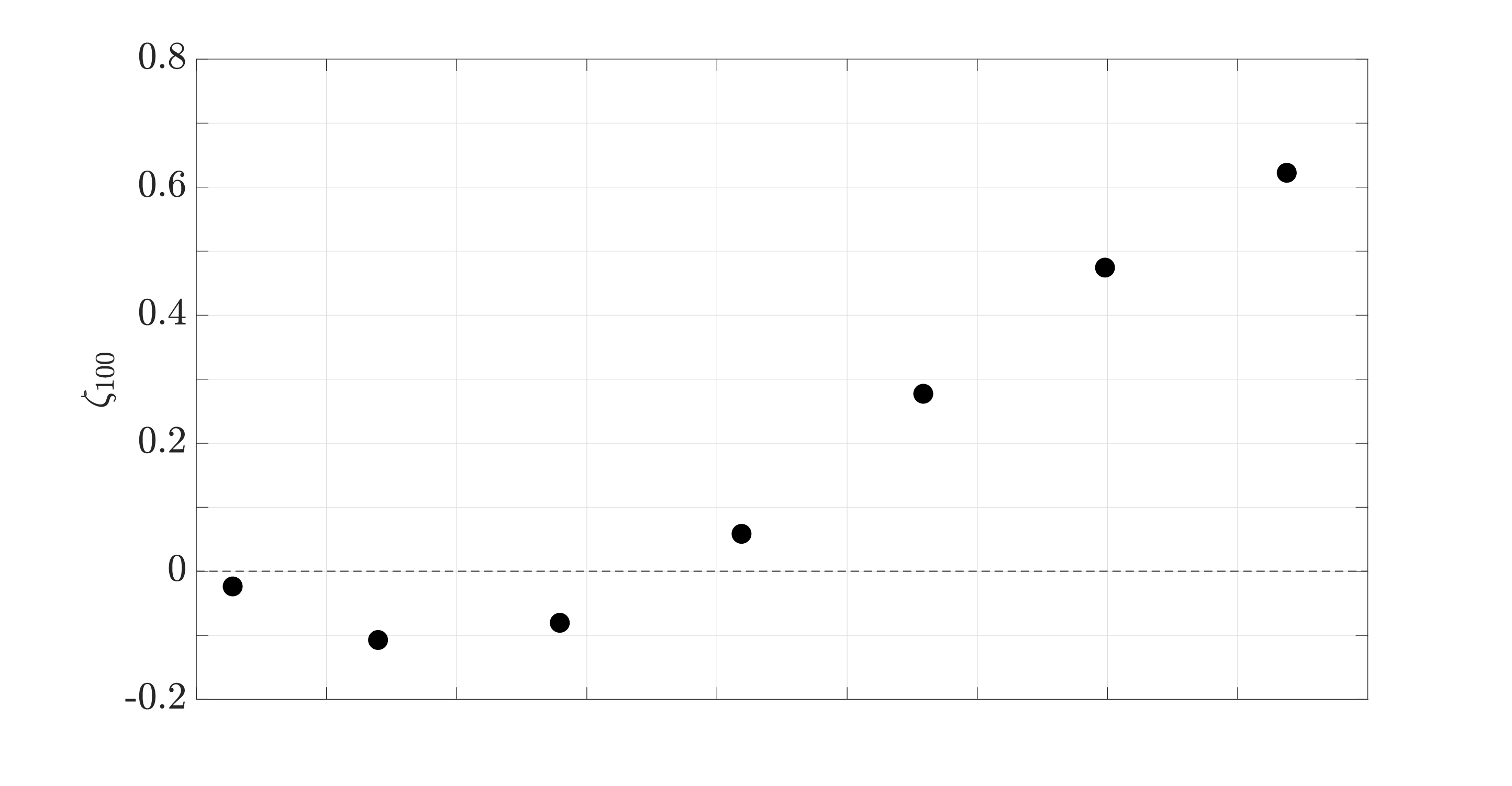}}{0.3cm}{0cm}
         \topinset{{\scriptsize (b)}}{\includegraphics[trim=2cm 0.1cm 3.1cm 1.3cm, clip,width=0.8\textwidth]{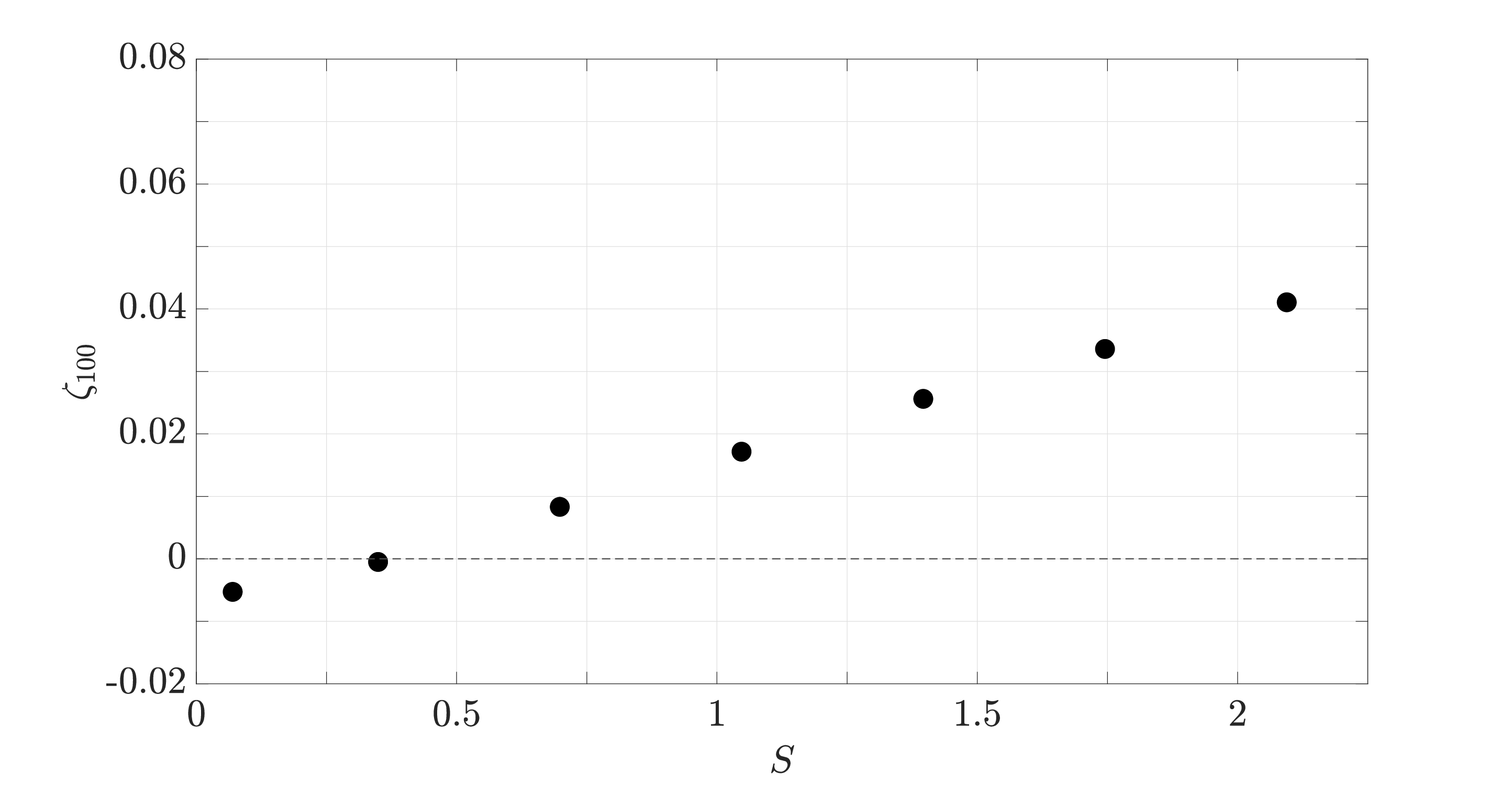}}{0.3cm}{0cm}
    \end{subfigure}
    \begin{tikzpicture}[overlay]
    \node[white,fill=white,inner sep=0] at (-2.9,11.4) {\includegraphics[width=0.175\textwidth]{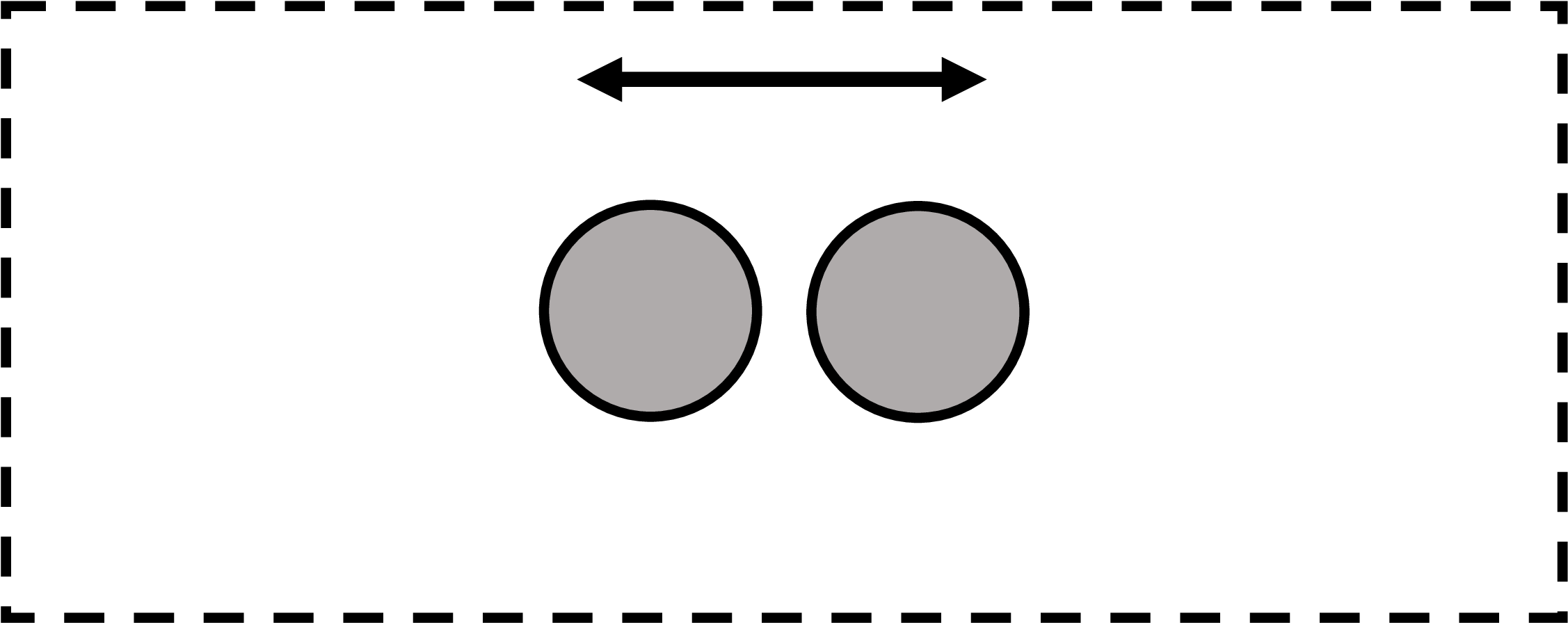} };  
    \node[white,fill=white,inner sep=0] at (-2.9,5.46) { % Adjust the coordinates (2,1) as needed
      \includegraphics[width=0.175\textwidth]{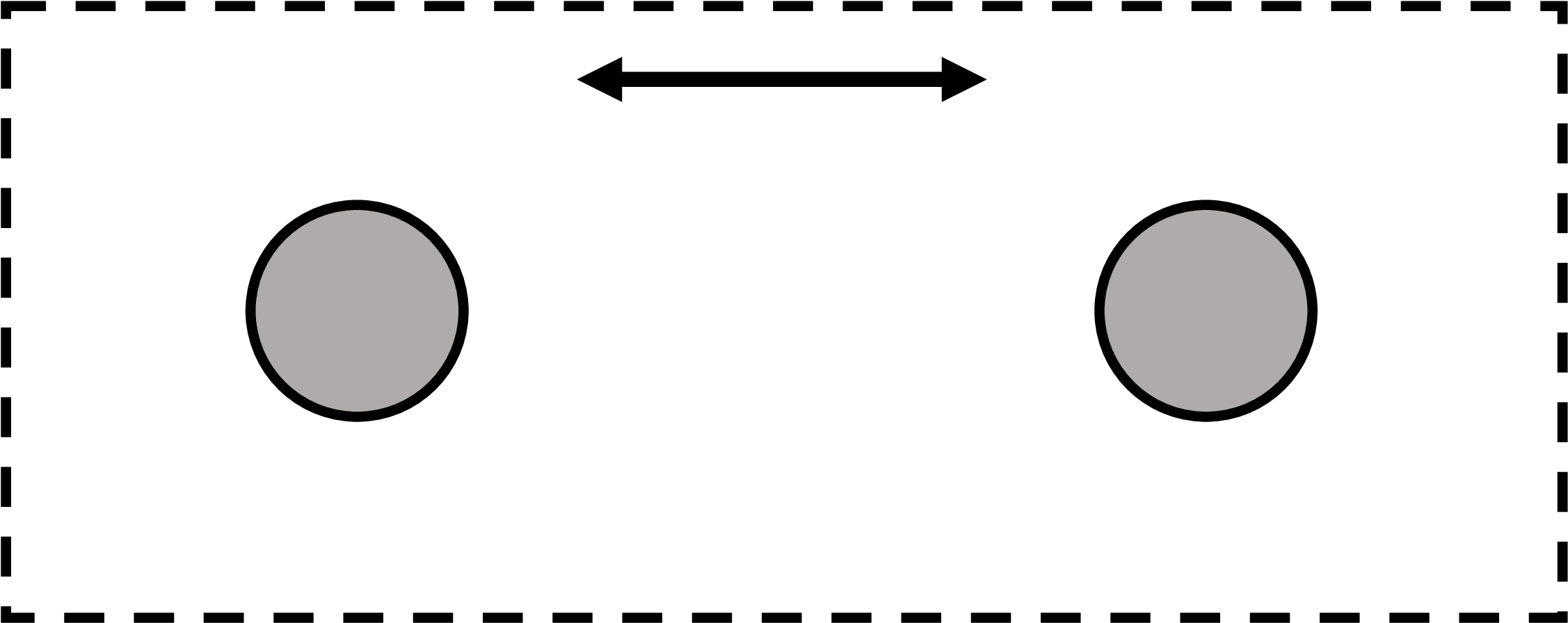} };
    \end{tikzpicture}
    \caption{$\zeta_{100}$ as a function of $S$ with $\rho_s=4.68$ for (a) $\zeta_i = 0.25$ and (b) $\zeta_i = 3.00$. The sketches in the subfigures illustrate the initial configuration and the direction of oscillation (black arrows). Note the difference in scale by one order of magnitude for the $y$-axes of the two figures. $\zeta_{100} < 0$ represents attraction and $\zeta_{100} > 0$ repulsion.}
\label{fig:zeta100_varyingS} 
\end{figure}

Since we have already seen that $S$ has a direct correlation to the resulting behavior of the particles, we will investigate $\zeta$ more systematically.
Here, $\zeta$ represents the change in the particle distance over time, computed as $\zeta = \zeta_{t} - \zeta_i$.
Therefore, we define $\zeta_{100}$ as the change of the particle distance $\zeta$ relative to its initial value after $100 \: T$, i.e. 100 oscillation periods, to provide a quantitative measure for the particle dynamics.
In this context, we also compare the results for different $\zeta_i$ to better quantify the effects of this configuration parameter.
In Figure \ref{fig:zeta100_varyingS}, we present $\zeta_{100}$ for the same values of $S$ as in figure \ref{fig:zeta_varyingS}.
In addition to the results for $\zeta_i = 0.25$ (figure \ref{fig:zeta100_varyingS}a), we also show $\zeta_{100}$ for $\zeta_i = 3.00$ in figure \ref{fig:zeta100_varyingS}(b).
Both figures reveal that the repulsive and attractive behavior becomes more pronounced for larger and smaller values of $S$, respectively. 
An exception can be seen for $S=0.07$ with $\zeta_i=0.25$ in figure \ref{fig:zeta100_varyingS}(a). 
This is caused by the work required to move the fluid inside the gap for particle pairs at very small $\zeta_t$, i.e. lubrication forces.  
Such behavior indicates that there is a peak for small initial gap sizes where the attracting behavior is most efficient. 
For the case of $\zeta_i=0.25$, that peak was determined at about $S = 0.35$.  
In addition, we found that the dynamics of the particles not only depend on $S$, but also on $\zeta_i$.
Comparing the two initial gap sizes $\zeta_i=0.25$ and $\zeta_i=3.00$ in figures \ref{fig:zeta100_varyingS}(a) and (b), we can conclude that the particle-interaction becomes less intense for larger $\zeta_i$.
Note the scale difference by one order of magnitude in figures \ref{fig:zeta100_varyingS}(a) and (b). 
For $\zeta_i=3.00$ in Figure \ref{fig:zeta100_varyingS}(b), the damping of particle motion toward each other by lubrication completely vanishes for low frequencies and the threshold for $S$ for the transition from attractive to repulsive behavior shifts to smaller values of $S$. 

\begin{figure}
    \centerline{\includegraphics[trim=2.0cm 0.1cm 3.1cm 1.3cm, clip,width=0.8\textwidth]{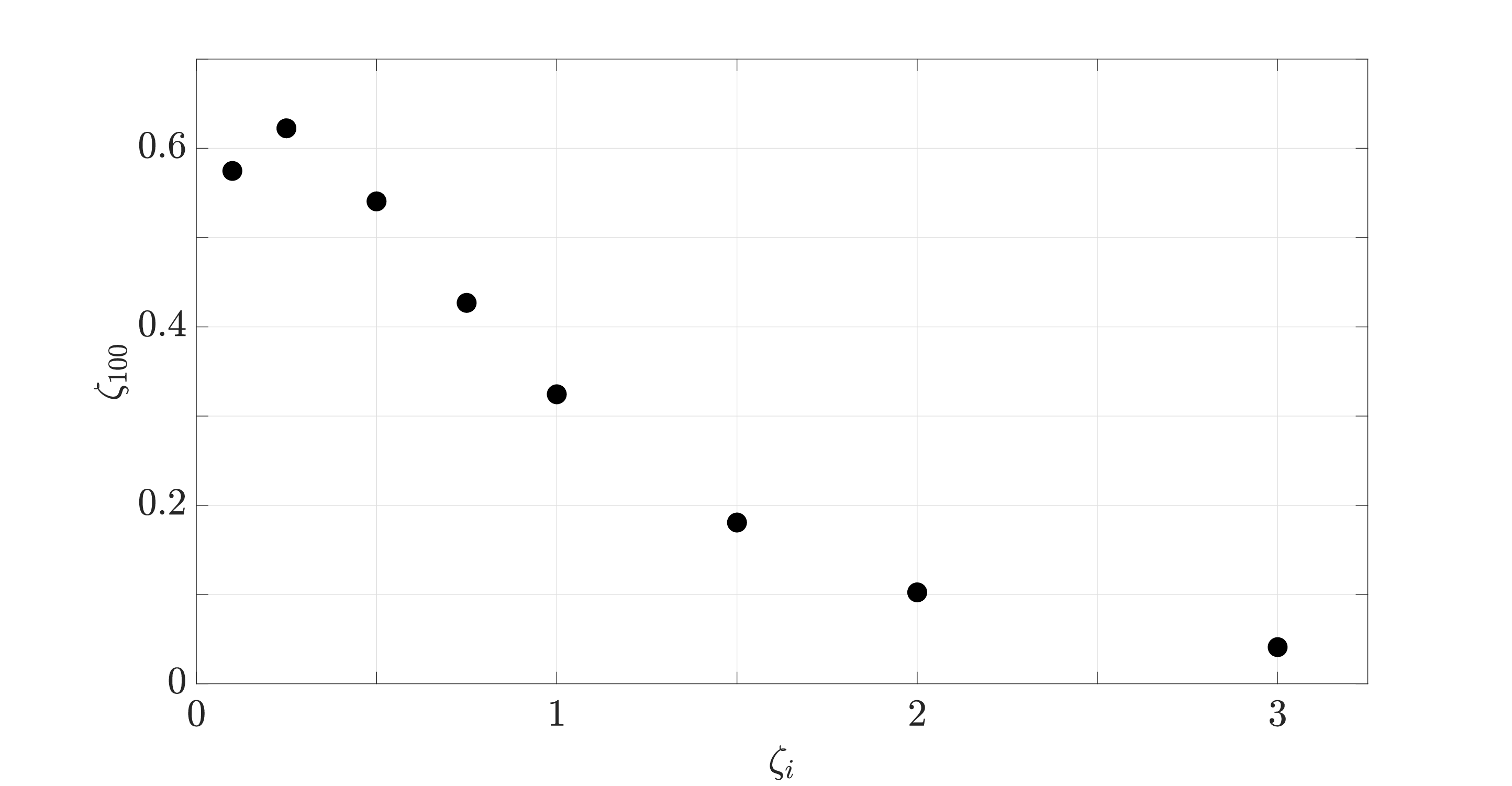}}% Images in 100% size
    \caption{$\zeta_{100}$ as a function of $\zeta_i$ with $\rho_s=4.68$ and for $S = 2.09$.}
    \label{fig:zeta100_overZeta_i}
\end{figure}

The analysis for particles of relative density $\rho_s=4.68$ presented so far suggests that there is not only an optimum for attractive particle interaction for $\zeta_i\approx0.25$, but  also a threshold condition for which the interaction transitions from attractive to repulsive behavior.
Both of these phenomena are a function of $S$ and $\zeta_i$. 
To elucidate the effect of $\zeta_i$ on the particle behavior further, we evaluate $\zeta_{100}$ for a constant frequency ($S=2.09$) and different values of $\zeta_i$ in figure~\ref{fig:zeta100_overZeta_i}.
Again, we use an initial orientation of $\theta_i = 0^{\circ}$ and a relative particle density of $\rho_s=4.68$ for this analysis.
For such high frequencies, all presented cases result in repulsion. 
These results demonstrate that the increase of the particle distance decays exponentially for larger values of $\zeta_i$. 
Again, the case of $\zeta_i = 0.10$ deviates from this behavior, because for such small gaps, lubrication forces become an important component of the particle interaction.
Nevertheless, for \mbox{$\zeta_i>0.1$}, the intensity of the mutual interaction decreases with increasing $\zeta_i$ and vanishes for large initial particle distances.
This is in line with the general assumption by \cite{2017_Fabre_etal} that the closer the particles are to each other, the more they are able to influence one another by interacting with each other's flow structures. 
Conversely, this means that the further apart the particles are, the less intense the interaction becomes and the particle motion resembles that of an individual particle as $\zeta_i$ increases. 

\begin{figure}
    \centerline{\includegraphics[trim=3.0cm 0.5cm 4.1cm 2.0cm, clip,width=0.8\textwidth]{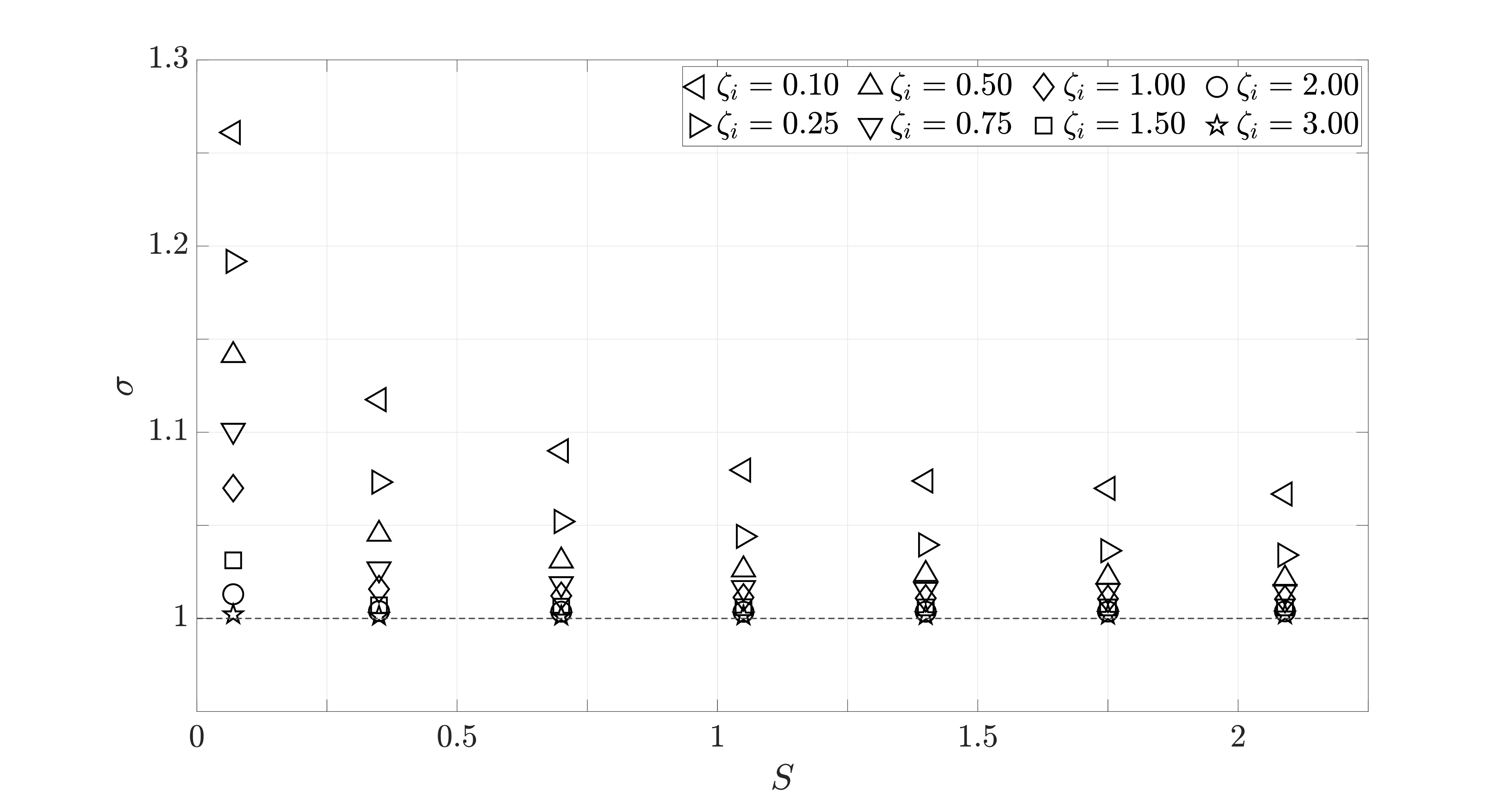}}% Images in 100% size
    \caption{Amplitude ratio $\sigma = A_{2p}(S,\zeta_i) / A_{1p}(S)$ between the amplitudes of the two-particle setups $A_{2p}(S,\zeta_i)$ and the respective individual-particle setups $A_{1p}(S)$ for \mbox{$\rho_s = 4.68$} and as a function of $S$ and $\zeta_i$.}
    \label{fig:ExcursionRatio}
\end{figure}

This behavior is exemplified by the comparison of the particle amplitudes of the two-particle setups $A_{2p}(S,\zeta_i)$, in dependence of $\zeta_i$, and the setups with individual particles $A_{1p}(S)$ for the same values of $S$ in the non-inertial frame of reference.
The comparison is shown by the ratio $\sigma = A_{2p}(S,\zeta_i) / A_{1p}(S)$ for $T = 11$ in figure \ref{fig:ExcursionRatio}.
The time for the calculation of $\sigma$ was chosen so that the particle arrangements still correspond as closely as possible to their initial state.
The decision for this instant is further explained in \S \ref{sec:totalSteadyStreaming}.
The comparison reveals that the deviation of the particle amplitudes between the two-particle and the individual-particle setups increases with decreasing $\zeta_i$.
This is in line with the observation and assumption made above that for the given setup the interaction of two particles increases with decreasing $\zeta_i$.
As an important consequence, this analysis also demonstrates that the closer the particles are, the more amplified their excursion lengths become. 

%%%%%%%%%%%%%%%%%%%%%%%%%%%%%%%%%
\subsection{Particle arrangements} \label{sec:particleArrangements}
%%%%%%%%%%%%%%%%%%%%%%%%%%%%%%%%%

\begin{figure}
    \def\stackalignment{l}
    \captionsetup[subfigure]{labelformat=empty}
    \begin{subfigure}[b]{\textwidth}
        \centering
        \topinset{}{\includegraphics[trim=0cm 0.5cm 2cm 2.2cm, clip,width=0.9\textwidth]{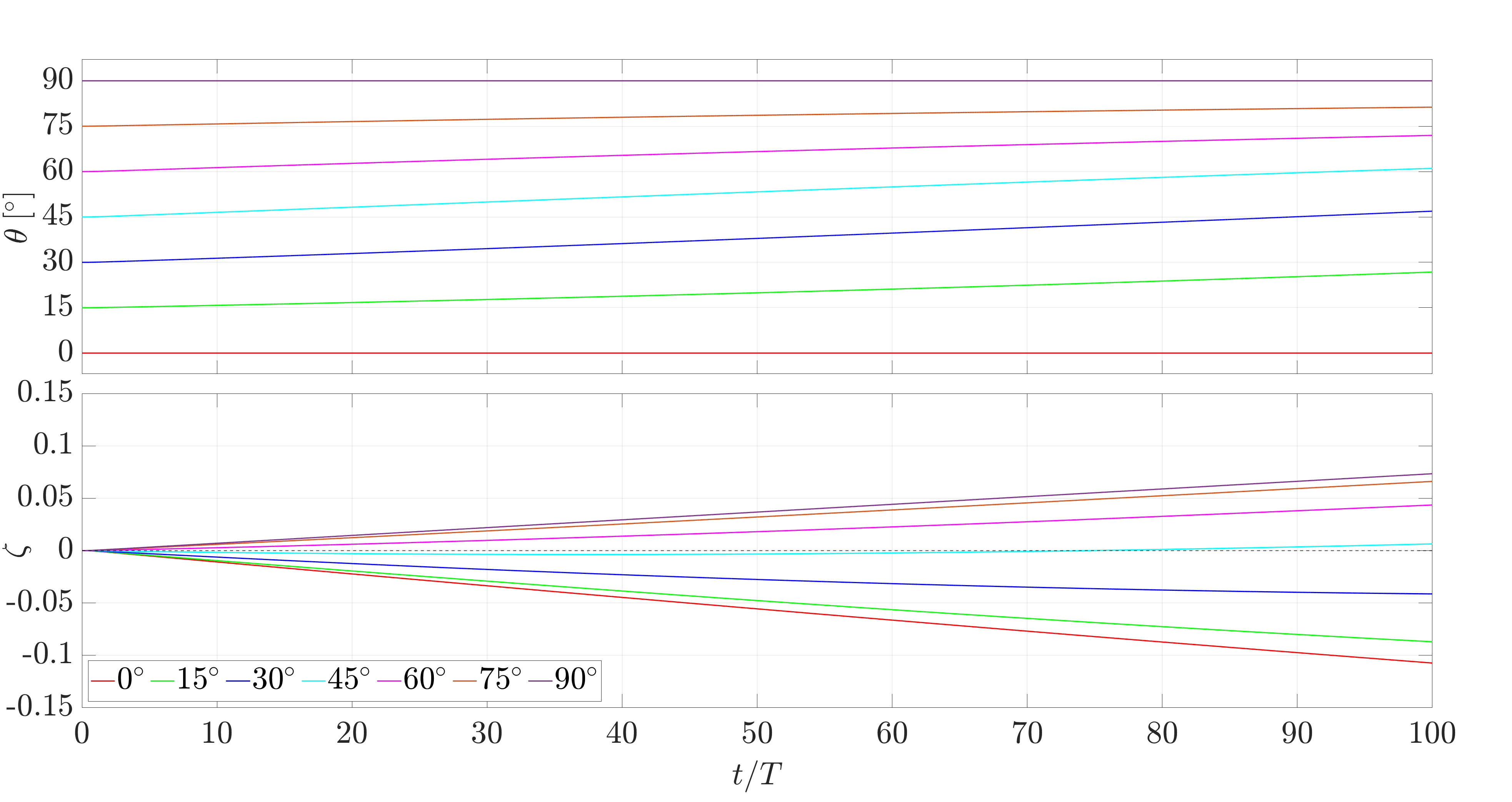}}{0cm}{0cm}% Images in 100% size
        \put(-12.45cm,5.75cm){{\scriptsize(a)}}
        \put(-12.45cm,3cm){{\scriptsize(b)}}
    \end{subfigure}
    \caption{Evolution of (a) $\theta$ and (b) $\zeta$ over $100$ oscillation periods for $\rho_s = 4.68$, $\zeta_i = 0.25$ and $S = 0.35$.}
    \label{fig:ShiftedParticles_S_0d35}
\end{figure}

\begin{figure}
    \def\stackalignment{l}
    \captionsetup[subfigure]{labelformat=empty}
    \begin{subfigure}[b]{\textwidth}
        \centering
        \topinset{}{\includegraphics[trim=0cm 0.5cm 2cm 2.2cm, clip,width=0.9\textwidth]{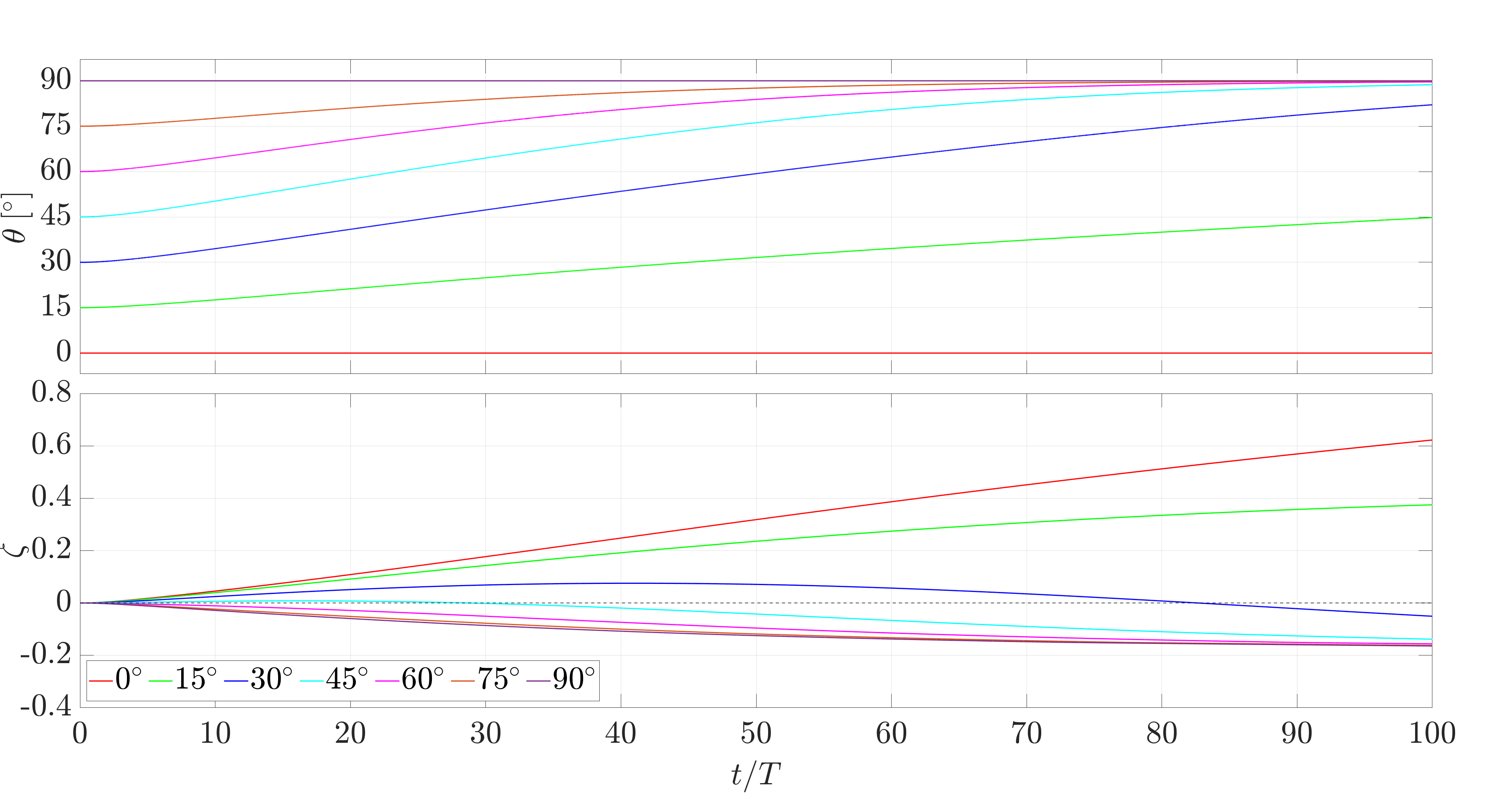}}{0cm}{0cm}% Images in 100% size
        \put(-12.45cm,5.75cm){{\scriptsize(a)}}
        \put(-12.45cm,3cm){{\scriptsize(b)}}
    \end{subfigure}
    \caption{Evolution of (a) $\theta$ and (b) $\zeta$ over $100$ oscillation periods for $\rho_s = 4.68$, $\zeta_i = 0.25$ and $S = 2.09$. Note the difference in scale for the $y$-axis in (b) compared to figure \ref{fig:ShiftedParticles_S_0d35}.}
    \label{fig:ShiftedParticles_S_2d10}
\end{figure}

In order to illustrate the influence of the arrangement orientation, we first look at the behavior of two monodisperse particles with $\rho_s = 4.68$, whose initial distance is $\zeta_i = 0.25$ and the initial angle of the orientation of the particle alignment with respect to the direction of oscillation $\theta_i$ is varied.
We modify $\theta_i$ in steps of $15^{\circ}$ in the range from $0^{\circ}$ to $90^{\circ}$ and analyze the change of $\theta$ and $\zeta$ over a period of $100$ oscillations.
Here, $\theta$ represents the alignment angle over time.
For all setups, the particles are arranged so that the midpoint between the particles coincides with the center of the domain.
For the current analysis, we focus exclusively on the two oscillation scenarios, $S = 0.35$ and $S=2.09$, which result in attraction and repulsion, respectively, for a horizontal alignment (cf. \S \ref{sec:inter-particleDistance}).
The results for different $\theta_i$ are shown in figure \ref{fig:ShiftedParticles_S_0d35} and figure \ref{fig:ShiftedParticles_S_2d10}, where the temporal evolution of the alignment angle $\theta$ is shown in the upper panel and the change in particle distance $\zeta$ is  displayed in the lower panel. 

The results illustrate that $\theta$ increases and converges toward a value of $90^{\circ}$ for particle configurations with $\theta_i \neq 0^{\circ}$ , whereas $\theta=\theta_i$ for $\theta_i = 0^{\circ}$.
The inter-particle distances~$\zeta$, on the other hand, develop significantly differently depending on $S$ and $\theta_i$.
In figure~\ref{fig:ShiftedParticles_S_0d35} ($S = 0.35$), the particles for initial arrangements of $\theta_i \le 30^{\circ}$ approach each other, meaning $\zeta < 0$, while particles move away from each other for $\theta_i \ge 60^{\circ}$.
Interestingly, the particles initially behave attractively and then repulsively for $\theta_i = 45^{\circ}$. 
For $S = 2.09$ (figure \ref{fig:ShiftedParticles_S_2d10}), similar phenomena occur, whereby in this case smaller angles, i.e. $\theta_i \le 15^{\circ}$, show a repulsive and larger angles, i.e. $\theta_i \ge 60^{\circ}$, an attractive behavior. 
Here, both $\theta_i = 30^{\circ}$ and $45^{\circ}$ show a reversal, both from repulsive to attractive.
An indication of the reversal can already be seen at $\theta_i = 15^{\circ}$, while the behaviour in the considered time frame is still repulsive.

The analysis presented above shows the following. 
On the one hand, our data confirms the observations  by \cite{2001_Lyubimov_etal}, \cite{2002_Voth_etal}, \cite{2002_Wunenburger_etal}, and \cite{2007_Klotsa_etal,2009_Klotsa_etal} that  two particles exposed to oscillatory flow align themselves perpendicularly to the direction of oscillation for any perturbation, e.g. in terms of $\theta_i$, that breaks the symmetry of the oscillation direction. 
The well developed stage for the alignment angle $\theta = 90^{\circ}$ was  further investigated by \cite{2007_Klotsa_etal}, \cite{2017_Fabre_etal}, \cite{2018_Jalal} and \cite{2022_vanOverveld_etal}, who all demonstrated that two particles continue to move at a relative distance which remains constant over time. 
On the other hand, \cite{2017_Fabre_etal} were able to prove that this equilibrium only occurs for $S < 2.23$, otherwise the particles approach each other until they come into contact. 
This is indeed the case for our simulations that do not exceed $S=2.09$. 
According to our results, the closer the initial alignment angle is to zero, the more the direction of motion tends to be aligned with the oscillation direction.
This behavior can be utilized in microfluidic devices, e.g., to induce targeted particle motion, as was investigated in a recent study by \cite{2019_Dietsche_etal}.
In the following, we will therefore focus on the specific configuration with an alignment angle $\theta = 0^{\circ}$ in order to investigate the parameter ranges for which particles can be brought into contact or separated from each other.

%%%%%%%%%%%%%%%%%%%%%%%%%%%%%%%%%
\subsection{Interaction behavior}\label{sec:InteractionRegimes}
%%%%%%%%%%%%%%%%%%%%%%%%%%%%%%%%%

The analysis presented in the previous section shows that there are threshold conditions for the interaction of two particles and the oscillating flow that change the particle dynamics from an attractive to a repulsive behaviour depending on the initial gap size and the dimensionless frequency of the oscillations.
It is now interesting to investigate whether a systematic behaviour can be identified that allows to determine these threshold conditions for a wide range of the parameters $S$, $\zeta_i$ and $\rho_s$. 
To this end, we conducted a set of simulations with $\epsilon = 0.1$ for $S = [0.07, 2.10]$ with increments of $\Delta S=0.035$, $\zeta_i = \{0.10, 0.25, 0.50, 0.75, 1.00, 1.50, 2.00, 3.00\}$ for the three density ratios considered, $\rho_s = \{0.47, 1.78, 4.68\}$.
Recall that these values of $\rho_s$ correspond to the same conditions investigated by \cite{2005_LEsperance_etal}. 

\begin{figure}
    \def\stackalignment{l}
    \centering
    \captionsetup[subfigure]{labelformat=empty}
    % Top
    \begin{subfigure}[b]{\textwidth}
        \centering
        \topinset{\scriptsize (a)}{\includegraphics[trim=3.5cm 1.5cm 4cm 2cm, clip,width=0.8\textwidth]{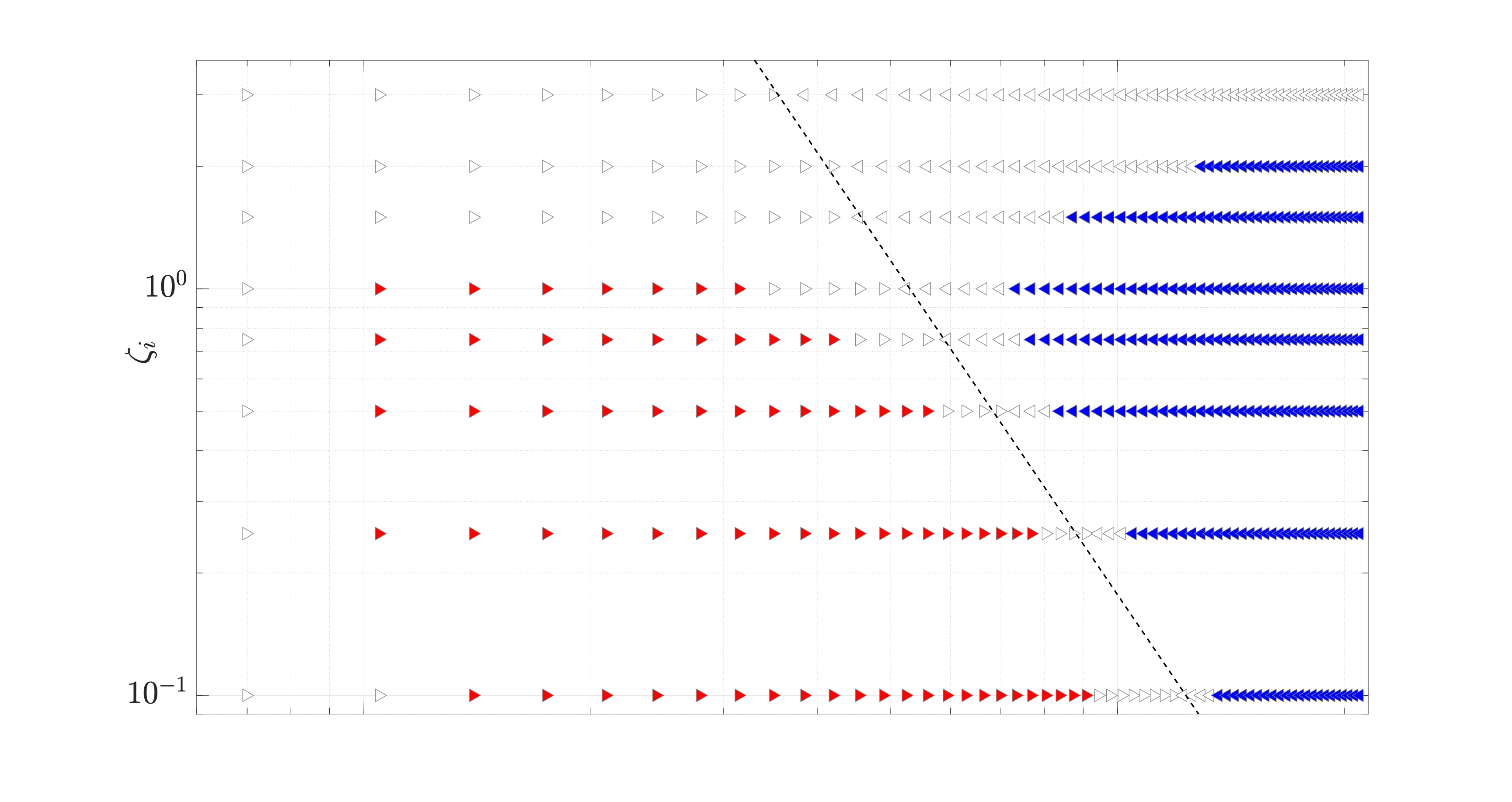}}{0.05cm}{0cm}
        \topinset{\scriptsize (b)}{\includegraphics[trim=3.5cm 1.5cm 4cm 2cm, clip,width=0.8\textwidth]{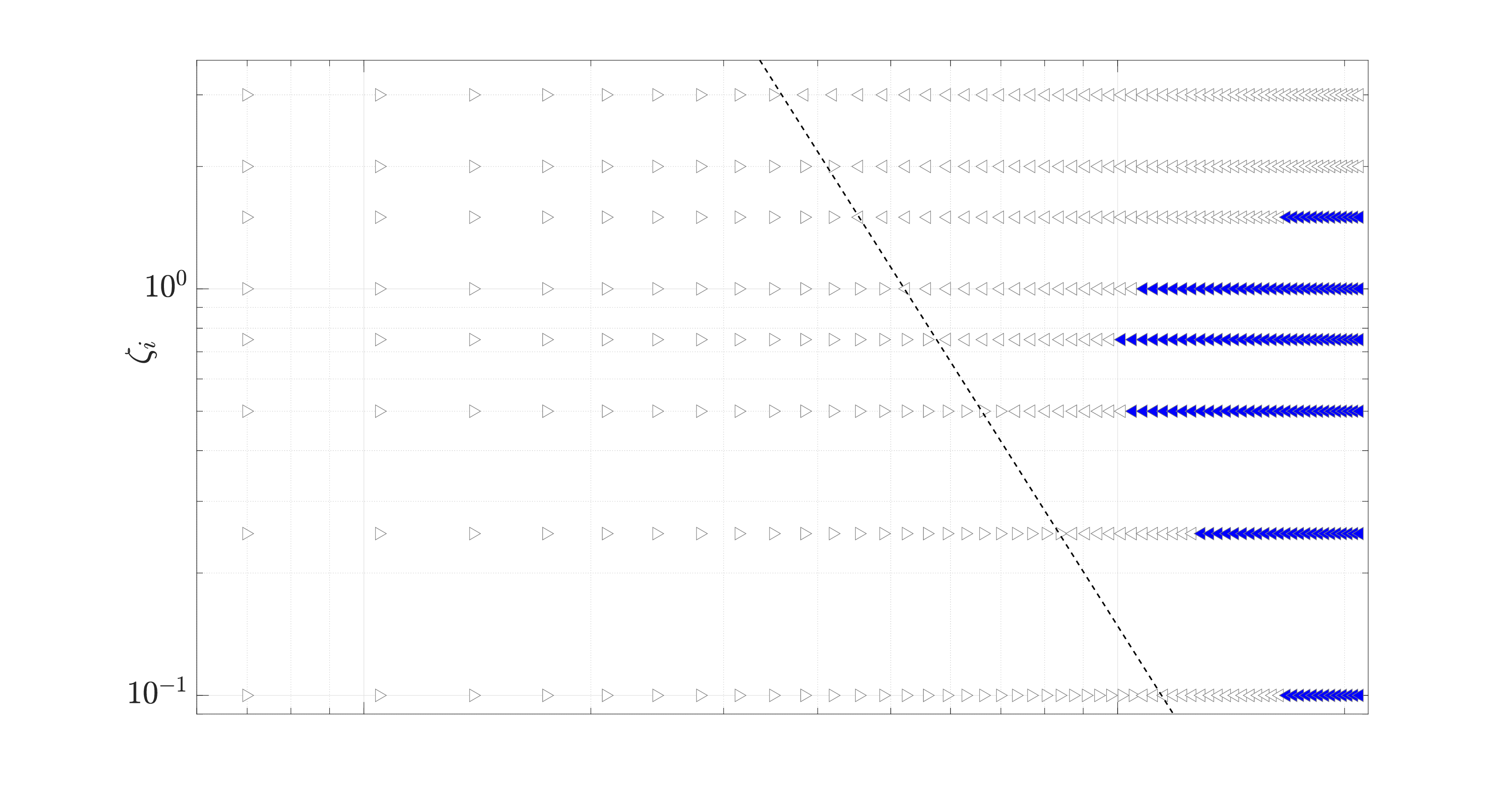}}{0.05cm}{0cm}
        \topinset{\scriptsize (c)}{\includegraphics[trim=3.5cm 0.5cm 4cm 2cm, clip,width=0.8\textwidth]{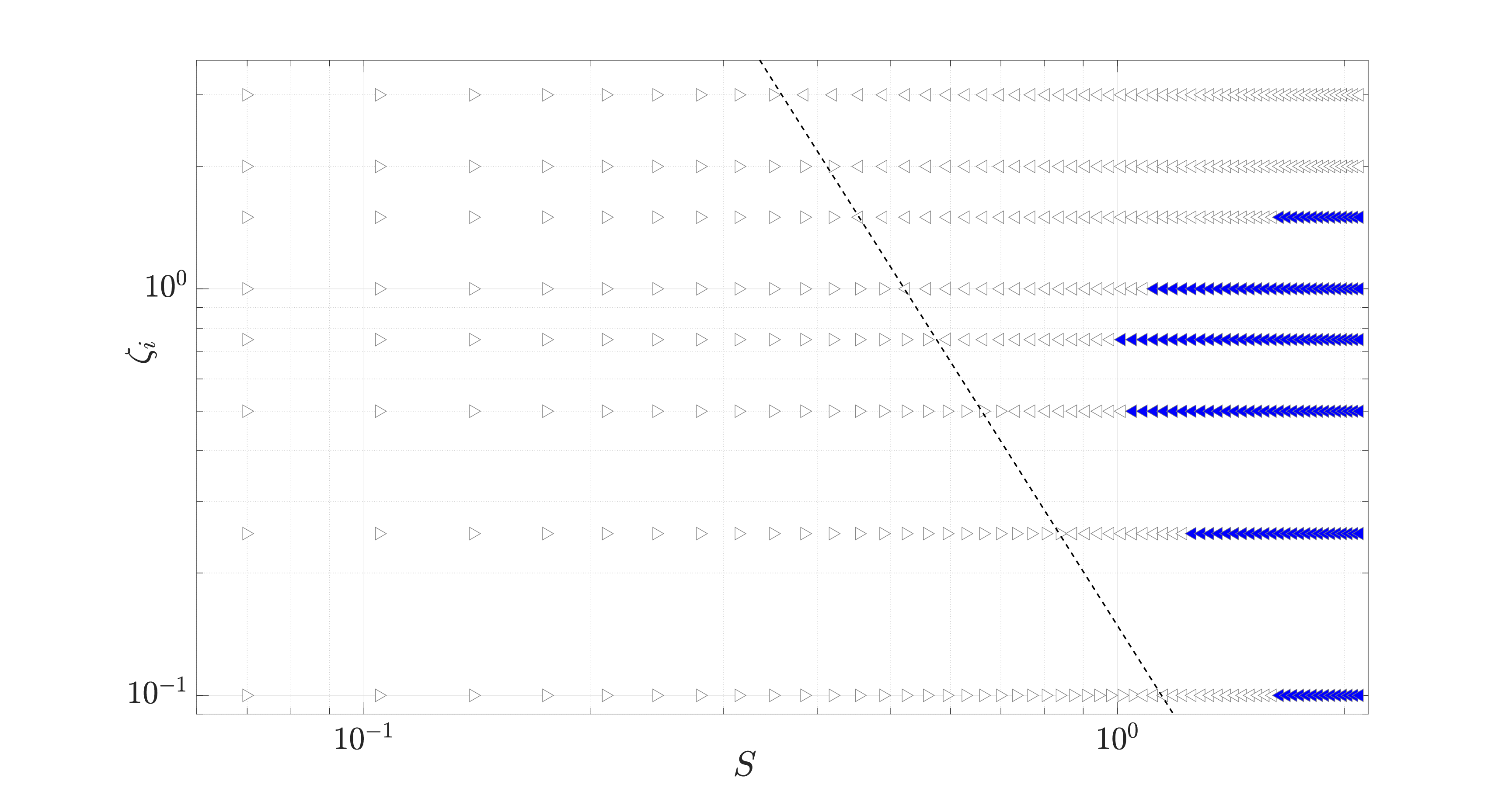}}{0.05cm}{0cm}
    \end{subfigure}
    \caption{Regime maps of the fluid-particle interactions as a function of $S$ and $\zeta_i$ in log-log plots for different particle density ratios: (a) $\rho_s = 4.68$, (b) $1.78$ and (c) $0.47$. Right pointing triangles ($\vartriangleright$) indicate $\zeta_{100} < 0$ and left pointing triangles ($\vartriangleleft$) $\zeta_{100} > 0$. The coloring of the symbols is according to the classification stated in \eqref{eq:classification_RegimeMap}, where {\color{red}$\blacktriangleright$} represents attraction, {\color{blue}$\blacktriangleleft$} repulsion and open symbols ($\vartriangleleft \vartriangleright$) transition.
    The dashed line has been determined by best fit of a power law and marks the threshold condition for which the particle interaction transitions from attractive to repulsive.}
    \label{fig:RegimeMap_loglog}
\end{figure}

This campaign allowed us to construct the regime maps shown in figure \ref{fig:RegimeMap_loglog}, where right pointing triangles ($\vartriangleright$) indicate that the particles are approaching each other, i.e. $\zeta_{100} < 0$, while left pointing triangles ($\vartriangleleft$) refer to setups where the particles are drifting apart, i.e. $\zeta_{100} > 0$. 
We furthermore categorize the mutual interaction as attractive ({\color{red}$\blacktriangleright$}), transitional (open symbols, $\vartriangleleft \vartriangleright$), and repulsive ({\color{blue}$\blacktriangleleft$}), depending on $S$ and $\zeta_i$.  
This categorization is derived from the analysis of $A_{2p}(S,\zeta_i)$ discussed in \S \ref{sec:inter-particleDistance}.
We define arrangements, in which \mbox{$| \zeta_{100} | \leq A_{2p}(S,\zeta_i)$} as transitional, whereas setups with exceeding \mbox{$| \zeta_{100} | > A_{2p}(S,\zeta_i)$} are classified as either attractive or repulsive, depending on the sign of $\zeta_{100}$. 
The setups classified as transitional represent cases in which the interaction of the flow fields has only minor effects on the respective particles. 
Although the particles move either toward or away from each other, such cases do not exhibit significant changes of the particle spacing over time and can therefore be described as a state of equilibrium. 
\begin{equation}
    | \zeta_{100} |
    \begin{cases}
        > A_{2p}(S,\zeta_i)
        \begin{cases}
            \text{= attraction, } & \text{for } \zeta_{100} < 0\\
            \text{= repulsion, } & \text{for } \zeta_{100} > 0\\
        \end{cases} 
        \vspace{0.2cm} \\
        \leq A_{2p}(S,\zeta_i)  \text{= transition}
        \vspace{0.1cm} \\
    \end{cases}
    \label{eq:classification_RegimeMap}
\end{equation}
Despite the low values of $\zeta_{100}$ that are assigned to transitional behavior, a more precise distinction into transitionally attractive and transitionally repulsive remained possible as will be shown in \S \ref{sec:circulationAnalysis} below. This is also expressed in the clear definition of the threshold condition shown as a dashed line in figure \ref{fig:RegimeMap_loglog}.

For $\rho_s = 4.68$ in figure \ref{fig:RegimeMap_loglog}(a), our classification scheme reveals that for all arrangements of $\zeta_i \le 1.00$, there is a clear trend that smaller $S$ lead to attraction and larger $S$ to repulsion. 
The threshold condition from attractive to repulsive results is covered by setups classified as transitional. 
With increasing $\zeta_i$, the threshold conditions shift to lower values of $S$ and the cases classified as transitional cover a wider value range of $S$. 
For initial particle gaps $\zeta_i = 3.00$, no significant attractive or repulsive behavior was observed and all cases fall into the transitional regime.
Again, this confirms that the effectiveness of the mutual interaction decreases with increasing particle distance.

The regime maps of $\rho_s = 1.78$ and $0.47$ in figures \ref{fig:RegimeMap_loglog}(b) and \ref{fig:RegimeMap_loglog}(c), respectively, are almost identical with only a few deviations. 
The similarity is due to the approximately similar deviation of their density from the neutrally buoyant case, i.e. $\rho_s=1$.  
However, comparing them to the case $\rho_s=4.81$ shown in figure~\ref{fig:RegimeMap_loglog}(a), we find that the magnitude, for which particles approach each other or drift apart, decreases substantially for the two cases that are closer to neutrally buoyant conditions. 
In fact, the small effect of particle inertia diminishes such that the particle interaction yields only marginally attractive cases that we classify as transitional according to the criterion given by \eqref{eq:classification_RegimeMap}.
In general, the majority of setups lead to transitional results for those two density ratios, where the number of repulsive cases are also less than for $\rho_s = 4.68$.
For example, the arrangements with larger initial gaps ($\zeta_i \geq 2.00$) do not lead to any significant repulsive behavior, so that all arrangements are classified as transitional. 
Nevertheless, even though the range of intensities of the particle interaction becomes  weaker as inertia effects of the particles are decreased, a very similar trend is observed for the region in which the scenario changes from attractive to repulsive as $S$ and $\zeta_i$ are increased.

As mentioned above, the dashed line in each regime map marks the transition from attractive to repulsive behavior as indicated by a change of sign in $\zeta_{100}$ shown in figure~\ref{fig:zeta_varyingS}. 
For figure~\ref{fig:RegimeMap_loglog}, this threshold condition was determined by best fit of a power law given by $\zeta = C_1 \: S^{\beta}$, for which we obtain an excellent correlation with our data. 
For the highest particle density ratio $\rho_s=4.81$ (figure \ref{fig:RegimeMap_loglog}a), we determine $C_1=0.18$ and $\beta=-2.73$. 
For the two cases $\rho_s=1.78$ and $\rho_s=0.47$, we obtain identical fitting parameters $C_1=0.15$ and $\beta=-2.93$. 
This illustrates two things: (i) an increase of inertia does not substantially change the threshold conditions to transition from attractive to repulsive  behavior and (ii) as long as the deviation from the neutrally buoyant condition remains the same in terms of particle inertia, the interaction regime is unaffected, even though we change particle properties to denser or lighter than the ambient fluid.

The deviation in particle behavior with respect to $\rho_s$ is hence insignificant. 
Therefore, our results can be directly compared with the force maps presented by \cite{2017_Fabre_etal}, which correlate forces in the context of two stationary particles in an oscillating flow. 
In their study, they investigated a wide range of oscillations ($S = [0.01, 11.10]$) and particle distances ($\zeta_i = [0, 2.0]$) and due to the immobilization of the particles, the parameter $\rho_s$ did not play a role.
In principle, our analyses agree in that small $S$ and small $\zeta_i$ generally lead to a decrease of $\zeta$, while large $S$ and large $\zeta_i$ result in an increase of $\zeta$ over time. 
In fact, one can convert our ($S,\zeta_i$)-regime map to the regime map of \cite{2017_Fabre_etal} (figure 6(b) in that reference) and we find very close agreement for the transition for $S=0.56$ and $S=0.89$ (termed $\Omega=5$ and $\Omega=8$ in that same reference).

%%%%%%%%%%%%%%%%%%%%%%%%%%%%%%%%%
%%%%%%%%%%%%%%%%%%%%%%%%%%%%%%%%%
\section{Flow field analysis}\label{sec:flow_field_analysis}
%%%%%%%%%%%%%%%%%%%%%%%%%%%%%%%%%
%%%%%%%%%%%%%%%%%%%%%%%%%%%%%%%%%

%%%%%%%%%%%%%%%%%%%%%%%%%%%%%%%%%
\subsection{Steady streaming}\label{sec:totalSteadyStreaming}
%%%%%%%%%%%%%%%%%%%%%%%%%%%%%%%%%

The previous results have shown that two axially arranged particles in an oscillating fluid flow either attract or repel each other, depending on the applied frequency and the initial distance. 
To better understand the mechanisms that lead to a decrease or increase in the particle distance, we average the flow field over one oscillation period to focus on the time-averaged component of the fluid flow that arises due to the presence of particles. 
In this regard, we apply a method of intrinsic averaging, where only the volume occupied by fluid is considered \citep{2017_Vowinckel_etal}.
Such an averaged flow field is commonly referred to as \textit{steady streaming} \citep{1966_Riley}.

As briefly introduced in \S\ref{sec:Introduction}, it represents the nonzero mean flow of the periodically averaged fluid field generated by the interaction of objects and the surrounding fluid, where either the fluid or the object can induce the oscillations \citep{2001_Riley}.
Such an analysis is particularly useful for high-frequency oscillations, because it allows to study the well-developed flow patterns of the steady streaming induced by the inertial particles in the oscillating domains.
However, two important aspects have to be taken into account when choosing the time for the calculation of the steady streaming: 
First, the chosen time should be at the initial phase of the simulation, so that the distance between the particles is as close as possible to its original arrangement.
Second, the effect of the initial displacement of the particles in a fluid at rest (\textit{viz.} $\S\ref{sec:initialMotion}$) must have already decayed and thus be negligible.
Consequently, a compromise must be made which best fulfills both requirements.
To this end, we have performed a convergence study in which we analyzed the impact of the initial conditions on the flow structures of the steady streaming.
This study showed that a well-developed state in the oscillatory motion is reached at  time $t/T\geq11$ for each simulation.
This is why all the flow characteristics shown hereafter refer to $t/T = 11$.
As mentioned earlier in \S \ref{sec:governingEqs}, the scenario under consideration does not clearly fall into the case distinction for a single particle proposed by \cite{1966_Riley}.
It is therefore important to note that all the following flow structures, which consider the setup parallel to the direction of oscillation, consist exclusively of what has been referred to as primary vortices surrounding the respective particle.

\begin{figure}
    \def\stackalignment{l}
    \centering
    \captionsetup[subfigure]{labelformat=empty}
    % Top
    \begin{subfigure}[b]{\textwidth}
         \centering
         \topinset{{\scriptsize (a)}}{\includegraphics[trim=11.5cm 0.4cm 13.8cm 1.5cm, clip,width=0.49\textwidth]{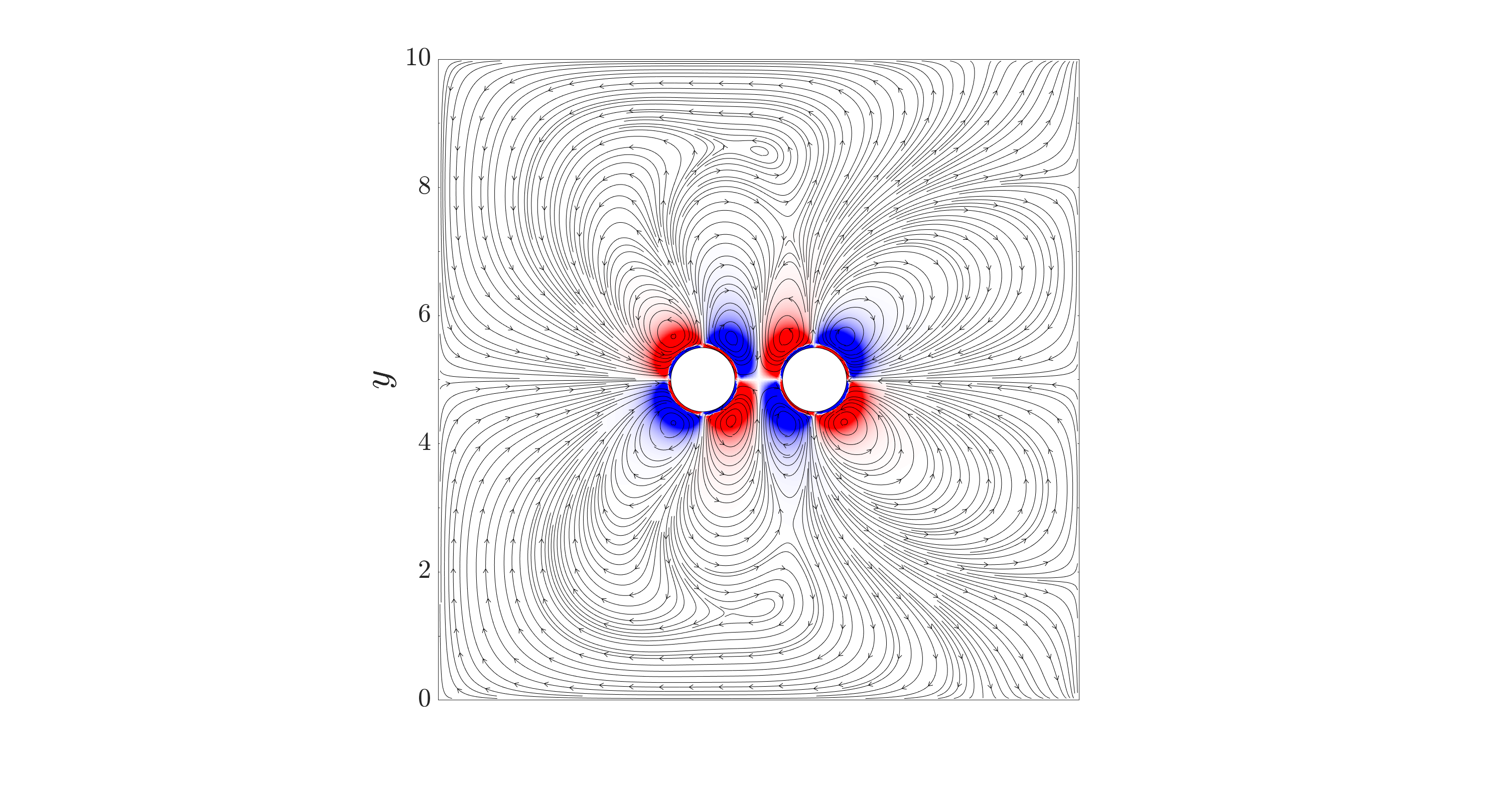}}{0.3cm}{0.05cm}
         \topinset{{\scriptsize (b)}}{\includegraphics[trim=11.5cm 0.4cm 13.8cm 1.5cm, clip,width=0.49\textwidth]{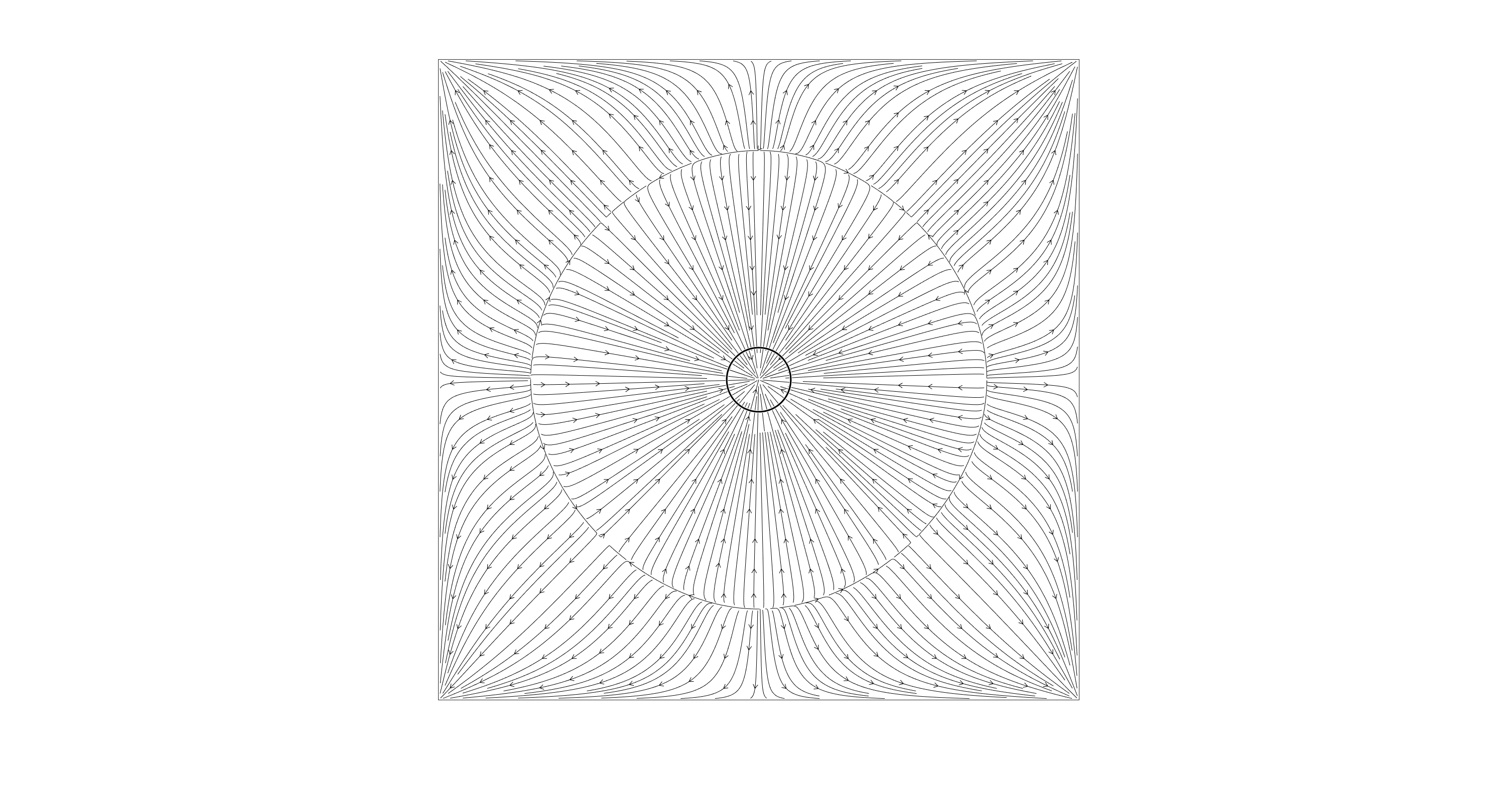}}{0.3cm}{0.2cm}
         \topinset{{\scriptsize (c)}}{\includegraphics[trim=11.5cm 0.4cm 13.8cm 1.5cm, clip,width=0.49\textwidth]{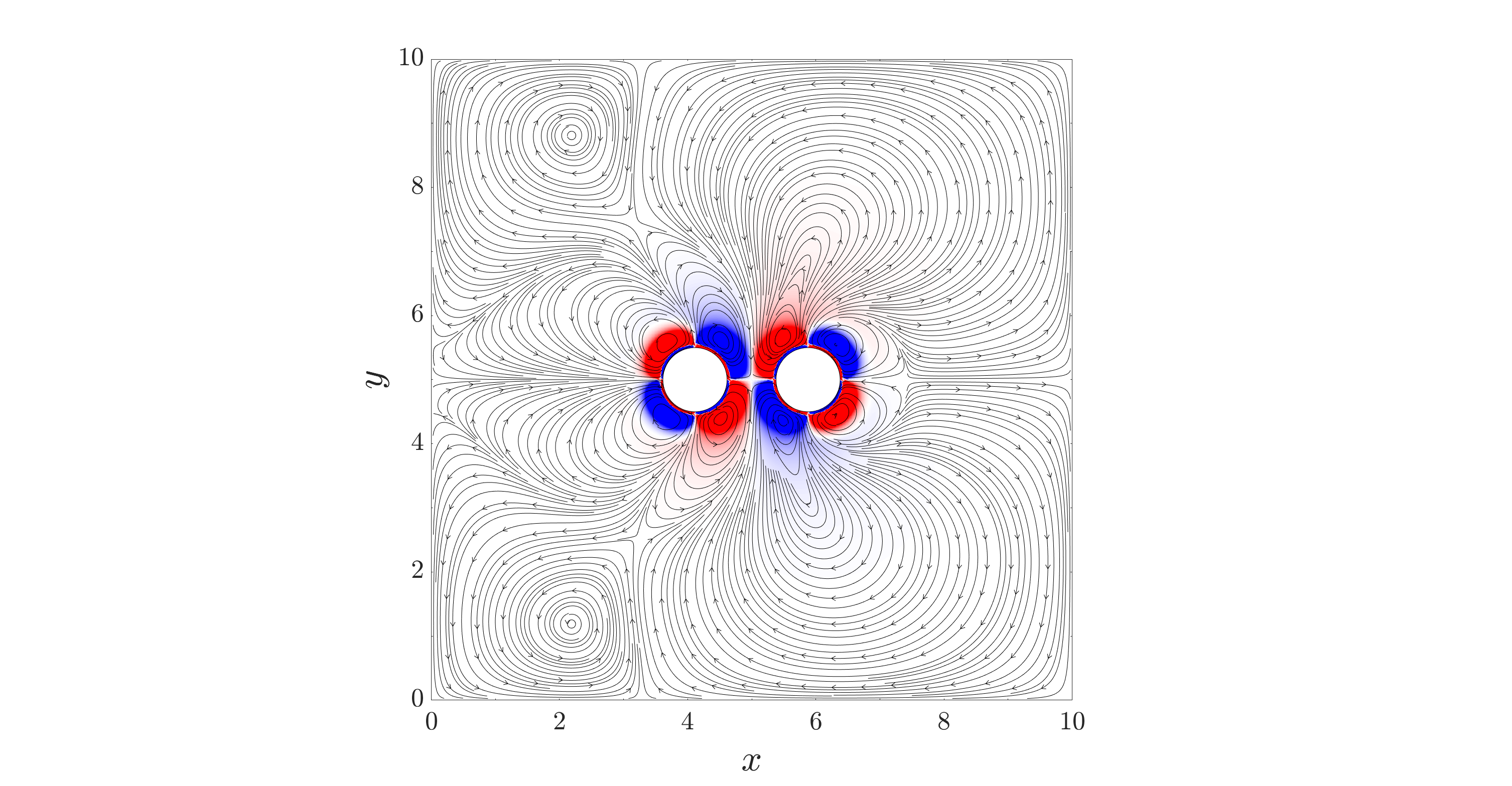}}{0.3cm}{0.05cm}
         \topinset{{\scriptsize (d)}}{\includegraphics[trim=11.5cm 0.4cm 13.8cm 1.5cm, clip,width=0.49\textwidth]{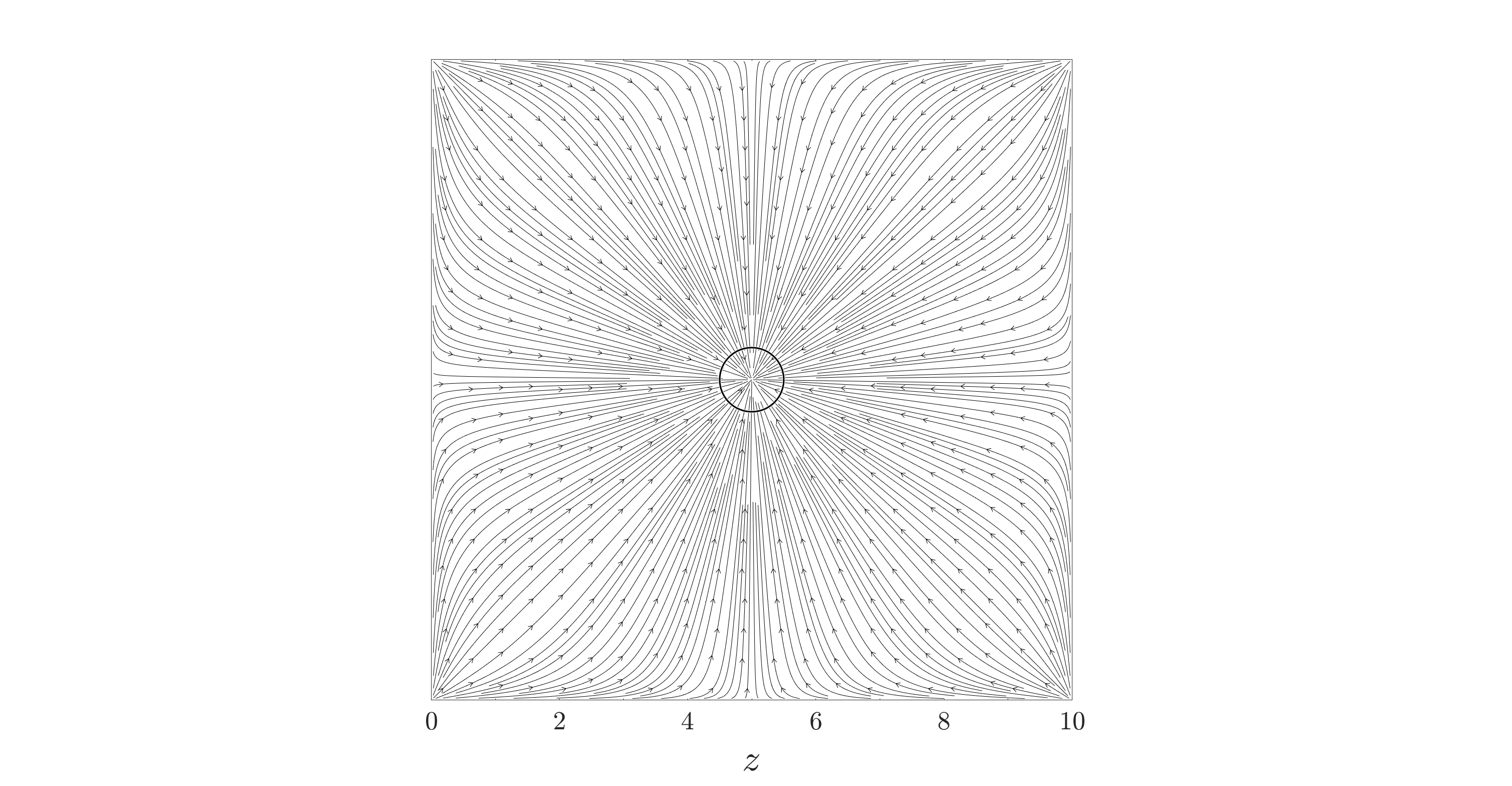}}{0.3cm}{0.2cm}
    \end{subfigure}
    \caption{Streamlines and vorticity contours of the steady streaming in $xy$- (figures a, c) and $zy$-planes (b, d). The top row shows an attractive case with $S = 0.35$ and the bottom row a repulsive behavior with $S = 1.05$. Both are arrangements with $\zeta_i = 0.75$ and $\rho_s = 4.68$. The black circles in the middle of the $zy$-planes indicate the particle positions, which are arranged in line along the $x$-axis. The outer circle in (b) is a stagnation line of the flow field and results from the averaging process of computing the steady streaming. Color scheme as in figure~\ref{fig:FlowCharacteristics_TwoParticles}.}
    \label{fig:SteadyStreaming_ZY-plane}
\end{figure}

In figure \ref{fig:SteadyStreaming_ZY-plane}, the flow structures of the steady streaming are shown for two representative cases of attractive ($S = 0.35$ in the upper panels) and repulsive behavior ($S = 1.05$ in the lower panels) using $\zeta_i = 0.75$ and $\rho_s = 4.68$. 
For this purpose, we show streamlines in the  $xy$- and $zy$-planes.
The planes cut through the center of the domain in the third respective dimension, i.e. at $z = 5$ for the $xy$-plane and at $x = 5$ for the $zy$-plane. 
Since the $zy$-plane cuts through a plane in between the particles, we show a hollow black circle at this location in \ref{fig:SteadyStreaming_ZY-plane}(b) and \ref{fig:SteadyStreaming_ZY-plane}(d) to indicate the particle positions aligned along the $x$-axis. 
We note, however, that this circle does not provide an obstacle to the flow for these figures. 
We also note that in all figures showing streamlines, these lines are uniformly distributed and serve only to qualitatively illustrate the flow patterns and directions. 
They do not represent a value of a streamfunction to visualize flow intensity. 
In addition, we calculate the polar component of the vorticity $\omega_z\left(x,y \right)$ to provide a quantitative measure. 
The vorticity contours are shown in blue $(-0.01)$ and red $(0.01)$, representing clockwise and counterclockwise rotation, respectively. 
This color scheme corresponds to the color scale in figure \ref{fig:FlowCharacteristics_TwoParticles} and is applied for all subsequent analyses.

The comparison of the $xy$-planes for the repulsive and the attractive case (figures \ref{fig:SteadyStreaming_ZY-plane}a and c) shows that in both cases, each individual particle is surrounded by four vortex structures, which are depicted by the tori of the streamlines and referred to as quadrupoles, as described in \S \ref{sec:validation}. 
Consequently, two quadrupole structures are formed in each setup.
However, instead of assigning the vortices to the particles, we examine the structures based on their spatial position and divide them into four central and four peripheral vortices (c.f. the sketch in figure \ref{fig:FlowCharacteristics_TwoParticles}b).
This distinction allows for a more detailed description and analysis of the flow mechanisms in the present and subsequent cases.
The central vortices, or more precisely, the vortex cores of the central vortices are located between the two particles.
The peripheral vortices, on the other hand, are located on the outward-facing sides.
Hence, for the setup considered here, the central vortices cause fluid to flow into the gap, tending to push the particles away from each other (repulsion), while the peripheral vortices direct the flow along the horizontal axis of symmetry towards the outward-facing sides of the particles, pushing the particles toward each other (attraction).
Although the double quadrupole structure of the steady streaming is formed in both the attractive (figure \ref{fig:SteadyStreaming_ZY-plane}a) and repulsive (figure \ref{fig:SteadyStreaming_ZY-plane}c) cases, the streamlines differ significantly in their spatial pattern.
Nevertheless, neither of the two cases shows a clear trend with respect to the formation of the central and peripheral vortices.
As was shown in \S\ref{sec:initialMotion}, this less distinct flow pattern is due to the decaying effects of the initial condition.
Qualitatively, however, these figures provide an indication that the size of the peripheral vortices is predominant for attraction and vice versa, the central vortices are larger for repulsion.
While the far-field flow patterns show the aforementioned differences, the vorticity contours in the near-field of the particles show a strong similarity.
Next to the particle surfaces, four vorticity contours emerge whose shapes mimic the curvature of the sphere.
These are followed by another four contours, each with reversed sign.
For each individual particle, the diagonally opposite central as well as peripheral contours have the same direction of rotation.

The $zy$-planes also differ in the structures of their streamlines.
In case of attraction (figure \ref{fig:SteadyStreaming_ZY-plane}b), there is a separation of the flow direction, which is marked by the outer circle. 
This circle is a result of the calculation of the steady streaming and represents a stagnation line where the flow is directed neither toward nor away from the center.
The inner part of the circle indicates an convergent inflow, where the streamlines point towards the center between the particles. 
For the part outside of the stagnation line, the flow is directed outwards away from the particles.
For the repulsive behavior (figure \ref{fig:SteadyStreaming_ZY-plane}d), the entire flow of the $zy$-plane is directed towards the center of the particles.
The vorticity in $zy$-plane is not discernible in neither of the two cases shown, which is related to the fact that the flow components in the $y-$ and $z-$directions, i.e., orthogonal to the direction of oscillation, are relatively minor and consequently the vorticity resulting from these components is very small as well.

The flow patterns shown in figures \ref{fig:SteadyStreaming_ZY-plane} show a strong axisymmetric structure. 
This is true for both, attraction as well as for repulsion.
We therefore conclude that the $xz$-plane (not shown here) and the $xy$-plane, both arranged through the center of the particles with $y=5$ and $z=5$, respectively, show the same flow properties and structures of the steady streaming.
For this reason we will limit our analysis to only one of these two planes, namely the $xy$-plane, as it provides sufficient information to study the fluid-particle interaction and the resulting particle behavior.

%%%%%%%%%%%%%%%%%%%%%%%%%%%%%%%%%
\subsection{Flow decomposition}\label{sec:flowDecomposition}
%%%%%%%%%%%%%%%%%%%%%%%%%%%%%%%%%

The detailed illustrations shown in figure \ref{fig:SteadyStreaming_ZY-plane} already indicate that there are significant differences in the flow structures of attractive and repulsive setups.
However, identifying and quantifying a clear physical mechanism that allows for conclusions regarding attractive or repulsive behavior of the particles is not possible due to the apparent overlap of different flow structures. 
For this reason, we perform a more detailed examination in terms of flow decomposition to better evaluate the vorticity generated by the fluid-particle interactions.

Following the argument of \cite{2003_Cerretelli_Williamson}, the {vorticity computed from  steady streaming $\omega_z\left(x,y \right)$, hereafter also referred to as total steady streaming, allows to decompose the flow field into a symmetric  and an antisymmetric part denoted as $\omega_{z,S}\left(x,y \right)$ and $\omega_{z,A}\left(x,y \right)$, respectively,  
\begin{equation}
    \omega_z \left(x,y \right) = \omega_{z,S} \left(x,y \right) + \omega_{z,A} \left(x,y \right) \qquad .
    \label{eq:total_vorticity}
\end{equation}
The symmetric part is generated by calculating the symmetry about the center of the distance between the particles, based on averaging the left and right components of the total steady streaming.
Subtracting this part from the total steady streaming yields the antisymmetric component. 
To this end, we introduce the transform $x' = x - x_c$, where $x_c$ is the x-coordinate at the center between the two particles.
Therefore, $\omega_z\left(x',y \right)$ can be written as
\begin{equation}
  \omega_z \left(x',y \right) = \underbrace{\frac{1}{2} \left[\omega_z \left(x',y \right) + \omega_z \left(-x',y \right) \right]}_{=\omega_{z,S}} + \underbrace{\frac{1}{2} \left[\omega_z \left(x',y \right) - \omega_z \left(-x',y \right) \right]}_{=\omega_{z,A}} \qquad .
  \label{eq:total_vorticity_expanded}
\end{equation}
\begin{figure}
    \def\stackalignment{l}
    \centering
    \captionsetup[subfigure]{labelformat=empty}
    % Top
    \begin{subfigure}[b]{\textwidth}
         \centering
         \topinset{{\scriptsize (a)}}{\includegraphics[trim=11.5cm 0.4cm 13.8cm 1.5cm, clip,width=0.49\textwidth]{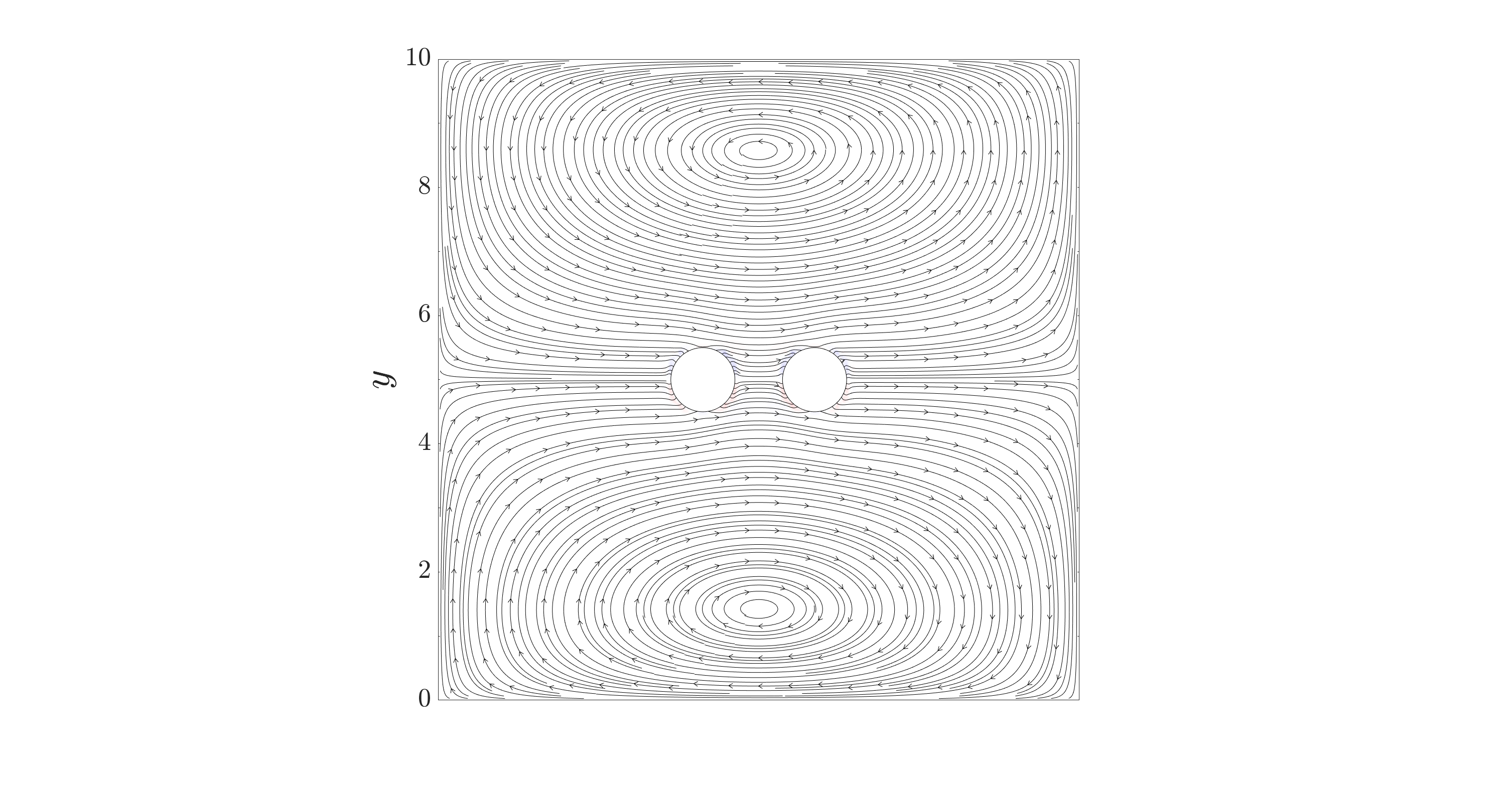}}{0.3cm}{0.05cm}
         \topinset{{\scriptsize (b)}}{\includegraphics[trim=11.5cm 0.4cm 13.8cm 1.5cm, clip,width=0.49\textwidth]{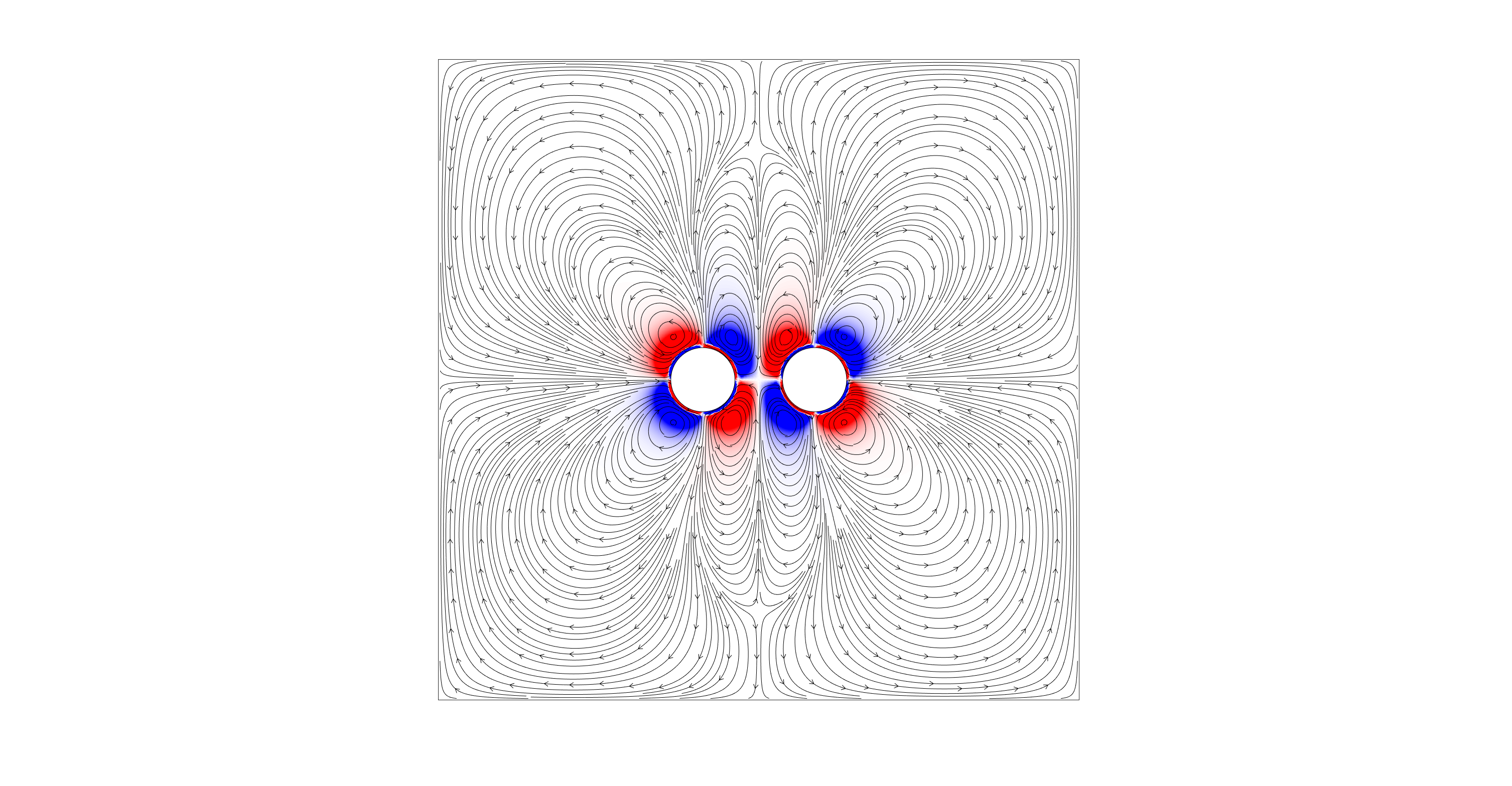}}{0.3cm}{0.2cm}
         \topinset{{\scriptsize (c)}}{\includegraphics[trim=11.5cm 0.4cm 13.8cm 1.5cm, clip,width=0.49\textwidth]{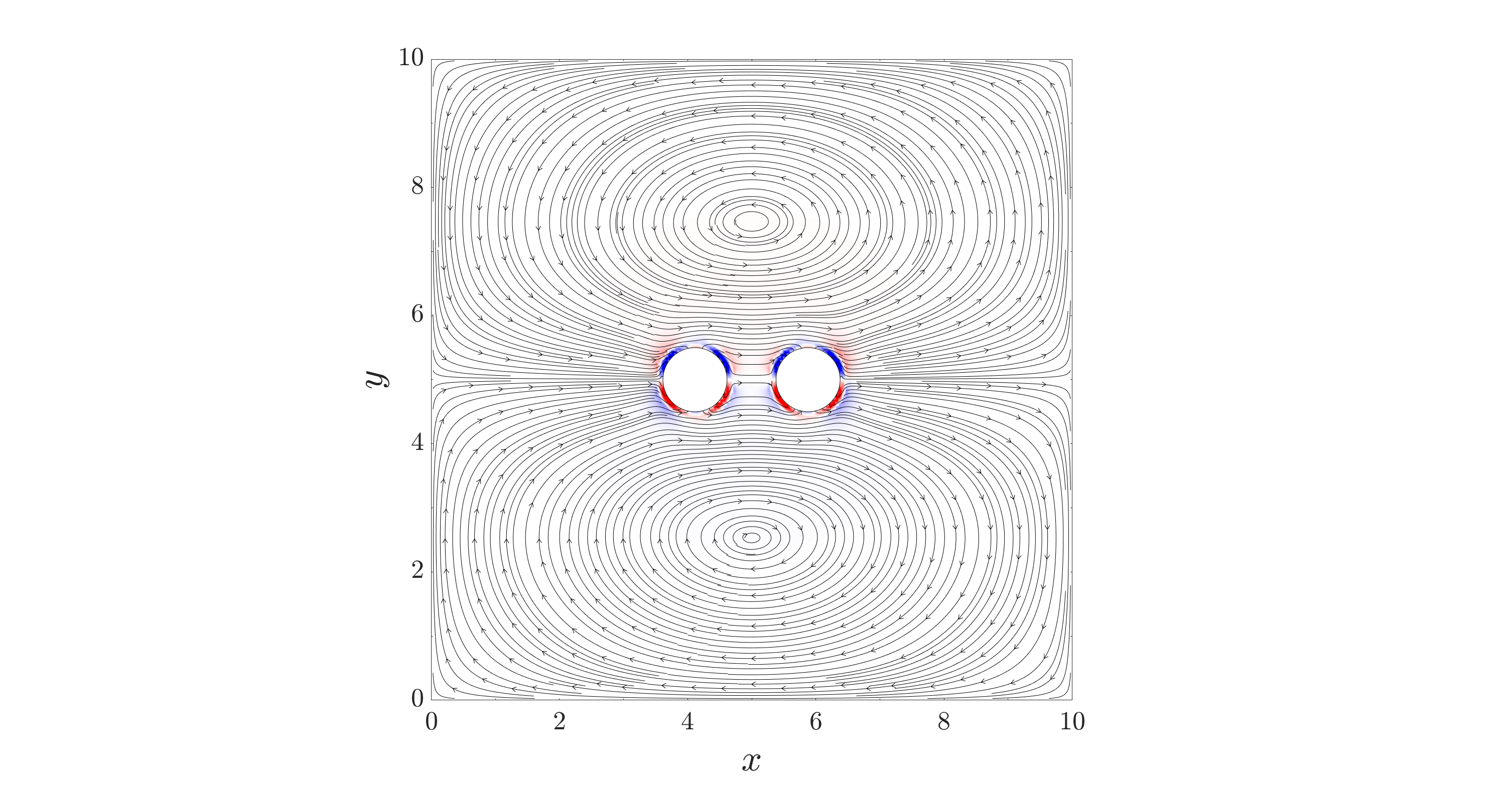}}{0.3cm}{0.05cm}
         \topinset{{\scriptsize (d)}}{\includegraphics[trim=11.5cm 0.4cm 13.8cm 1.5cm, clip,width=0.49\textwidth]{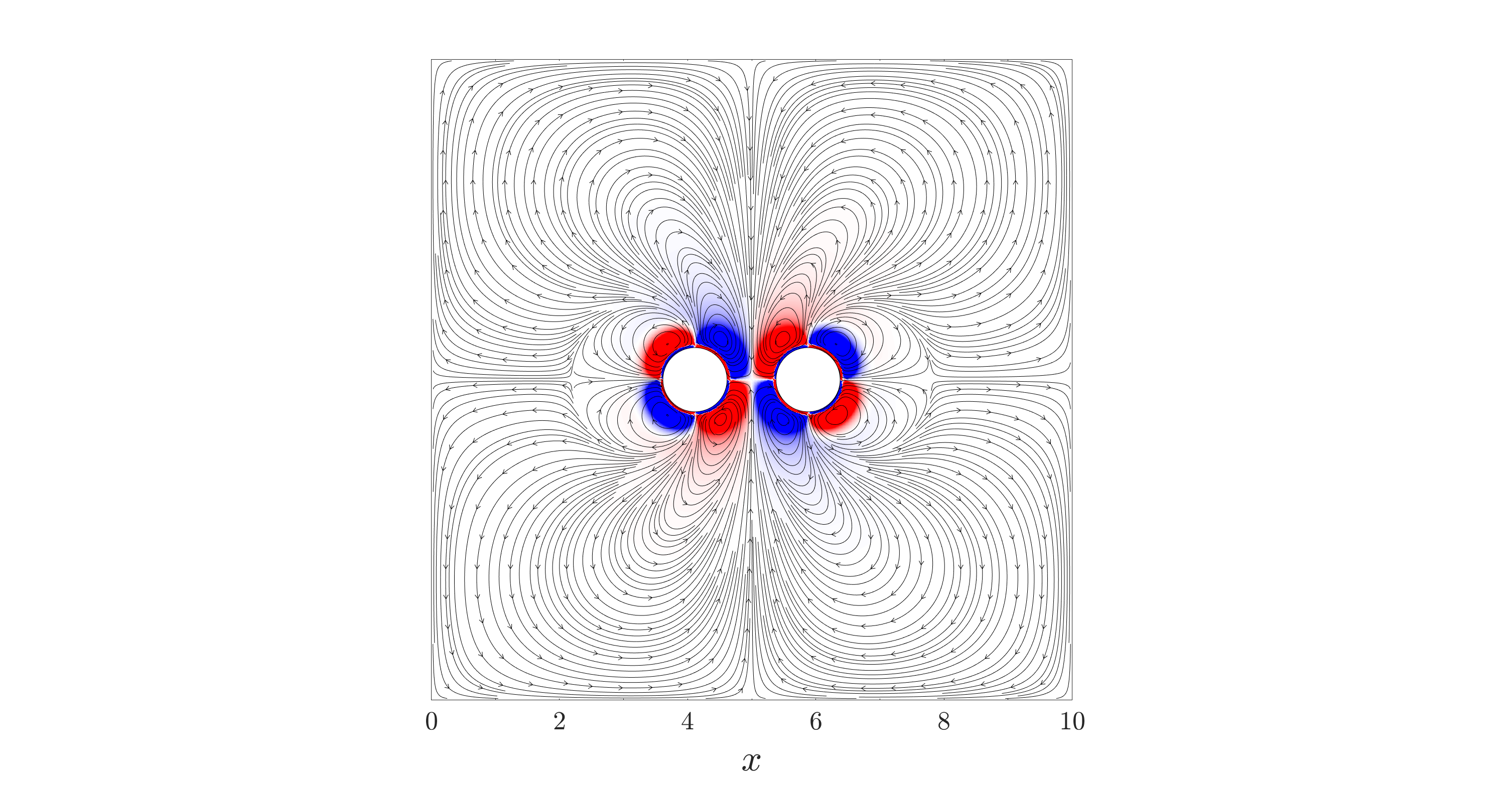}}{0.3cm}{0.2cm}
    \end{subfigure}
    \caption{Streamlines and vorticity contours of the symmetric (a, c) and antiymmetric components (b, d) for $\zeta_i = 0.75$ and $\rho_s = 4.68$. The figures in the top row, \mbox{(a, b)}, represent $S = 0.35$ resulting in attraction, and in the bottom row, \mbox{(c, d)}, $S = 1.05$ leading to repulsion. The total steady streaming used for decomposition is given in figures \ref{fig:SteadyStreaming_ZY-plane}(a, c). Color scheme as in figure~\ref{fig:FlowCharacteristics_TwoParticles}.}
    \label{fig:SteadyStreaming_SymmAntisymm}
\end{figure}

In figure \ref{fig:SteadyStreaming_SymmAntisymm}, the streamlines and vorticity contours of the symmetric and antisymmetric components are shown in the panels of the left and the right column, respectively. 
The setups are the same as given in  figure \ref{fig:SteadyStreaming_ZY-plane}. 
In this regard, figures \ref{fig:SteadyStreaming_ZY-plane}(a) and (c) represent the total steady streaming used for the decomposition \eqref{eq:total_vorticity}.
When considering the streamlines of the symmetric part (left two panels of figure \ref{fig:SteadyStreaming_SymmAntisymm}), two vortex structures with center points in the far field orthogonal to the particle arrangement and in line with the center of the particle distance can be seen for the attractive as well as repulsive behavior. 
In both cases, the upper vortex rotates in counterclockwise and the lower one in clockwise direction, but the position of the stagnation points is at smaller distance to the particle surfaces for repulsion (figure \ref{fig:SteadyStreaming_SymmAntisymm}c) than it is for attraction (figure \ref{fig:SteadyStreaming_SymmAntisymm}a). 
The vortex contours also differ in magnitude and spatial extent. 
While almost no contours are visible for attraction (figure \ref{fig:SteadyStreaming_SymmAntisymm}a),  a distinct layer near the particles is present for repulsion.
In general, however, the symmetric component discloses the resulting motion of the particle pairs and, along with it, the basic direction of motion of the flow field, which is from left to right in both arrangements presented (cf. figure \ref{fig:SteadyStreaming_SymmAntisymm}a and c).
The net motion of the particles is due to the transients that are still present, caused by the initial condition of the still fluid, as described in \S\ref{sec:initialMotion}.
Since the initial displacement is larger for $S=1.05$ than for $S=0.35$, there is a higher deviation from the center of the domain at the time of calculating the steady streaming.
This results in an stronger motion towards the center of the domain for $S=1.05$, which in turn produces more intense vortex contours.
Even though this has no influence on the respective particle behavior regarding attraction or repulsion, it is responsible for the flow field of the symmetric component of the steady streaming.

The antisymmetric fields (figures \ref{fig:SteadyStreaming_SymmAntisymm}b and d) represent the deviations between the flow fields of the total steady streaming and the symmetric component.
It thus illustrates the filtered flow field that is generated solely by the pairwise fluid-particle interaction. 
This flow component actually dictates the governing motion of the particles with respect to one another, because this is the motion that deviates from the underlying oscillatory dynamics.
In this regard, the vorticty contours of the antisymmetric parts are very similar to the contours of the total steady streaming. 
This is due to the fact that the vorticty of the previously described symmetric component is small for both cases, attraction and repulsion. 
This analysis already reveals that the particle behavior of attraction or repulsion is caused by small flow conditions deviating from the oscillating main flow. These small flow deviations are clearly visible in the streamlines of the antisymmetric flow fields. As mentioned above, these flow structures cause either repulsive or attractive behavior and they compete in size depending on $S$. 
A qualitative analysis of the shape and size of the vortex structures underlines the findings of \S\ref{sec:totalSteadyStreaming}, in which the peripheral vortex structures dominate over the central ones for attraction and vice versa for repulsion.

%%%%%%%%%%%%%%%%%%%%%%%%%%%%%%%%%
\subsection{Effect of particle inertia}\label{sec:EffectParticleInertia}
%%%%%%%%%%%%%%%%%%%%%%%%%%%%%%%%%

Since the previous figures have shown results for  $\rho_s = 4.68$ only, we now turn our attention to the steady streaming observed for two additionally considered particle density ratios $\rho_s = 1.78$ and $0.47$ that were already analyzed in terms of their particle interaction in figure \ref{fig:RegimeMap_loglog}. 
We therefore performed the same analysis of flow decomposition as for $\rho_s = 4.68$ given in the previous section. 
Figure \ref{fig:SteadyStreaming_SymmAntisymm_densities} shows the corresponding results of steady streaming for $S = 0.35$ and $\zeta_i = 0.75$ for these two density ratios.
In this way, we can compare the flow structures and vortex contours of these two setups ($\rho_s = 1.78$ in figures \ref{fig:SteadyStreaming_SymmAntisymm_densities}a, c \& e and $\rho_s = 0.47$ in figures \ref{fig:SteadyStreaming_SymmAntisymm_densities}b, d \& f) with the previous one ($\rho_s = 4.68$ in figures \ref{fig:SteadyStreaming_ZY-plane}a and \ref{fig:SteadyStreaming_SymmAntisymm}a,b) and can thus infer the behavior of two particles in oscillating flow that are both denser and lighter than the surrounding fluid.
Since we neglect gravity, the particle dynamics are affected by their mass only in terms of their inertia, but not their weight.

\begin{figure}
    \def\stackalignment{l}
    \centering
    \captionsetup[subfigure]{labelformat=empty}
    % Top
    \begin{subfigure}[b]{\textwidth}
         \centering
         \topinset{{\scriptsize (a)}}{\includegraphics[trim=11.5cm 0.4cm 13.8cm 1.5cm, clip,width=0.49\textwidth]{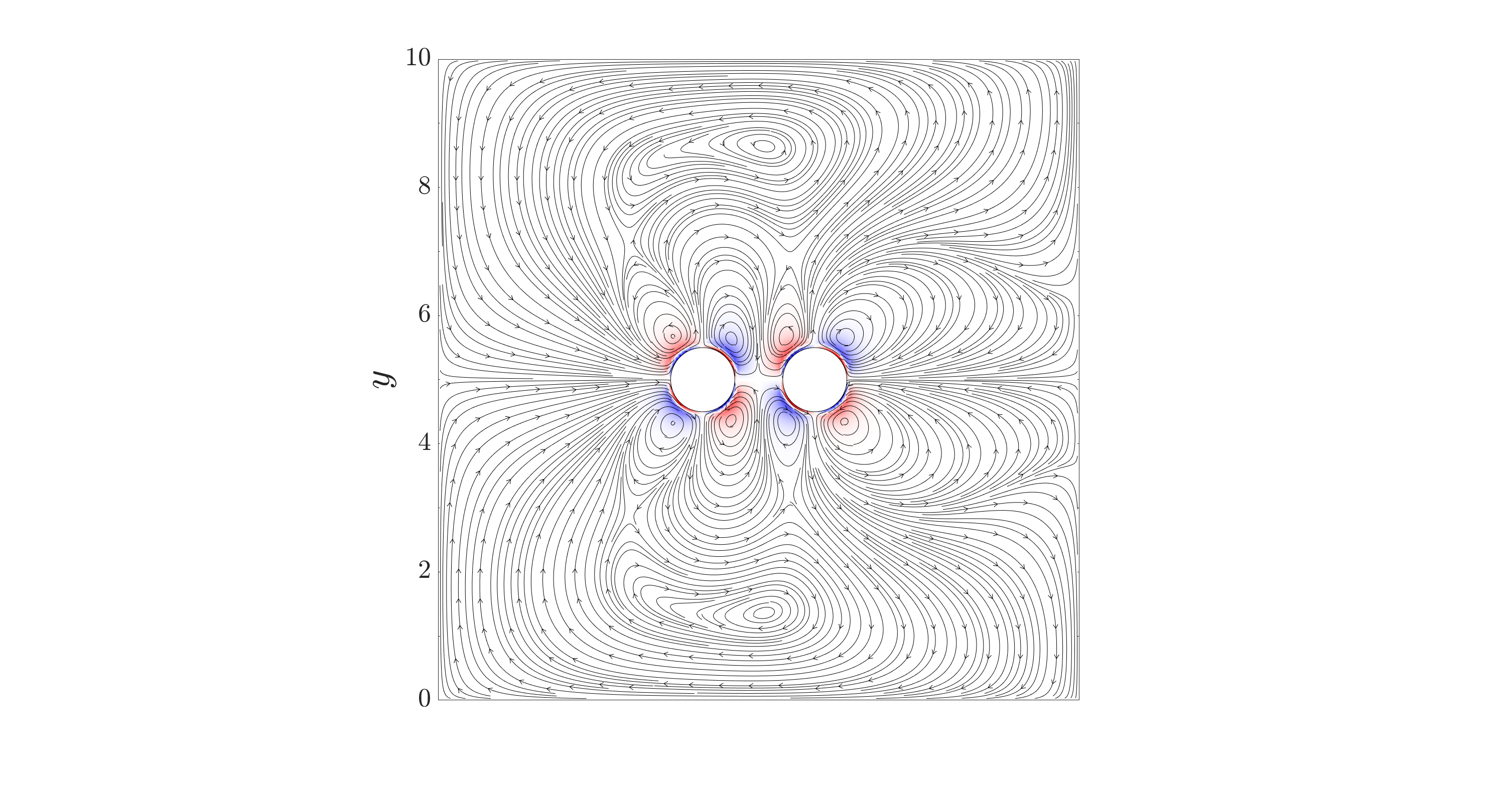}}{0.3cm}{0.05cm}
         \topinset{{\scriptsize (b)}}{\includegraphics[trim=11.5cm 0.4cm 13.8cm 1.5cm, clip,width=0.49\textwidth]{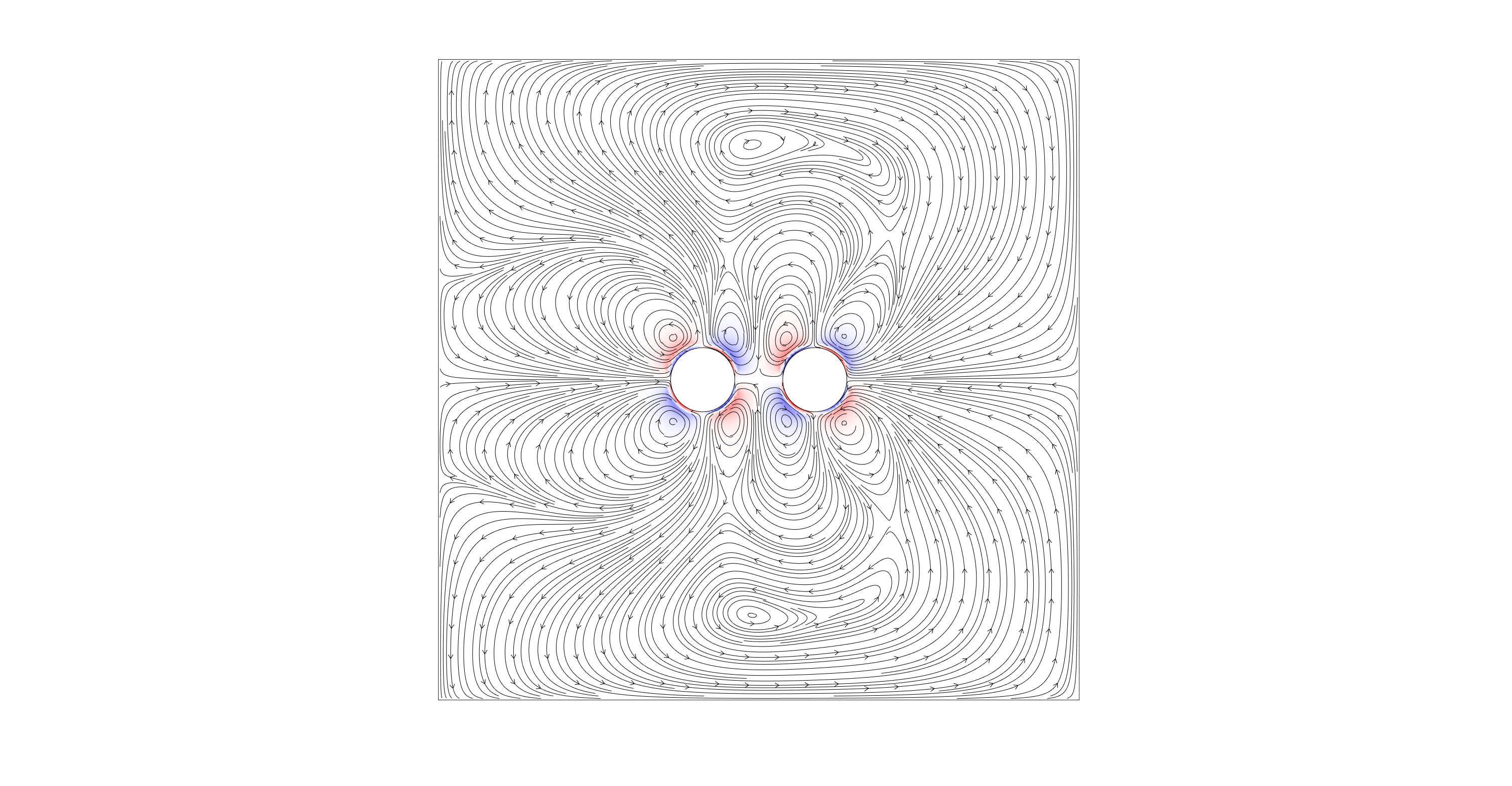}}{0.3cm}{0.3cm}
         \topinset{{\scriptsize (c)}}{\includegraphics[trim=11.5cm 0.4cm 13.8cm 1.5cm, clip,width=0.49\textwidth]{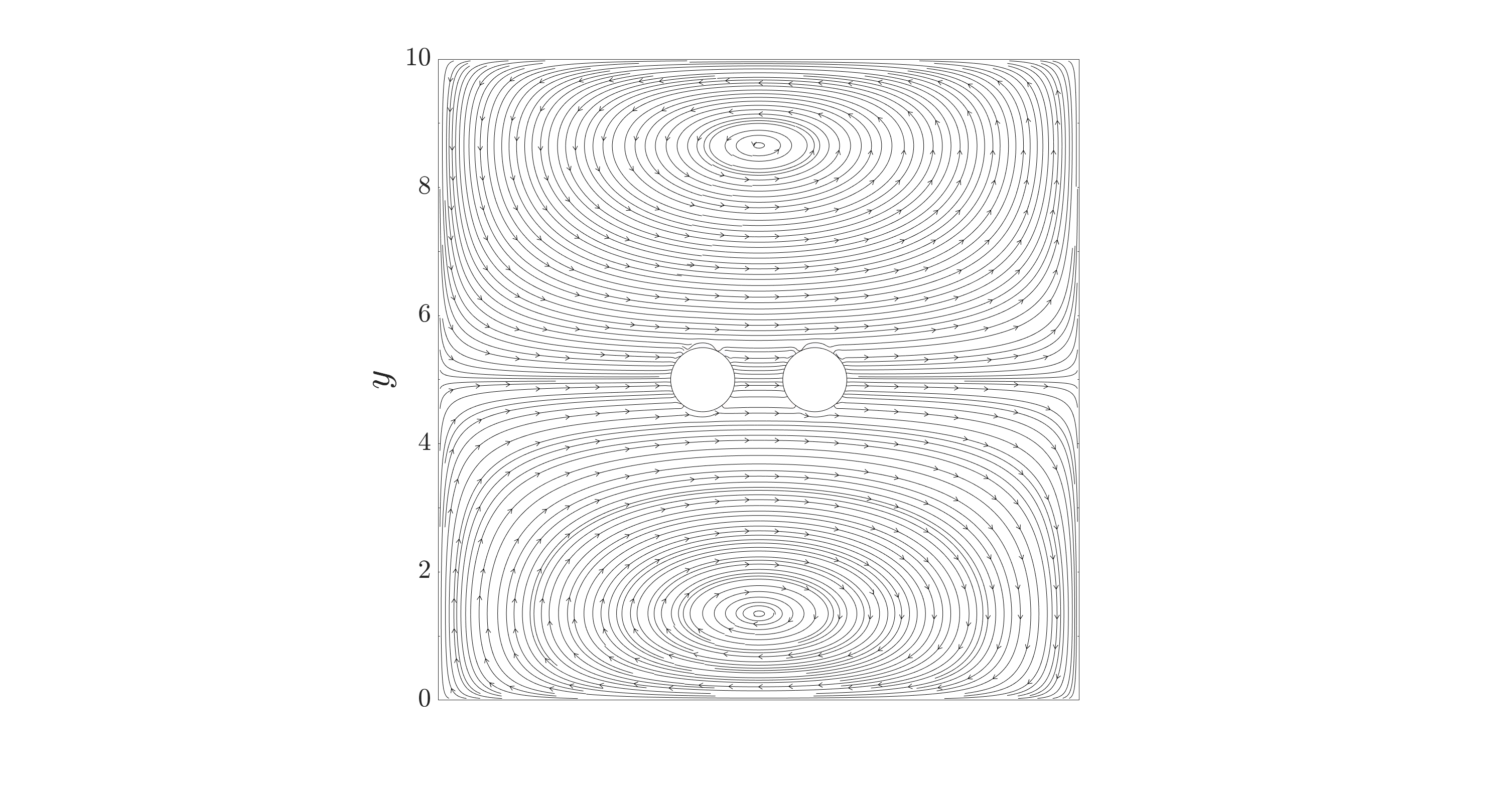}}{0.3cm}{0.05cm}
         \topinset{{\scriptsize (d)}}{\includegraphics[trim=11.5cm 0.4cm 13.8cm 1.5cm, clip,width=0.49\textwidth]{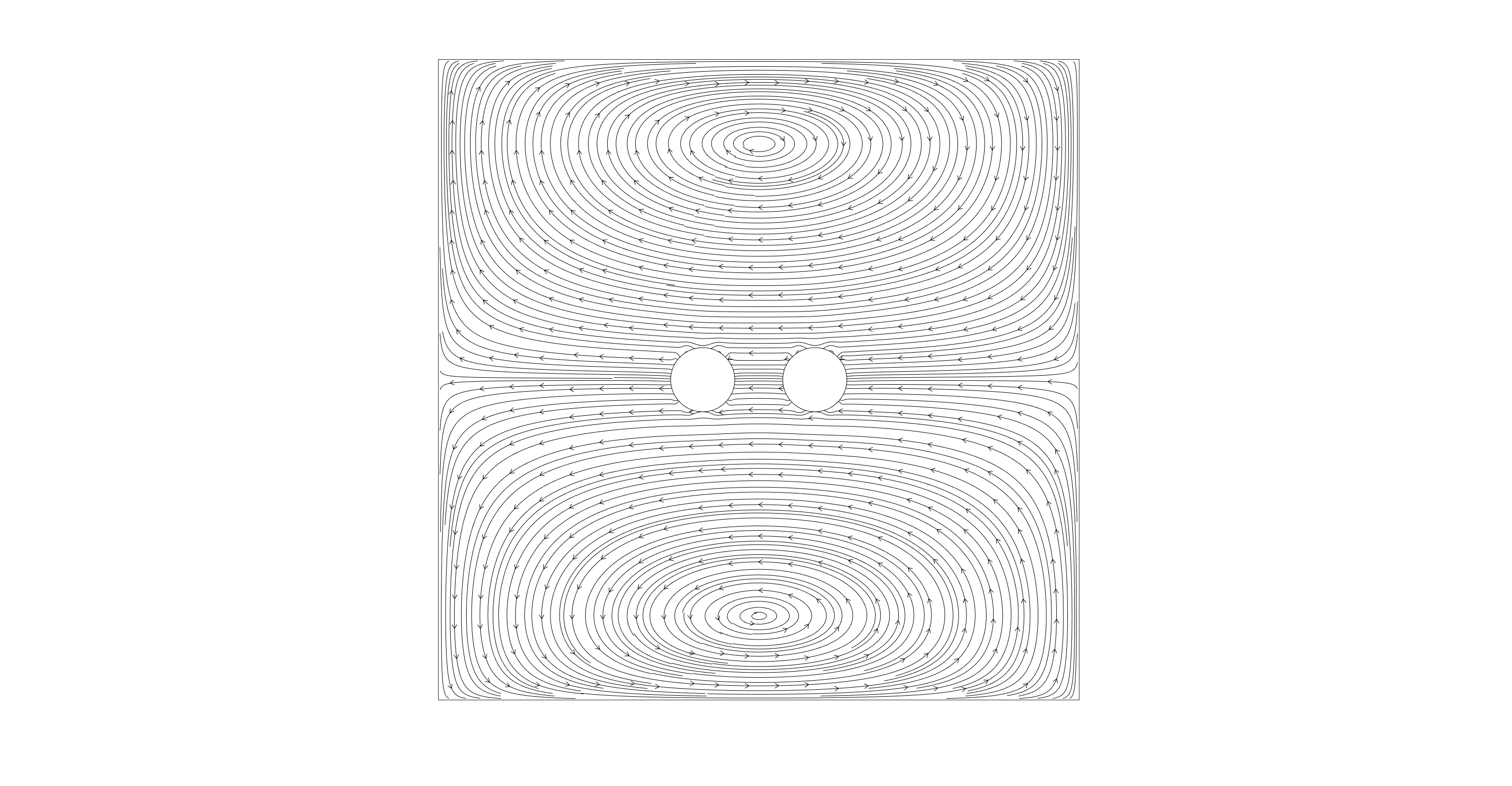}}{0.3cm}{0.3cm}
         \topinset{{\scriptsize (e)}}{\includegraphics[trim=11.5cm 0.4cm 13.8cm 1.5cm, clip,width=0.49\textwidth]{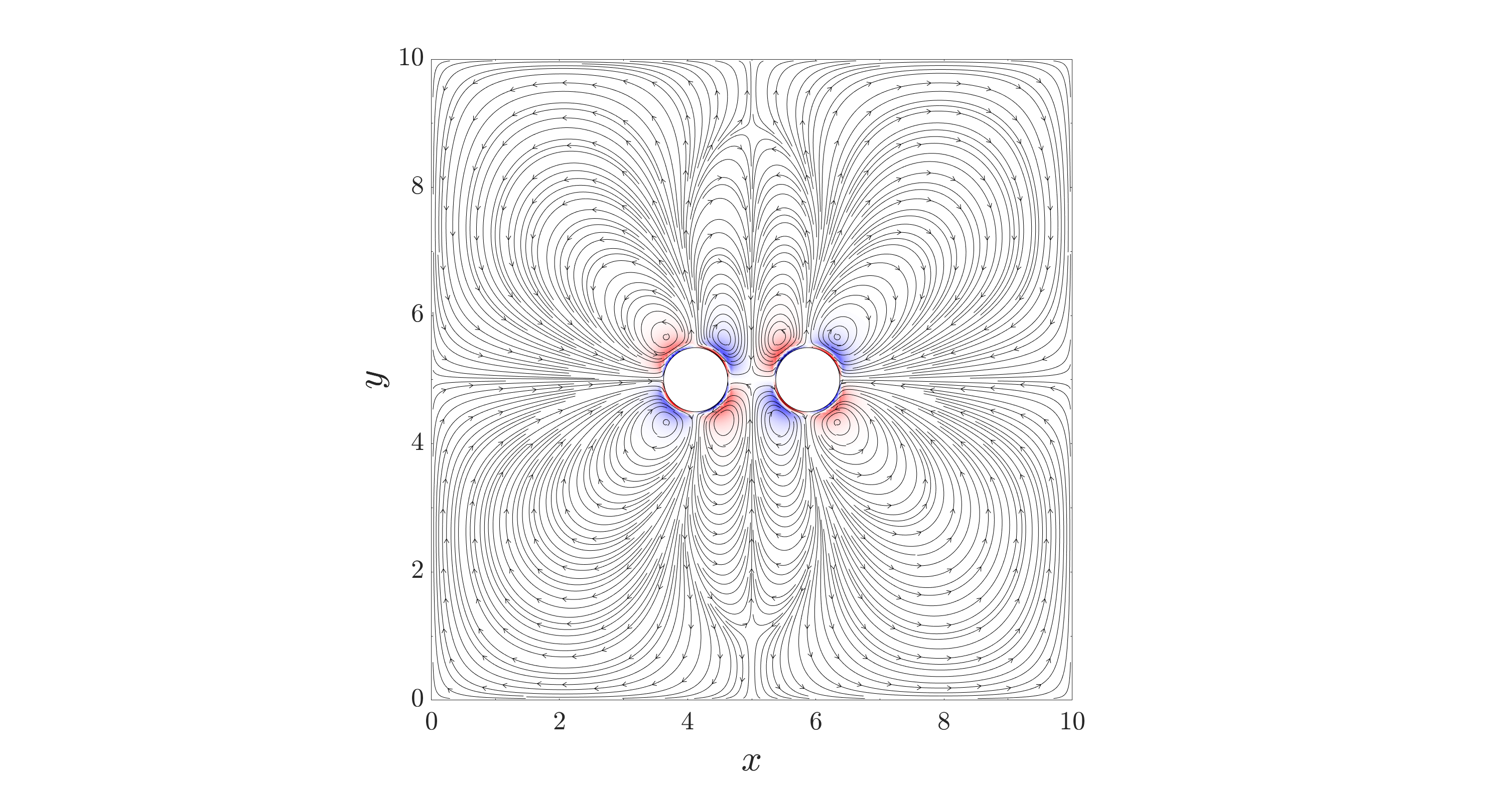}}{0.3cm}{0.05cm}
         \topinset{{\scriptsize (f)}}{\includegraphics[trim=11.5cm 0.4cm 13.8cm 1.5cm, clip,width=0.49\textwidth]{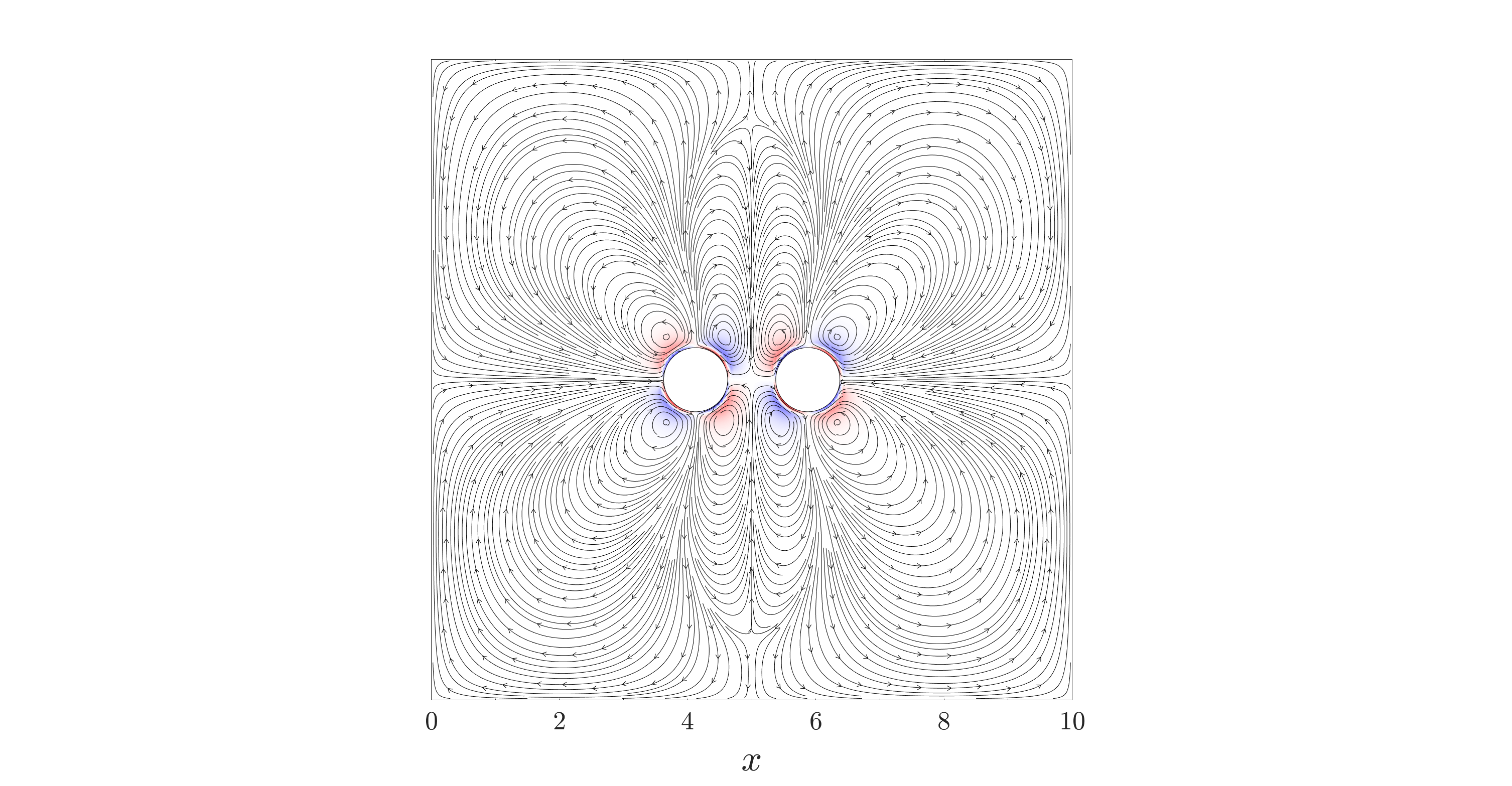}}{0.3cm}{0.3cm}
    \end{subfigure}
    \caption{Streamlines and vorticity contours of the total steady streaming (a) \& (b), its symmetric (c) \& (d) and antiymmetric components (e) \& (f) for a setup with $\zeta_i = 0.75$ and $S = 0.35$. Here, the left column represents $\rho_s = 1.78$  \mbox{(a), (c) \& (e)} and   $\rho_s = 0.47$ for the right column \mbox{(b), (d) \& (f)}. Color scheme as in figure~\ref{fig:FlowCharacteristics_TwoParticles}.}
    \label{fig:SteadyStreaming_SymmAntisymm_densities}
\end{figure}

According to the regime maps shown in figure \ref{fig:RegimeMap_loglog}, for $S = 0.35$ and $\zeta_i = 0.75$, all $\rho_s$ result in an attractive behavior of the particles.
Only $\rho_s = 4.68$ is assigned to the category of attraction, while $\rho_s = 1.78$ and $0.47$ are associated with transitional behavior according to \eqref{eq:classification_RegimeMap}.
The comparison of the total steady streaming (figures \ref{fig:SteadyStreaming_ZY-plane}a, \ref{fig:SteadyStreaming_SymmAntisymm_densities}a and \ref{fig:SteadyStreaming_SymmAntisymm_densities}b) indicates the same formations of vorticity contours with respect to the clockwise and counterclockwise rotation directions.
For each $\rho_s$, each respective particle is surrounded by four adjacent and four far-field contours.
The intensity of the vorticity decreases with decreasing $| \rho_s - 1 |$, which serves as a measure of the difference between the fluid and particle inertia and corresponds to the strength of the fluid-particle interaction.
As expected, these results show that the smaller the difference between the particle density and the fluid density, the smaller the inertial effect, and thus the particle and fluid motion are more in agreement.
Consequently, less intense flow develops, which is recognizable by lower values of vorticity as indicated by the blue and red contours.
Despite the lower vorticity, an examination of the streamlines of the total steady streaming velocity field yields that $\rho_s = 4.68$ and $1.78$ reveal very similar patterns of the respective vortex structures.
The vortex structures of $\rho_s = 0.47$ show the same patterns as for $\rho_s = 1.78$, but in a mirror-inverted arrangement. 

The same observation holds for the streamlines of the symmetric component for the three density ratios shown in figures \ref{fig:SteadyStreaming_SymmAntisymm}(a), \ref{fig:SteadyStreaming_SymmAntisymm_densities}(c) and \ref{fig:SteadyStreaming_SymmAntisymm_densities}(d), respectively.
For $\rho_s = 4.68$ and $1.78$, i.e. particles with higher density than the fluid, the upper vortex rotates counterclockwise and the lower one clockwise.
For particles less dense than the fluid, i.e. $\rho_s = 0.47$, it is the other way around.
This is due to the fact that the direction of action changes sign for particles with $\rho_s > 1$ and $\rho_s < 1$.
Consequently, particles denser than the surrounding fluid and, hence, larger inertia tend to follow the fluid motion in a reduced and delayed manner, resulting in an initial movement to the left in a non-inertial frame for the present scenario.
Less dense particles, on the other hand, have a lower inertia than the fluid and accelerate faster than the fluid, so that they overshoot and move to a direction opposite to those of denser particles.
After the initial displacement, particles slowly migrate back to the center of the domain, which causes the background flow in the symmetric component (\textit{cf.} \S \ref{sec:initialMotion}).
The vorticity for the symmetric component, however, remains  small for all $\rho_s$ so that their contours do not show up in figures \ref{fig:SteadyStreaming_SymmAntisymm}(a), \ref{fig:SteadyStreaming_SymmAntisymm_densities}(c) and \ref{fig:SteadyStreaming_SymmAntisymm_densities}(d).

Looking at the antisymmetric component shown in figures \ref{fig:SteadyStreaming_SymmAntisymm}(b), \ref{fig:SteadyStreaming_SymmAntisymm_densities}(e) and \ref{fig:SteadyStreaming_SymmAntisymm_densities}(f), we find the vorticity contours to be identical for all three density ratios. This was already observed for the total steady streaming (\textit{cf.} figures \ref{fig:SteadyStreaming_ZY-plane}a, \ref{fig:SteadyStreaming_SymmAntisymm_densities}a and \ref{fig:SteadyStreaming_SymmAntisymm_densities}b), but here the mirror-inverted arrangement was filtered out by the symmetric component.
This suggests that even though $\rho_s = 1.78$ and $0.47$ are classified as transitional, because they are only marginally attractive, and $\rho_s=4.68$ is strongly attractive, the mechanism that drives particle motion is dictated by the non-dimensional frequency $S$ and remains unaffected by the particle density ratio.
This result is consistent with the findings of \cite{2022_vanOverveld_etal}.

%%%%%%%%%%%%%%%%%%%%%%%%%%%%%%%%%
\subsection{Effect of oscillation frequency and initial particle distance}\label{sec:EffectFrequencyDistance}
%%%%%%%%%%%%%%%%%%%%%%%%%%%%%%%%%

The analysis of the decomposition of the total steady streaming demonstrated that the antisymmetric component governs the flow structures that drive the pairwise fluid-particle interactions.
Therefore, all subsequent illustrations and analyses refer to the antisymmetric part only.
While the flow structures seem to be less influenced by the density ratio $\rho_s$, the applied frequency $S$ appears to have a major effect.
This could already be seen in figures \ref{fig:SteadyStreaming_SymmAntisymm} and \ref{fig:SteadyStreaming_SymmAntisymm_densities} as well as in the regime maps, presented in figure \ref{fig:RegimeMap_loglog}.
To analyze whether the change of $S$ has an impact on the resulting flow patterns of the antisymmetric part, we show a comparison of the flow field for three different characteristic frequencies.
The chosen frequencies ${S = \{0.35, 0.70, 1.05\}}$ correspond to attraction, transition, and repulsion, respectively, for the given $\zeta_i = 0.50$ and $\rho_s = 4.68$. 
The resulting streamlines and vorticity contours are depicted in figure \ref{fig:SteadyStreaming_increasingFrequency}, with $S = 0.35$ in figure \ref{fig:SteadyStreaming_increasingFrequency}(a), $S = 0.70$ in \ref{fig:SteadyStreaming_increasingFrequency}(b), and $S = 1.05$ in \ref{fig:SteadyStreaming_increasingFrequency}(c) .
The flow structures of the attractive and repulsive setups (figures \ref{fig:SteadyStreaming_increasingFrequency}a and c), respectively, are again very similar to the corresponding flow structures observed for larger initial gap size $\zeta_i = 0.75$ (figures \ref{fig:SteadyStreaming_SymmAntisymm}c and f). 
Hence, for attraction (repulsion) the peripheral (central) vortices dominate. 
\begin{figure}
    \def\stackalignment{l}
    \centering
    \captionsetup[subfigure]{labelformat=empty}
    \begin{subfigure}[b]{\textwidth}
         \centering
         \topinset{\scriptsize (a)}{\includegraphics[trim=11.5cm 0.4cm 13.8cm 1cm, clip,width=0.49\textwidth]{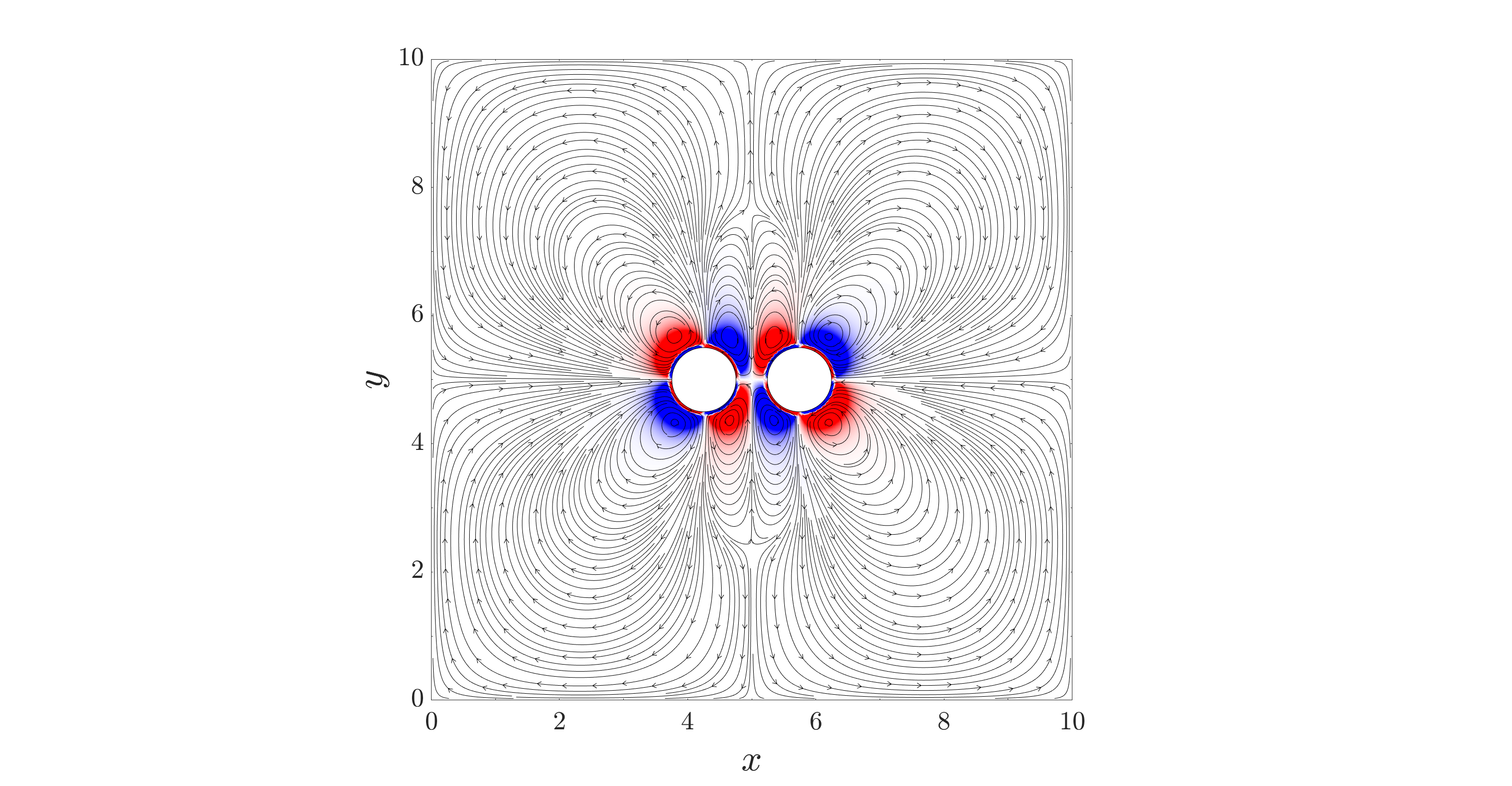}}{0.3cm}{0.05cm}
         \topinset{\scriptsize (b)}{\includegraphics[trim=11.5cm 0.4cm 13.8cm 1cm, clip,width=0.49\textwidth]{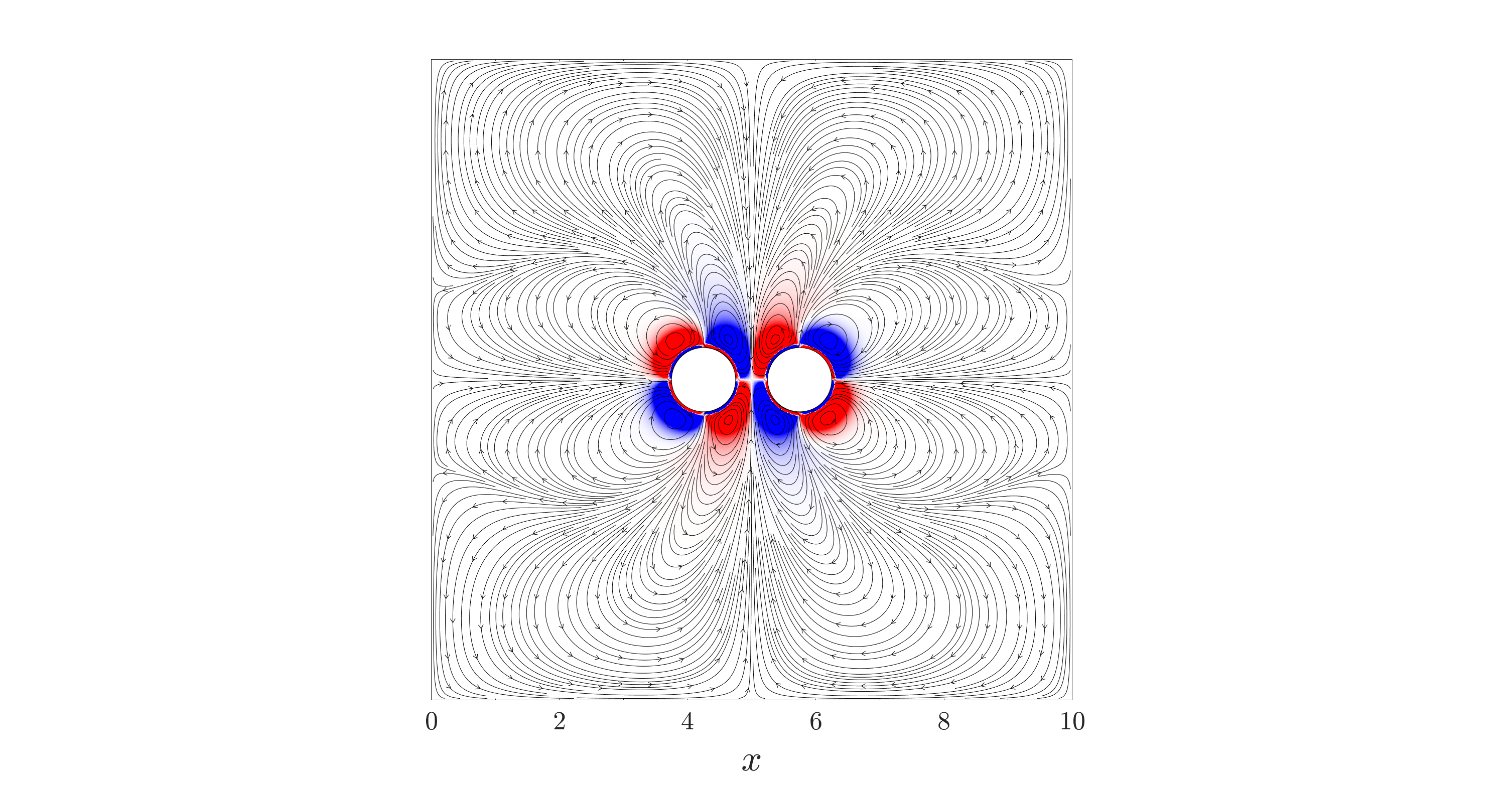}}{0.3cm}{0.4cm}
         \topinset{\scriptsize (c)}{\includegraphics[trim=11.5cm 0.4cm 13.8cm 1cm, clip,width=0.49\textwidth]{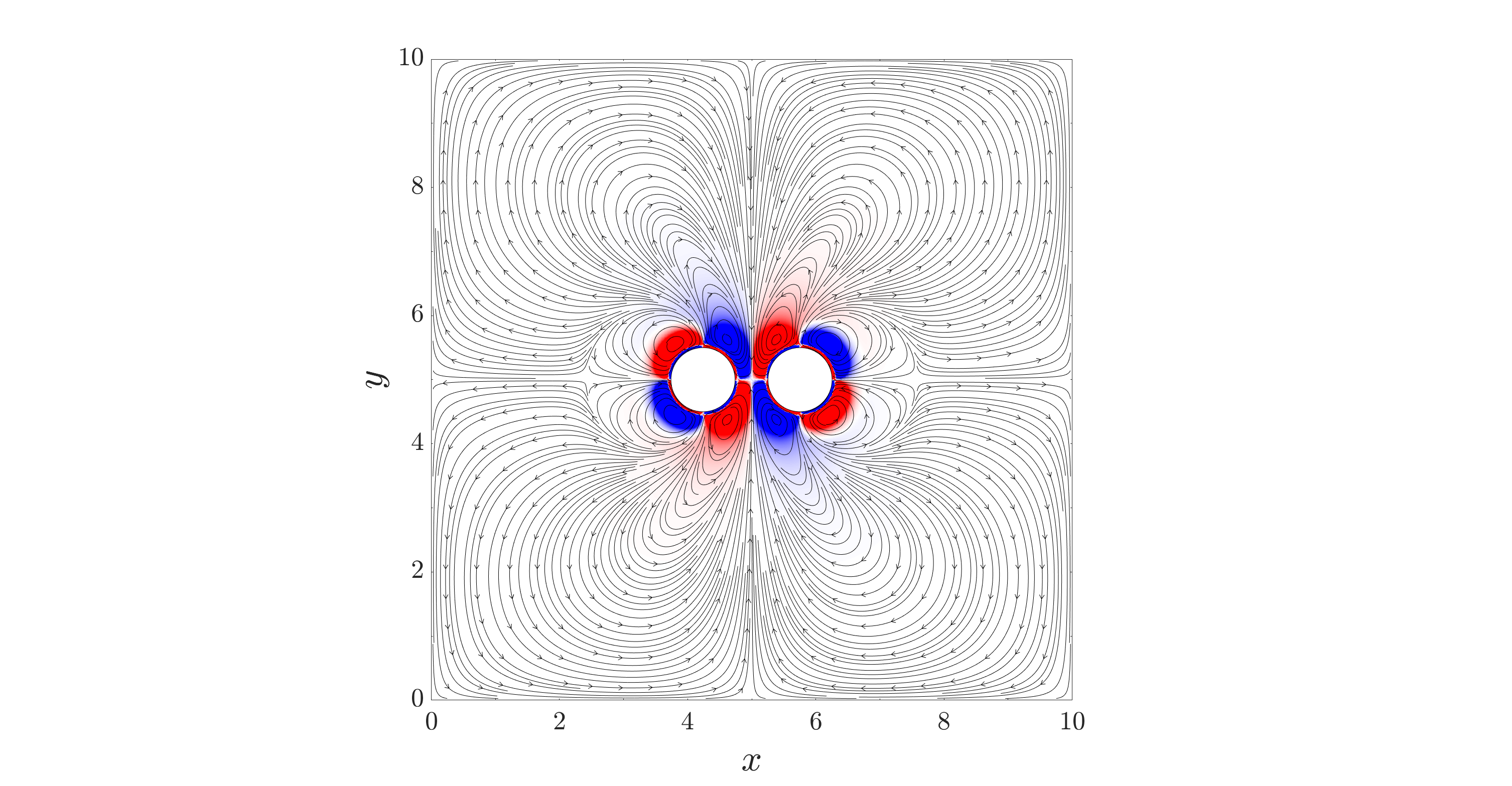}}{0.3cm}{0.05cm}
    \end{subfigure}
    \caption{Streamlines and vorticity contours of the antisimmetric part of the flow field for a constant $\zeta_i = 0.50$ and varying $S$. Figure (a) represents $S = 0.35$ with attractive behavior, figure (b) shows $S = 0.70$ (transition), and figure (c) illustrates $S = 1.05$ (repulsion). Color scheme as in figure~\ref{fig:FlowCharacteristics_TwoParticles}.}
    \label{fig:SteadyStreaming_increasingFrequency}
\end{figure}

Another interesting aspect arises if one considers the transition of the respective vortices from attraction via the transition to repulsion. 
While, qualitatively, the central vortex structures are completely suppressed in case of attraction, they appear to be in equilibrium, when the transition is established, and become clearly superior in case of repulsion. 
Naturally, we find the corresponding opposite trend for the peripheral vortices. 
The alteration of the central and peripheral vortex sizes can also be observed on the basis of the vorticity contours.
Even though the changes of the contours vary only slightly in the course of the increase of $S$, it can be seen that the size of the central vortices increases while the peripheral decrease.

As already indicated by the regime map given by figure \ref{fig:RegimeMap_loglog}, the behavior of the two particles in oscillation is dictated by frequency and the initial distance of the two particles.
We, therefore, also examine the impact of $\zeta_i$ on the flow characteristics in figure \ref{fig:SteadyStreaming_increasingGap} by comparing $\zeta_i = \{0.10, 0.25, 0.50\}$ for $S = 0.94$ and $\rho_s = 4.68$. 
Here, $\zeta_i = 0.10$ results in attraction (figure \ref{fig:SteadyStreaming_increasingGap}a), $\zeta_i = 0.25$ leads to a transitional state (figure \ref{fig:SteadyStreaming_increasingGap}b), and \mbox{$\zeta_i = 0.50$} yields repulsion (figure \ref{fig:SteadyStreaming_increasingGap}c).
At first glance, the vorticity contours of the three arrangements appear to be almost identical, and the flow patterns also show a close resemblance with predominant central and suppressed peripheral vortex structures, respectively.
A closer look at the peripheral flow structures, however, shows a slight decrease of peripheral vortex size with increasing $\zeta_i$.
This is accompanied by the fact that with increasing $\zeta_i$, more fluid can flow into the gap, increasing the strength of the flow into the gap trying to push the particles apart.
\begin{figure}
    \def\stackalignment{l}
    \centering
    \captionsetup[subfigure]{labelformat=empty}
    % Top
    \begin{subfigure}[b]{\textwidth}
        \centering
        \topinset{\scriptsize (a)}{\includegraphics[trim=11.5cm 0.4cm 13.8cm 1cm, clip,width=0.49\textwidth]{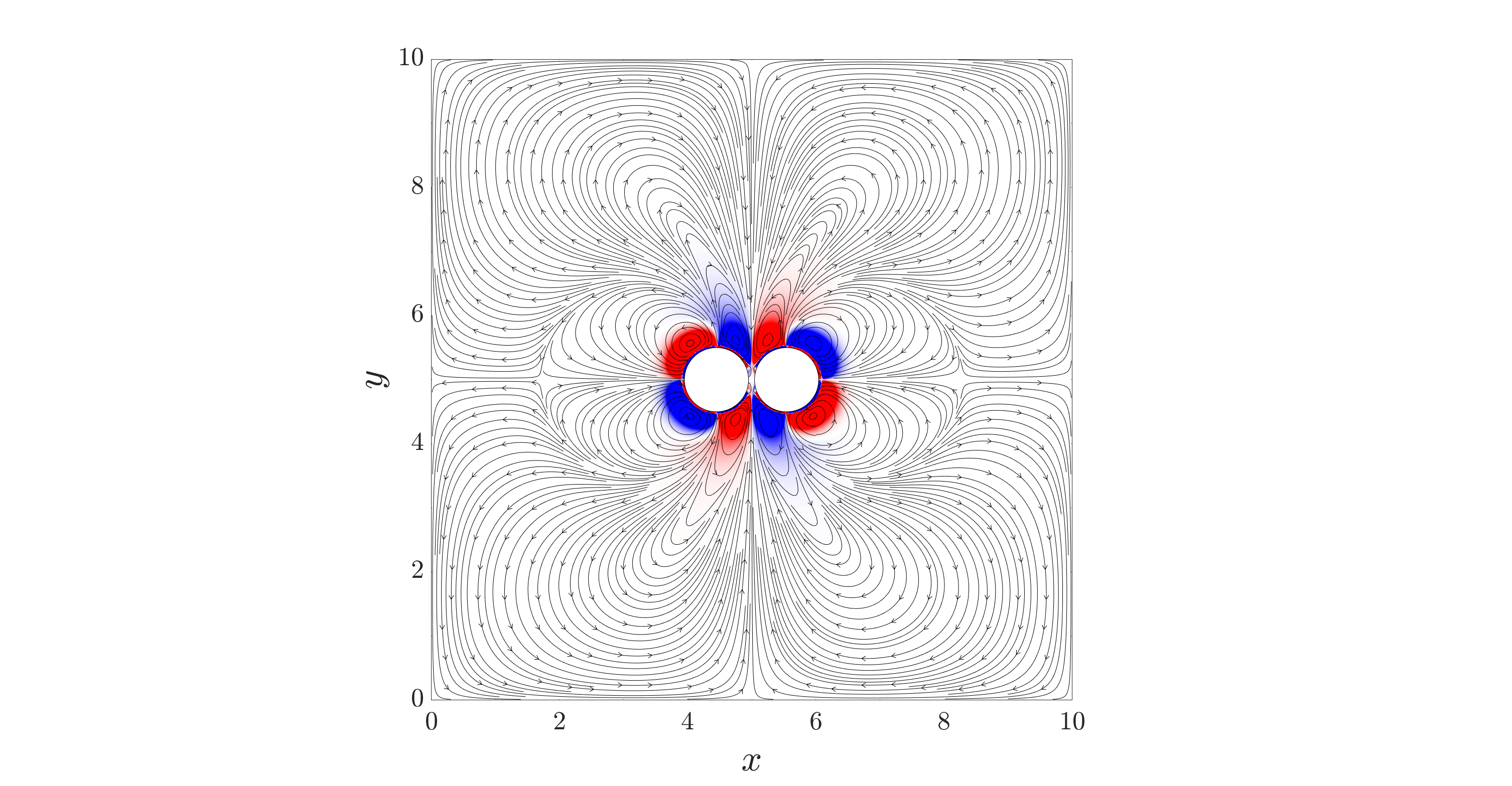}}{0.3cm}{0cm}
         \topinset{\scriptsize (b)}{\includegraphics[trim=11.5cm 0.4cm 13.8cm 1cm, clip,width=0.49\textwidth]{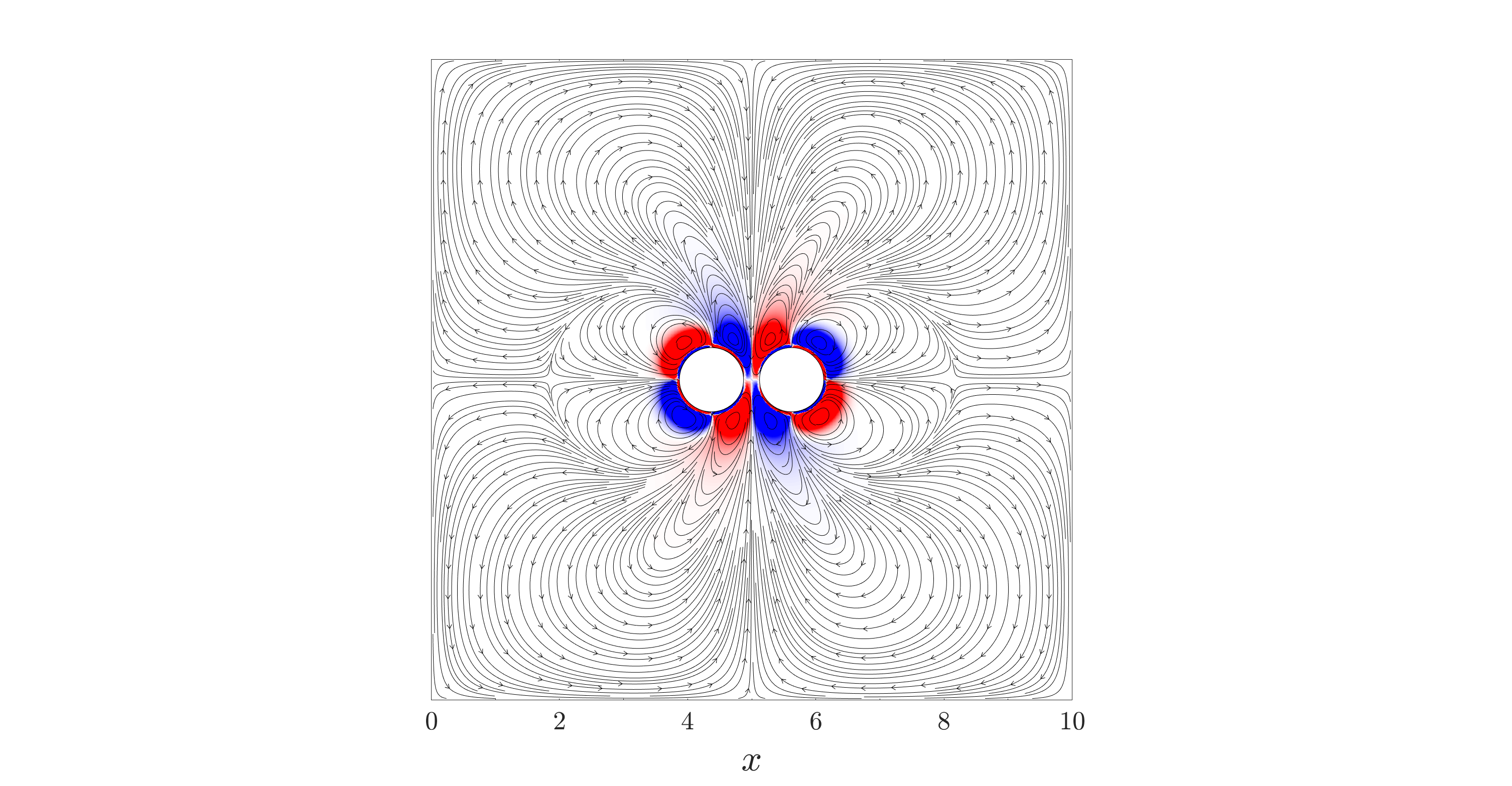}}{0.3cm}{0.05cm}
         \topinset{\scriptsize (c)}{\includegraphics[trim=11.5cm 0.4cm 13.8cm 1cm, clip,width=0.49\textwidth]{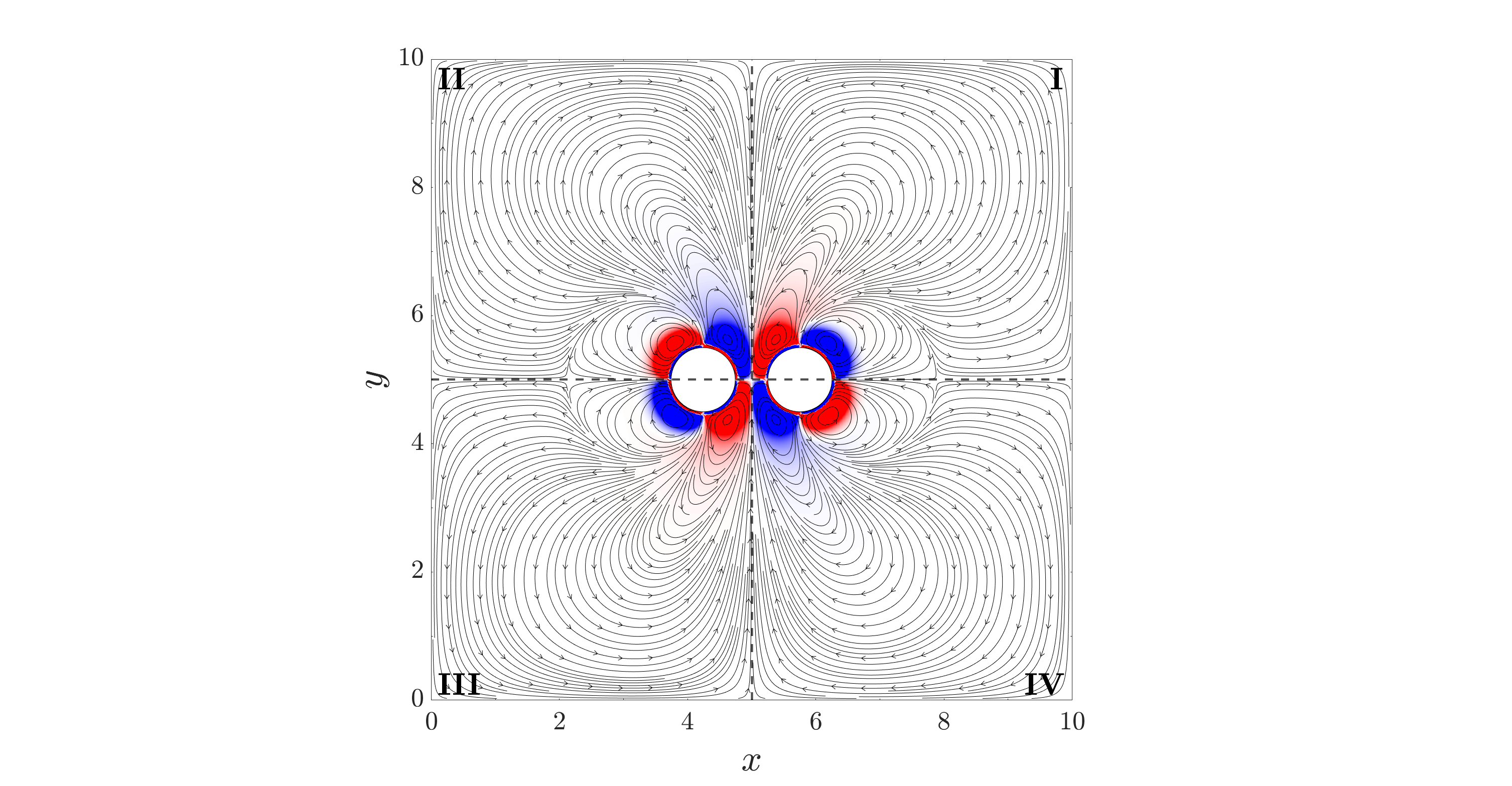}}{0.3cm}{0.05cm}
    \end{subfigure}
    \caption{Streamlines and vorticity contours of the antisimmetric part of the fluid field for constant $S = 0.94$ and varying $\zeta_i$. (a)  $\zeta_i = 0.10$ resulting in attraction,  (b)  $\zeta_i = 0.25$ leading to transition, and  (c)  $\zeta_i = 0.50$ causing repulsion. Color scheme as in figure~\ref{fig:FlowCharacteristics_TwoParticles}. The dashed lines in (c) indicate the symmetry lines used for the division into four quadrants to compute $\omega_{z,\alpha}$ \eqref{eq:Circulation}.}
    \label{fig:SteadyStreaming_increasingGap}
\end{figure}

%%%%%%%%%%%%%%%%%%%%%%%%%%%%%%%%%
\subsection{A circulation-based criterion to determine the particle interaction}\label{sec:circulationAnalysis}
%%%%%%%%%%%%%%%%%%%%%%%%%%%%%%%%%

The qualitative examination of the antisymmetric component of the steady streaming shows that the visualization of the flow structures can be helpful in determining the behavior of two particles in oscillatory flows. 
Owing to the low flow intensities, however, it is challenging to make a clear distinction of repulsion and attraction. 
This was already demonstrated in the context of analyzing different density ratios, where we find the same flow patterns for strongly and weakly attractive behavior for dense and less dense particles, respectively. 
To come to a more general conclusion, a more precise and quantitative measure for the interactions is, therefore, needed. 
For this purpose, we compute the circulation of the antisymmetric part of the flow field separately for central and peripheral vortices.

\begin{figure}
    \def\stackalignment{l}
    \centering
    \captionsetup[subfigure]{labelformat=empty}
    % Top
    \begin{subfigure}[b]{\textwidth}
        \centering
        \topinset{\scriptsize (a)}{\includegraphics[trim=4cm 3cm 6cm 1.4cm, clip,width=0.8\textwidth]{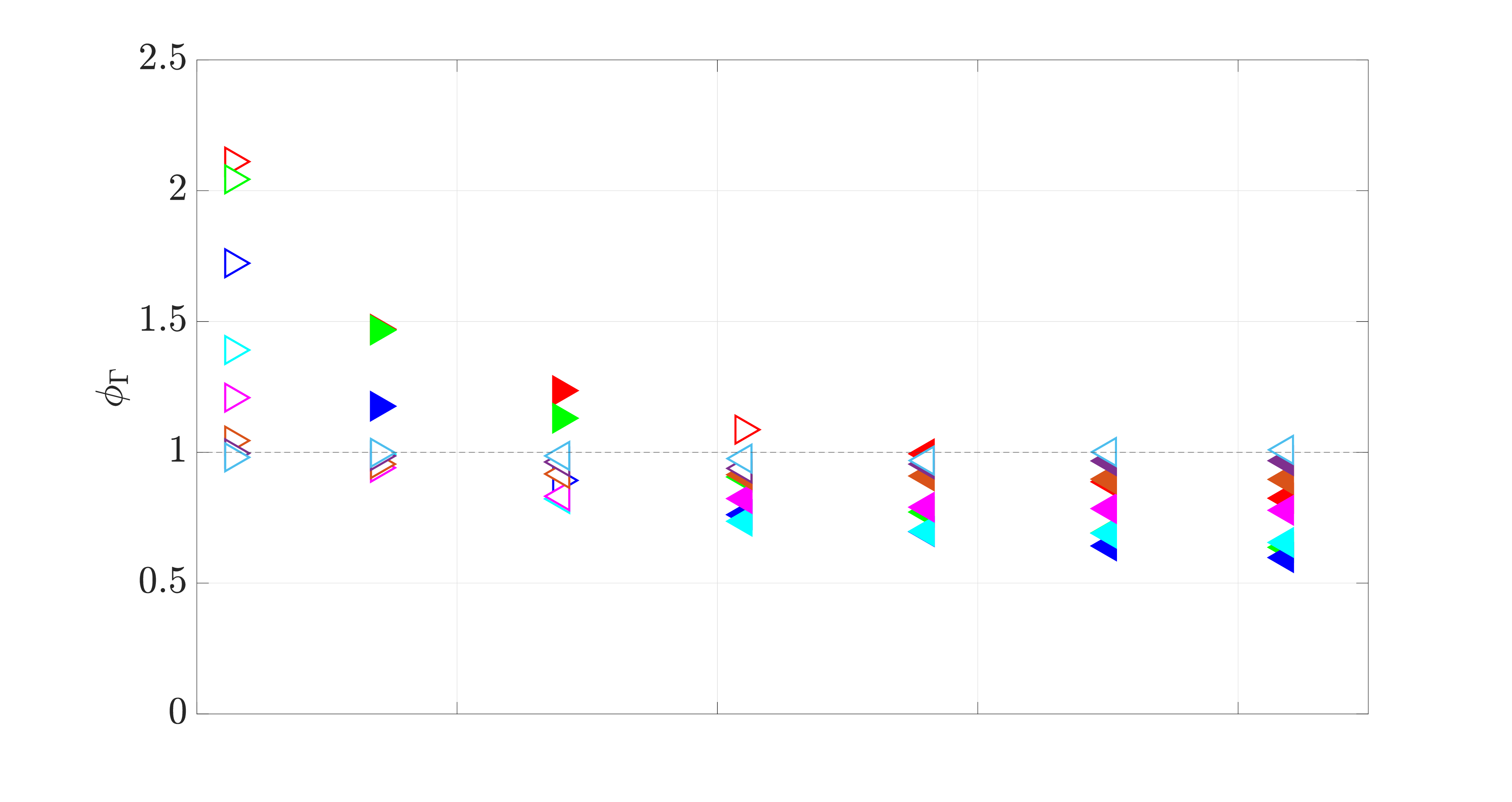}}{0.35cm}{0cm}
        \topinset{\scriptsize (b)}{\includegraphics[trim=4cm 3cm 6cm 1.4cm, clip,width=0.8\textwidth]{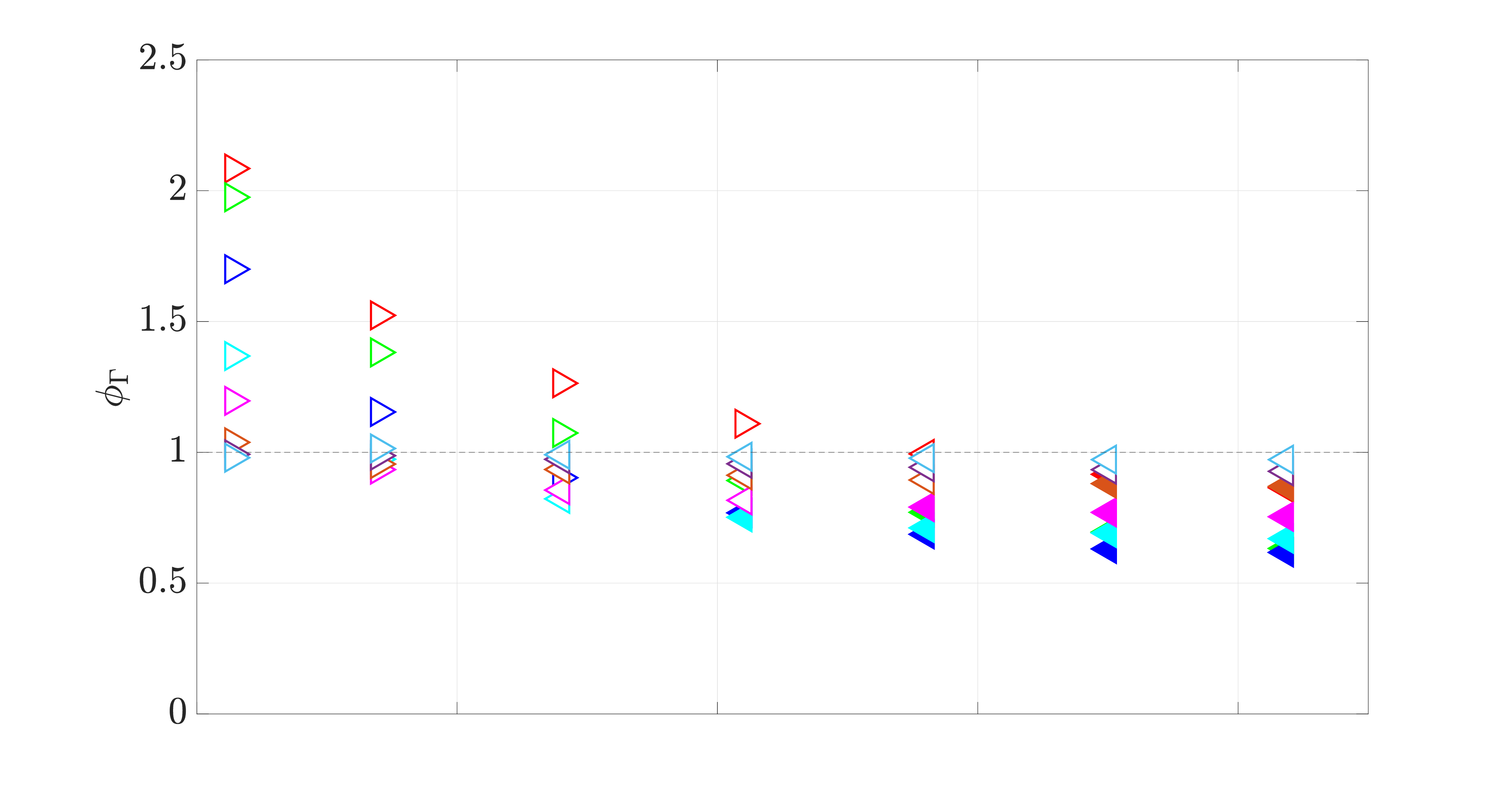}}{0.35cm}{0cm}
        \topinset{\scriptsize (c)}{\includegraphics[trim=4cm 0.55cm 6cm 1.4cm, clip,width=0.8\textwidth]{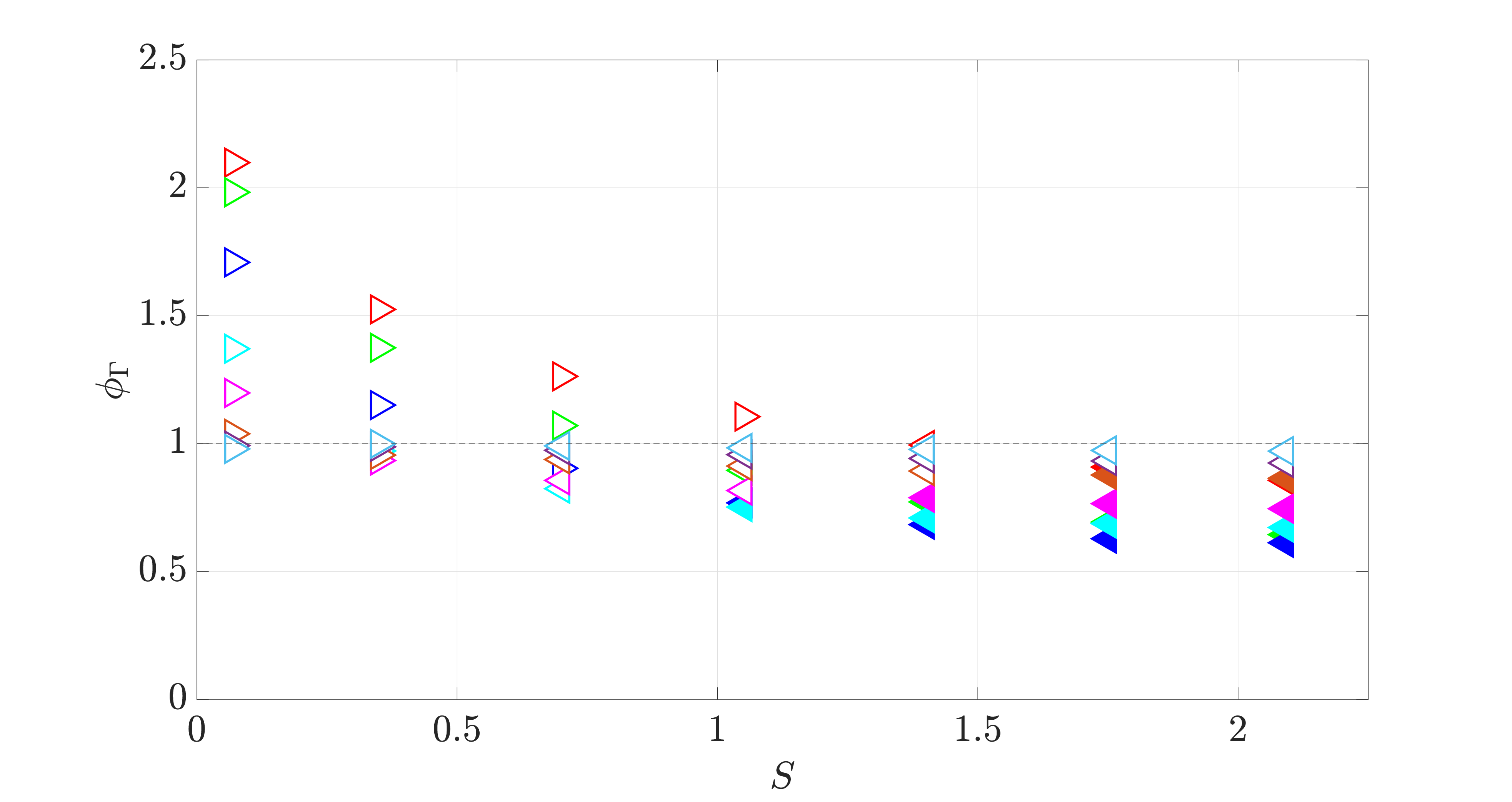}}{0.35cm}{0cm}
    \end{subfigure}
    \caption{Circulation ratio $\phi_\Gamma = \Gamma_{attr} / \Gamma_{rep}$ in dependence of $\zeta_i$ and $S$. The color scheme represents the initial particle distances, where 
    \mbox{\textcolor{color_1}{$\vartriangleleft \vartriangleright$} $\zeta_i = 0.10$}, \mbox{\textcolor{color_2}{$\vartriangleleft \vartriangleright$} $\zeta_i = 0.25$},
    \mbox{\textcolor{color_3}{$\vartriangleleft \vartriangleright$} $\zeta_i = 0.50$},
    \mbox{\textcolor{color_4}{$\vartriangleleft \vartriangleright$} $\zeta_i = 0.75$},
    \mbox{\textcolor{color_5}{$\vartriangleleft \vartriangleright$} $\zeta_i = 1.00$},
    \mbox{\textcolor{color_6}{$\vartriangleleft \vartriangleright$} $\zeta_i = 1.50$},
    \mbox{\textcolor{color_7}{$\vartriangleleft \vartriangleright$} $\zeta_i = 2.00$},
    \mbox{\textcolor{color_8}{$\vartriangleleft \vartriangleright$} $\zeta_i = 3.00$}.
    Right pointing triangles ($\vartriangleright$) indicate $\zeta_{100} < 0$ and left pointing triangles ($\vartriangleleft$) $\zeta_{100} > 0$. According to the classification stated in \eqref{eq:classification_RegimeMap}, filled symbols either represent attraction or repulsion depending on the orientation of the symbol, and open symbols indicate transition.}
    \label{fig:circulationRatios}
\end{figure}

In a first step, we assign each vorticity contour value of the $xy$-plane to contribute to either attractive or repulsive behavior based on the considerations explained in \S \ref{sec:totalSteadyStreaming} and sketched in figure \ref{fig:FlowCharacteristics_TwoParticles}(b).
Since this cannot be done on the basis of the direction of rotation only, we divide the flow field into quadrants as shown in figure \ref{fig:SteadyStreaming_increasingGap}(c), where the dashed lines separate the quadrants and the quadrant numbering is introduced by the roman numbers.
The intersection of the two lines coincides with the stagnation point between the particles.
Starting from this point, the subdivision of the field follows the axes of symmetry.
The assignment of the vorticity contours is done  for each quadrant separately and takes the contribution of the respective vorticity, $\omega_{z,\alpha}$, into account.
Here, $\alpha$ is used as an index for either attraction or repulsion.
Where, according to figure \ref{fig:FlowCharacteristics_TwoParticles}(b), central vortices push fluid inside the gap and act repulsively whereas peripheral vortices push against the particles from the outside so that particles tend to approach each other.
With the help of this assignment, we calculate the circulation $\Gamma_\alpha$ contributing to attraction or repulsion, respectively, as
\begin{equation}
    \Gamma_\alpha = \iint_{A} \omega_{z, \alpha} \,dA \quad .
    \label{eq:Circulation}
\end{equation}
On this basis, we can compute the ratio of attractive to repulsive circulation ${\phi_\Gamma = \Gamma_{attr} / \Gamma_{rep}}$ and, thus, draw conclusions about the behavior of two particles in oscillating fluid flow.
A ratio of $\phi_\Gamma = 1$ would represent a balance of the central and peripheral vortices, $\phi_\Gamma > 1$ indicates  a predominantly attractive circulation leading to the approach of the two particles, and $\phi_\Gamma < 1$ refers to repulsive behavior due to dominant circulation pushing the particles apart. 

Figure \ref{fig:circulationRatios} presents $\phi_\Gamma$ for the considered density ratios $\rho_s$.
The same set of $S$ is used as in figures \ref{fig:zeta_varyingS}.
The dashed lines highlight the balance of the circulations in each graph, i.e. $\Gamma_\alpha=1$, and the symbols correspond to the regime map in figure \ref{fig:RegimeMap_loglog}.
As already shown in \S\ref{sec:EffectParticleInertia}, the influence of particle inertia on the behavior of the two particles in terms of the qualitative distinction of either attraction or repulsion is small.
Consequently, the ratios for the three densities $\rho_s = 4.68$ (figure \ref{fig:circulationRatios}a), $1.78$ (figure \ref{fig:circulationRatios}b) and $0.47$ (figure \ref{fig:circulationRatios}c) do not differ much, if we choose to normalize the attractive circulation by the respective repulsive circulation of the same case.
The trend that small $S$ together with small $\zeta_i$ lead to attractive results is reflected by the fact that these frequencies lead to $\phi_\Gamma > 1$, whereas we find $\phi_\Gamma < 1$ for repulsive setups for large $S$ and large $\zeta_i$.

In the context of Figure \ref{fig:RegimeMap_loglog}, it was already shown that the main difference between the considered density ratios is the assignment of different classes, with density ratios closer to neutrally buoyant conditions leading to more cases that we classify as transitional.
The circulation-based criterion depicted in figure \ref{fig:circulationRatios} now provides a solid measure for distinguishing the different behavior of repulsion and attraction. 
Indeed, the finding in \S\ref{sec:inter-particleDistance} that the interaction of the flow fields of the respective particles increases with decreasing $\zeta_i$ is clearly evident by the increase in $\phi_\Gamma$ with decreasing $\zeta_i$. 
For larger $\zeta_i$, $\phi_\Gamma$ decreases to values below unity. 
This observation now holds for all density ratios considered, where the different cases shown in figure \ref{fig:circulationRatios} appear to be almost identical.

By means of this quantitative analysis, the conclusion from \S \ref{sec:EffectParticleInertia} can be confirmed that the inertia of the particles, characterized by the density, has no effect on the respective behavior with regard to attraction or repulsion.
Thus, this is solely determined by the non-dimensional oscillation frequency $S$ and the initial particle distance $\zeta_i$.

%%%%%%%%%%%%%%%%%%%%%%%%%%%%%%%%%
%%%%%%%%%%%%%%%%%%%%%%%%%%%%%%%%%
\section{Conclusions}\label{sec:conclusions}
%%%%%%%%%%%%%%%%%%%%%%%%%%%%%%%%%
%%%%%%%%%%%%%%%%%%%%%%%%%%%%%%%%%

To investigate the potential hydrodynamic mechanism that induces flocculation of primary particles in oscillatory flows, a systematic simulation campaign was conducted to analyze the mutual particle behavior of two freely moving particles subjected to oscillatory fluid flow.
In such flows, particles tend to drift based on their inertia, which can cause particle trajectories to deviate from the background oscillatory motion.
To this end, particle-resolved simulations were performed for the setup of two monodisperse mobile particles submerged in a viscous fluid, where gravity was neglected to focus on particle inertia.
Three particle density ratios were considered: a large value with pronounced inertia, and two corresponding less inertial density values being denser and lighter than the ambient fluid, but approximately equally far from neutrally buoyancy conditions.

In a first step, different orientations of the particle arrangement in relation to the oscillation direction were investigated.
The results showed that any arrangement whose orientation angle deviates from the direction of oscillation will be aligned perpendicular to it over time.
Previous studies have shown that in the frequency range we investigated, particles whose arrangement is perpendicular to the oscillation reach a state of equilibrium in which the distance between the particles remains constant \citep{2017_Fabre_etal, 2018_Jalal}.
An arrangement in line with the direction of oscillation, on the other hand, remains in its orientation, in which the particles move horizontally towards or away from each other, respectively.
On this basis, we have focused in detail on horizontally aligned arrangements and investigated the impact of the oscillation frequency, the initial distance and the particle density on the mutual particle behavior.
To this end, we analyzed the distance between the particles over a period of 100 oscillations and created a regime map that defines threshold conditions to distinguish between attractive and repulsive behavior in dependence on the above-mentioned parameters.
The regime map was found to be self-similar for the three different density ratios, whereby the threshold conditions can be described by a power law of approximately the same coefficients.
It was also shown that small distances and low frequencies tend to lead to an attraction of the two particles, while in contrast, large distances and higher frequencies lead to a repulsion.

The mutual particle behavior was then further analyzed by visualizing the flow patterns initiated by the fluid-particle interaction.
In this context, we have calculated the steady streaming, i.e. the averaging over an oscillation period, and filtered out the impact of the initial state of a fluid at rest by decomposing the flow field into its symmetric and antisymmetric components.
The latter component is decisive for the individual motion of the particles and characterized by a quadrupole flow structure surrounding each particle when analyzing the plane through the torroidal vortices.
Based on this decomposition, the competing effect of the quadrupole structures acting on the particles was revealed and identified as the decisive factor for the respective particle behavior.
On the one hand, the central vortices act to pump fluid into the gap to drive the particles apart. 
On the other hand, the peripheral vortices act to push particles together from the outside. 
We have also demonstrated this effect quantitatively by calculating the circulation of the steady streaming and comparing the respective components, i.e. vortex structures that promote either attraction or repulsion of the particles.
Our results show that this effect is the governing mechanism for all density ratios investigated in this study and yielded identical results in terms of the competing circulation patterns in all cases. 
Given the self-similar results of the circulation produced for the three density ratios considered in the present study, the threshold condition for attractive and repulsive behavior that can be expressed by a power-law appears to be universal and solely dependent on the oscillation frequency and the initial particle distance.

Our analysis could therefore open up opportunities to control flocculation of particles based on hydrodynamic effects in oscillating devices with high precision.
On the foundation presented here, this would primarily find application in microfluidics, where the number of particles can be controlled and spatial constraints can induce an orientation of the particles that is approximately aligned with the oscillation direction.
However, future analyses of this kind should extend the parameter space to oscillation amplitudes and particles of different size.
In addition, a larger container size with more particles would extend the presented framework and could provide insights into particle interactions over both shorter and longer ranges in more complex arrangements.

\vspace{.1in}
\noindent
{\bf Funding}\\
FK and BV gratefully acknowledge support through German Research Foundation (DFG) grant VO2413/2-1. 
FK also gratefully acknowledges the support of a scholarship from the German-American Fulbright Commission. 
EM and PLF were supported through NSF grant CBET-1638156. 
EM furthermore acknowledges support through U.S. Army ERDC grant W912HZ22C0037, U.S. ARO grant  W911NF-23-2-0046, and through NSF grant HS EAR 2100691.

\vspace{.1in}
\noindent
{\bf Acknowledgements}\\
The authors thank Roger H. Rangel and Carlos F. M. Coimbra for fruitful discussions on the analytical solution of the excursion ratio.
In addition, the authors express their gratitude to the anonymous reviewers for their valuable comments.
This work used the supercomputer Phoenix and was supported by the Gauß-IT-Zentrum of the University of Braunschweig (GITZ). 
The authors gratefully acknowledge the Gauss Centre for Supercomputing e.V. (www.gauss-centre.eu) for funding this project by providing computing time on the GCS Supercomputer SUPERMUC-NG at Leibniz Supercomputing Centre (www.lrz.de). 
Further computational resources are supported by XSEDE grant TG-CTS150053.
This research was also supported in part by the National Science Foundation under Grant No. NSF PHY-1748958.

\vspace{.1in}

\noindent
{\bf Declaration of Interests}\\
\noindent
The authors report no conflict of interest.

\vspace{.1in}
\noindent{\bf Author ORCIDs}\\
F. Kleischmann, \url{https://orcid.org/0009-0009-7193-1586}\\
P. Luzzatto-Fegiz, \url{https://orcid.org/0000-0003-3614-552X} \\
E. Meiburg, \url{https://orcid.org/0000-0003-3670-8193}\\
B. Vowinckel, \url{https://orcid.org/0000-0001-6853-7750}\\

%%%%%%%%%%%%%%%%%%%%%%%%%%%%%%%%%
%%%%%%%%%%%%%%%%%%%%%%%%%%%%%%%%%
\appendix
%%%%%%%%%%%%%%%%%%%%%%%%%%%%%%%%%
%%%%%%%%%%%%%%%%%%%%%%%%%%%%%%%%%

%%%%%%%%%%%%%%%%%%%%%%%%%%%%%%%%%
\section{Benchmark tests for single-particle setups} \label{app:appendixA}
%%%%%%%%%%%%%%%%%%%%%%%%%%%%%%%%%

As pointed out in \S\ref{sec:validation}, several validation runs were performed with respect to a single-particle setup. 
In \ref{app:SingleParticle_Excursion}, we present the excursion of the trajectory of a freely moving particle in response to the oscillating fluid and in \ref{app:SingleParticle_StreamLines}, we analyze the flow characteristics based on the steady streaming generated by an oscillating particle in a fluid at rest.

%%%%%%%%%%%%%%%%%%%%%%%%%%%%%%%%%
\subsection{Excursions of particle trajectory}\label{app:SingleParticle_Excursion}
%%%%%%%%%%%%%%%%%%%%%%%%%%%%%%%%%

To validate the fluid-particle interaction of a single particle subjected to oscillating background flow, we compare our numerical results to the analytical and experimental data of \cite{2005_LEsperance_etal}.
As introduced in \S\ref{sec:Introduction}, these authors investigated the effect of inertia on a single sphere in a sinusoidally oscillating fluid.
To this end, \cite{2005_LEsperance_etal} performed an experimental study of the response of the particle to the background flow.
The authors attached the particle to a thin tether and installed it either at the top or bottom of the fluid container, depending on the density ratio between the particle and the fluid. 
In this way, the particle was prevented from rising (sinking) under gravity to the top (bottom) of the container. 
To minimize the effects of the tether drag on the particle motion, different tether thicknesses and lengths were examined \citep{2004_Coimbra_etal}. 
Based on this, particles of different materials with differing densities were investigated for a range of frequencies and compared to the analytical solution of \cite{2001_Coimbra_Rangel}.
The analytical solution uses the response coefficient $\eta$ as the excursion ratio to compare the particle amplitude $A_p$ to the amplitude of the background oscillations $A_f$ given by
\begin{equation}\label{eq:excursionFunction}
    \eta \left( S, \rho_s \right) = \frac{A_p}{A_f} = \left|1 + \frac{2 i S \left( \frac{1}{\rho_s} - 1 \right)}{i S \left(\frac{1}{\rho_s} + 2 \right) + \frac{1}{\rho_s} + \frac{3}{\rho_s} \sqrt{S} \, e^{\left( i \frac{\pi}{4} \right)}} \right| \qquad ,
\end{equation}
where $i=\sqrt{-1}$ indicates the imaginary unit.
The theoretical predictions have been confirmed by the experiments of \cite{2004_Coimbra_etal} and \cite{2005_LEsperance_etal, 2006_LEsperance_etal}. 
As described in the introduction, in both experimental studies the response of an individual particle suspended in a container filled with liquid and oscillating sinusoidally at was investigated. 
Particles of different materials with differing densities were applied and investigated for a range of frequencies. 
For the present study, we focus on the same density ratios investigated by \cite{2005_LEsperance_etal}, \textit{viz.} \mbox{$\rho_s = 0.47$}, $1.78$, and $4.68$. 
These ratios represent a strongly inertial case and two less inertial cases of densities larger and smaller than the fluid density. 

The dimensions of the numerical domain are the same as described in \S \ref{sec:numerical_setup}. 
A cubic box of size \mbox{$L_{x,y,z} = L = 10$} is considered and a single particle of size $D_p$ is initially placed at the domain center. 
To apply the oscillations, \cite{2005_LEsperance_etal} used the parameter ranges $0.06\leq \epsilon \leq 0.10$ and $1.40\leq S\leq4.90$ for the non-dimensional oscillation amplitude and frequency, respectively. 
The corresponding $\Rey$ of the oscillating container ranges from $5.03$ to $17.59$. 
For our numerical simulations, we chose a constant amplitude of $\epsilon = 0.10$ and extended the range of $S$ to $[0.07, 4.90]$, which yields $\Rey = [0.25,17.59]$ and $\Rey_s = [0.03,1.76]$ to validate even smaller values of $S$ with the analytical solution as used by \cite{2005_LEsperance_etal} for the experiments.
Here, $A_p$ represents half the distance between the high and low point of the particle's excursion within one oscillation period in the inertial reference frame which we determined at $t = 50T$ to assure well-developed conditions.

Figure \ref{fig:LEsperance_validation} shows the comparison of the numerical \mbox{results $\left( \ast \right)$} with the experimental \mbox{data $\left( \square \right)$} and the analytical \mbox{predictions $\left( \full \right)$} by showing $\eta$ for various $S$ for all three values of $\rho_s$. 
Note that \eqref{eq:excursionFunction} is only valid for the inertial reference frame. 
We, therefore, transformed the results by $\eta = 1 - A_p'/A_f$, to subtract the amplitude of the domain oscillation.
Here, $A_p'$ denotes the particle motion in a non-inertial reference frame, which consequently represents the relative deviation of the particle motion from the fluid motion.
The comparison demonstrates excellent agreement of our results and the experimental as well as the analytical outcomes. 
Hence, our method is well suited to analyze the fluid-particle interaction for more complex scenarios in oscillatory, low-Reynolds number flows.
\begin{figure}
  \centerline{\includegraphics[trim=3.4cm 0.3cm 4cm 1.4cm, clip,width=\textwidth]{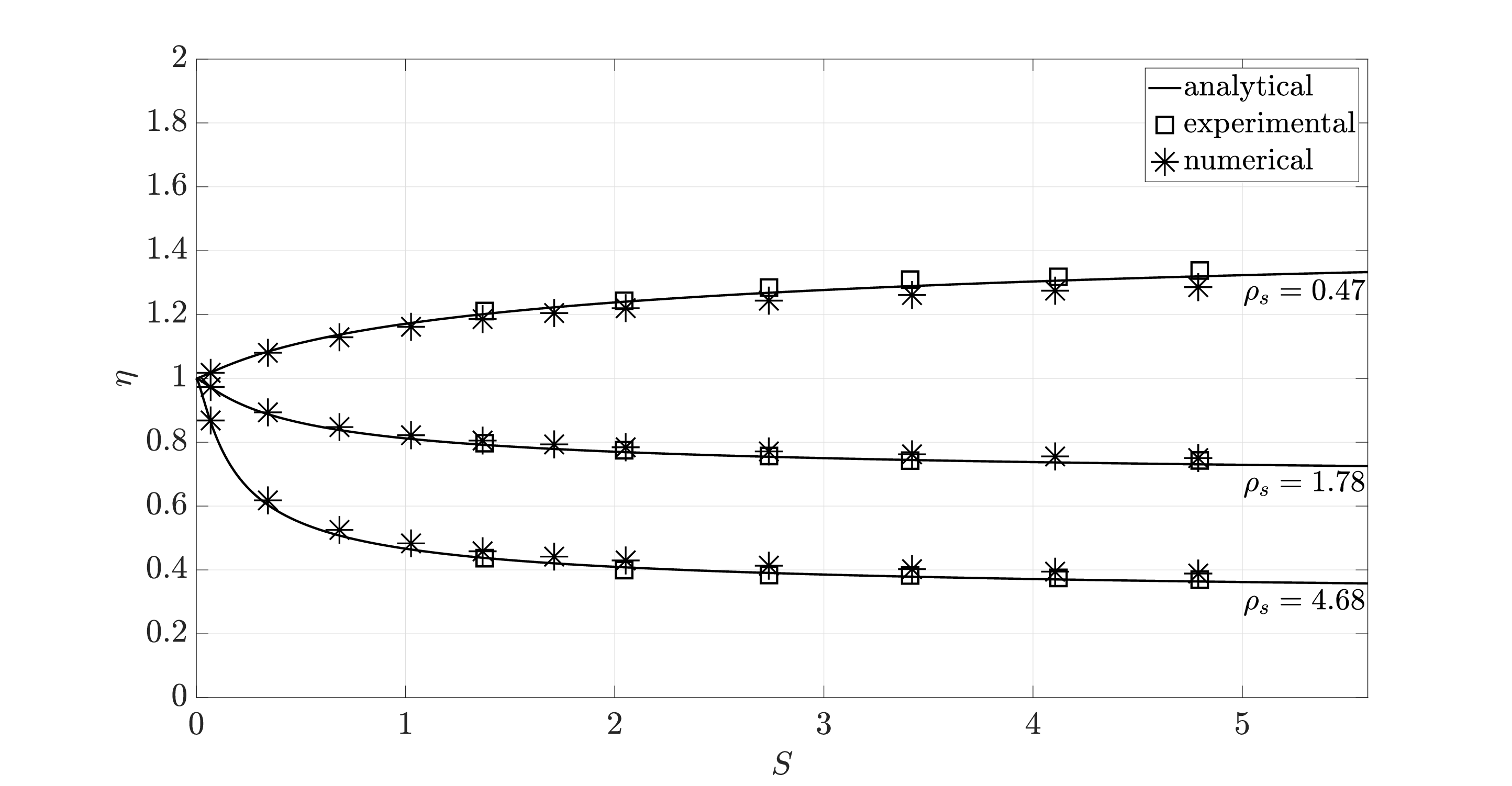}}% Images in 100% size
  \caption{Comparison of the numerical simulation results of the amplitude ratio $\eta$ for various dimensionless frequencies $S$ with the analytical predictions and experimental results of \citet{2005_LEsperance_etal} for a constant amplitude $A_f = 0.1$ and three different density ratios $\rho_s = 0.47$, $1.78$, and $4.68$.}
\label{fig:LEsperance_validation}
\end{figure}

%%%%%%%%%%%%%%%%%%%%%%%%%%%%%%%%%
\subsection{Flow characteristics}\label{app:SingleParticle_StreamLines}
%%%%%%%%%%%%%%%%%%%%%%%%%%%%%%%%%

In addition to the comparison of particle dynamics presented above, we analyze the flow characteristics of a single particle oscillating in a fluid at rest. 
This is done both qualitatively (by examining the flow patterns in terms of shapes and flow directions), and quantitatively (by analyzing the distance from the stagnation point of the flow perpendicular to the axis of oscillation to the particle surface).
Such detailed analysis is necessary to ensure that the fluid-particle interaction is not affected by numerical effects.
For this purpose, we compare the streamlines emerging due to steady streaming, which represents the averaging of the flow field over one oscillation period and which has been thoroughly described in \S\ref{sec:totalSteadyStreaming}.

This comparison is based on the work of \cite{2007_Kotas_etal} and \cite{1994_Chang_Maxey}.
\cite{2007_Kotas_etal} experimentally investigated  a single sphere oscillating in an otherwise quiescent fluid.
To this end, the sphere was mounted on a steel rod and subjected to oscillatory motion while immersed in a container of viscous fluid.  
The resulting steady streaming flow field around the sphere was visualized by averaging the records of phase-locked particle pathline images over multiple oscillation periods.  
The experiment run of \cite{2007_Kotas_etal} used for comparison is characterized by the non-dimensional numbers $S = 9.33$, $\Rey = 16.79$ and $\Rey_s = 0.84$.
Using the same parameters, \cite{1994_Chang_Maxey} conducted DNS to investigate the steady streaming generated by a sphere oscillating with low amplitude. 

We chose the same characteristics of the setup as in \S \ref{sec:numerical_setup} and prescribed a particle velocity to match the non-dimensional parameters given above.
Figure \ref{fig:Kotas_validation}a shows our simulation results for which we calculated the steady streaming after $100$ oscillation periods.
The figure represents the part above the symmetry axis, which is simultaneously the oscillation axis of the particle.
Two vortex structures evolve close to the particle surface, replicating its spherical structure, followed by two more distant vortex structures.
The near-surface structures, often referred to as shear-wave or Stokes layer, rotate counterclockwise in the left and clockwise in the right vortex, respectively.
The respective adjacent structures further away rotate in exactly the opposite direction.
In consideration of the case distinctions according to \cite{1966_Riley}, shown in \S \ref{sec:governingEqs}, the present case can be assigned to case two, which in turn is indicated by the characteristic flow numbers and the illustrated flow patterns consisting of primary and secondary vortices.

\cite{2007_Kotas_etal} and \cite{1994_Chang_Maxey} focused mainly on the shape, size, and rotational direction of the primary vortices, which are very well matched by our simulations. 
No further information is given about the secondary structures except for the directions of rotation. 
This is probably due to the fact that the extent and shapes of these structures strongly depend on the domain size and form. 
A sketch of the close-up of the general streaming structures is given in figure \ref{fig:Kotas_validation}b.
The shift of the further distant flow structures to the left in figure \ref{fig:Kotas_validation}a is related to the transients caused by initial motion of the particle, which is explained in \S \ref{sec:initialMotion}.
\begin{figure}
    \def\stackalignment{l}
    \centering
    \captionsetup[subfigure]{labelformat=empty}
    % Top
    \begin{subfigure}[b]{\textwidth}
         \centering
         \topinset{{\scriptsize (a)}}{\includegraphics[trim=6cm 5cm 6cm 4cm, clip,width=0.8\textwidth]{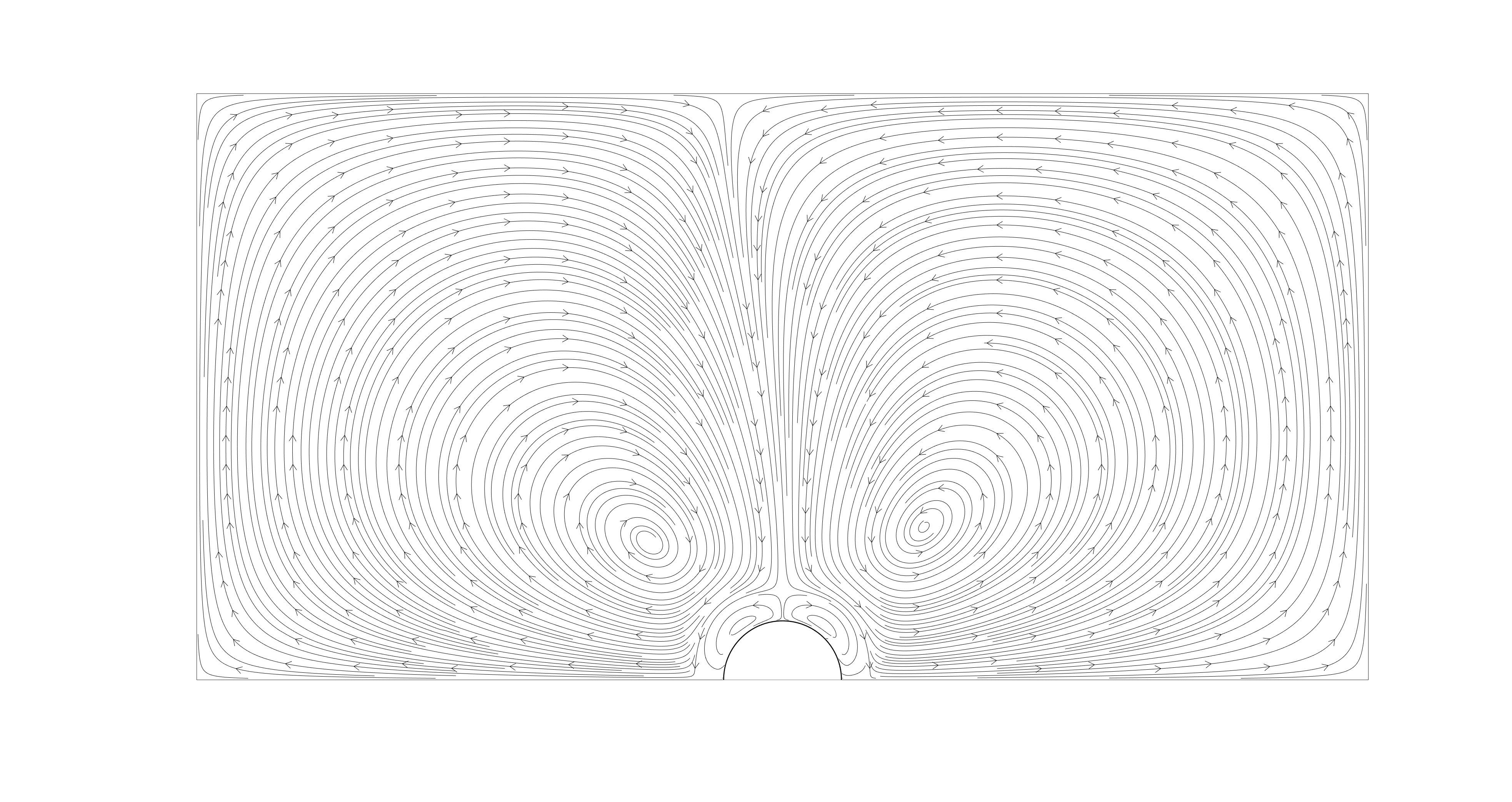}}{0.2cm}{0.05cm}
         \topinset{{\scriptsize (b)}}{\includegraphics[trim=7cm 5cm 7.6cm 8.8cm, clip,width=0.8\textwidth]{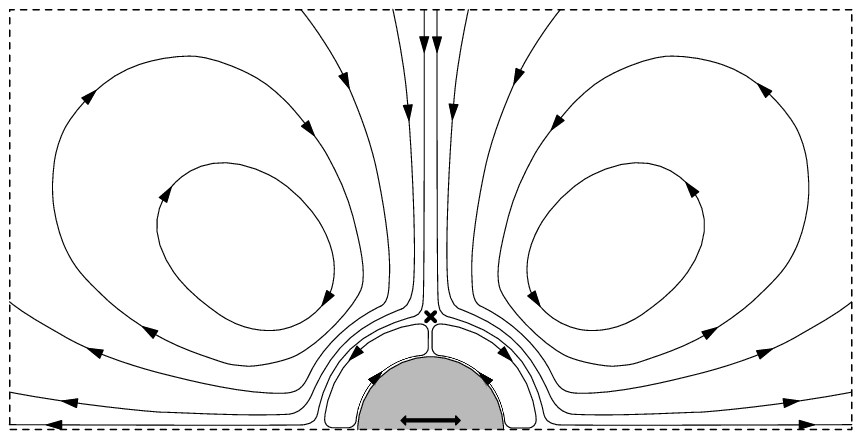}}{0.2cm}{0.05cm}
    \end{subfigure}
    \caption{Flow characteristics illustrated by streamlines of the steady streaming of a single particle oscillating horizontally (black arrows) in a fluid at rest. (a) represents the numerical results of the setup with $S = 9.33$, $\Rey = 16.79$ and $\Rey_s = 0.84$ and (b) a simplified sketch of a close-up to highlight the flow characteristics as well as the stagnation point indicated by the cross.}
    \label{fig:Kotas_validation}
\end{figure}

The cross above the particle in figure \ref{fig:Kotas_validation}b indicates the stagnation point, which represents the extent of the inner streaming region close to the particle surface.
The distance of the stagnation point from the particle surface $\delta_{SP}$ was measured experimentally by \cite{2007_Kotas_etal} and calculated numerically by \cite{2009_Klotsa_PhD} in dependence of the oscillation properties.
The results of $\delta_{SP}$ are shown in figure \ref{fig:StagnationPoint_validation}, where we compare our numerical simulations $\left( \ast \right)$ with the experimental data of \cite{2007_Kotas_etal} $\left( \square \right)$ and the numerical outcomes of \cite{2009_Klotsa_PhD} ({\larger[2]$\circ$}) for a variety of $S$.
The solid and dashed lines represent the best fits of \cite{2007_Kotas_etal} and \cite{2009_Klotsa_PhD}, respectively.
Both, \cite{2007_Kotas_etal} and \cite{2009_Klotsa_PhD} point out that numerical and experimental results may differ due to transient effects. 
Such effects are primarily caused by the smaller number of oscillation periods used for numerical simulations to calculate $\delta_{SP}$ than for experiments.
For example, \cite{2007_Kotas_etal} states that the extent of the inner flow region is $10 \%$ larger after $10 T$  than after $100 T$.
In this regard, \cite{2009_Klotsa_PhD} argues that the deviation of their results is caused by the fact that their simulations were limited to $50 T$.
In our case, we analyzed the flow properties after $100 T$ and consequently obtain a better agreement with the experimental data.
However, we also want to note that the spatial discretization has an effect on the magnitude of $\delta_{SP}$.
Nevertheless, since a higher spatial resolution is associated with a considerably increase of computational cost, we consider our results sufficient to support the accuracy of the computation of the flow characteristics.
\begin{figure}
    \def\stackalignment{l}
    \captionsetup[subfigure]{labelformat=empty}
    \begin{subfigure}[b]{\textwidth}
        \centering
        \topinset{}{\includegraphics[trim=3.5cm 0.5cm 5.2cm 2.0cm, clip,width=\textwidth]{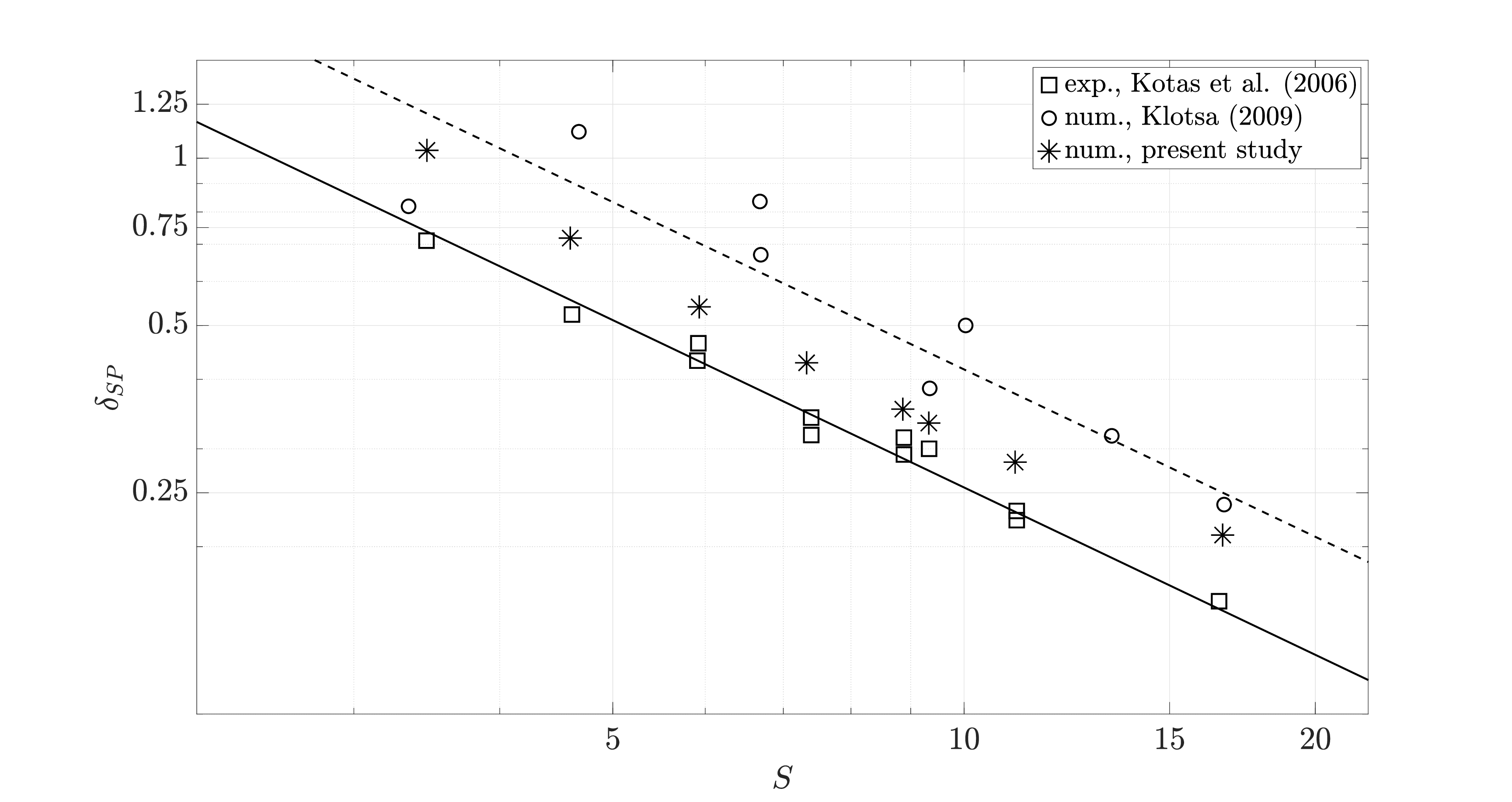}}{0.2cm}{0.05cm}% Images in 100% size
        \put(-12.48cm,1.55cm){ \tikz{\draw[solid](0,1)--(0.35,1)}}
        \put(-12cm,1.5cm){{\fontsize{7}{7} $y = 46/(18S)$}}
        \put(-12.48cm,1.1cm){ \tikz{\draw[dashed](0,1)--(0.375,1)}}
        \put(-12cm,1.05cm){{\fontsize{7}{7} $y = 75/(18S)$}}
    \end{subfigure}
    \caption{Logarithmic plot of the comparison of $\delta_{SP}$ for various $S$ between the numerical simulations and experimental and numerical data of \cite{2007_Kotas_etal} and \cite{2009_Klotsa_PhD}, respectively. The solid and dashed line represent the respective best fits.}
    \label{fig:StagnationPoint_validation}
\end{figure}

We conclude that our results are in close agreement with the experimental and numerical comparison data with respect to the shape and size of the vortices as well as to the size of the inner streaming region for an individual particle oscillating in a quiescent fluid.

\bibliographystyle{jfm}
%\bibliography{jfm-instructions}
\bibliography{main}

\begin{thebibliography}{58}
\expandafter\ifx\csname natexlab\endcsname\relax\def\natexlab#1{#1}\fi
\def\au#1{#1} \def\ed#1{#1} \def\yr#1{#1}\def\at#1{#1}\def\jt#1{\textit{#1}} \def\bt#1{#1}\def\bvol#1{\textbf{#1}} \def\vol#1{#1} \def\pg#1{#1} \def\publ#1{#1}\def\arxiv#1{#1}\def\org#1{#1}\def\st#1{\textit{#1}}

\bibitem[Alassar(2008)]{2008_Alassar}
{\sc \au{Alassar, R.~S.}} \yr{2008}  \at{Acoustic streaming on spheres}.  \jt{International Journal of Non-Linear Mechanics}  \bvol{43}~(9),  \pg{892--897}.

\bibitem[Alassar \& Badr(1997)]{1997_Alassar_Badr}
{\sc \au{Alassar, R.~S.} \& \au{Badr, H.~M.}} \yr{1997}  \at{Oscillating viscous flow over a sphere}.  \jt{Computers \& Fluids}  \bvol{26}~(7),  \pg{661--682}.

\bibitem[Biegert {\em et~al.\/}(2017)Biegert, Vowinckel \& Meiburg]{2017_Biegert_etal}
{\sc \au{Biegert, E.}, \au{Vowinckel, B.} \& \au{Meiburg, E.}} \yr{2017}  \at{A collision model for grain-resolving simulations of flows over dense, mobile, polydisperse granular sediment beds}.  \jt{J. Comp. Phys.}  \bvol{340},  \pg{105--127}.

\bibitem[Blackburn(2002)]{2002_Blackburn}
{\sc \au{Blackburn, H.~M.}} \yr{2002}  \at{Mass and momentum transport from a sphere in steady and oscillatory flows}.  \jt{Physics of Fluids}  \bvol{14}~(11),  \pg{3997--4011}.

\bibitem[Cerretelli \& Williamson(2003)]{2003_Cerretelli_Williamson}
{\sc \au{Cerretelli, C.} \& \au{Williamson, C. H.~K.}} \yr{2003}  \at{The physical mechanism for vortex merging}.  \jt{J. Fluid Mech.}  \bvol{475},  \pg{41–77}.

\bibitem[Chang \& Maxey(1994)]{1994_Chang_Maxey}
{\sc \au{Chang, E.~J.} \& \au{Maxey, M.~R.}} \yr{1994}  \at{Unsteady flow about a sphere at low to moderate reynolds number. part 1. oscillatory motions}.  \jt{J. Fluid Mech.}  \bvol{277},  \pg{347--379}.

\bibitem[Coimbra {\em et~al.\/}(2004)Coimbra, L'esperance, Lambert, Trolinger \& Rangel]{2004_Coimbra_etal}
{\sc \au{Coimbra, C. F.~M.}, \au{L'esperance, D.}, \au{Lambert, R.~A.}, \au{Trolinger, J.~D.} \& \au{Rangel, R.~H.}} \yr{2004}  \at{An experimental study on stationary history effects in high-frequency stokes flows}.  \jt{J. Fluid Mech.}  \bvol{504},  \pg{353--363}.

\bibitem[Coimbra \& Rangel(2001)]{2001_Coimbra_Rangel}
{\sc \au{Coimbra, C. F.~M.} \& \au{Rangel, R.~H.}} \yr{2001}  \at{Spherical particle motion in harmonic stokes flows}.  \jt{AIAA Journal}  \bvol{39}~(9),  \pg{1673--1682}.

\bibitem[Dietsche {\em et~al.\/}(2019)Dietsche, Mutlu, Edd, Koumoutsakos \& Toner]{2019_Dietsche_etal}
{\sc \au{Dietsche, C.}, \au{Mutlu, B.~R.}, \au{Edd, J.~F.}, \au{Koumoutsakos, P.} \& \au{Toner, M.}} \yr{2019}  \at{Dynamic particle ordering in oscillatory inertial microfluidics}.  \jt{Microfluid. Nanofluid.}  \bvol{23}~(6).

\bibitem[Fabre {\em et~al.\/}(2017)Fabre, Jalal, Leontini \& Manasseh]{2017_Fabre_etal}
{\sc \au{Fabre, D.}, \au{Jalal, J.}, \au{Leontini, J.~S.} \& \au{Manasseh, R.}} \yr{2017}  \at{Acoustic streaming and the induced forces between two spheres}.  \jt{J. Fluid Mech.}  \bvol{810},  \pg{378–391}.

\bibitem[Gupta {\em et~al.\/}(2019)Gupta, Chen, Yu, Prhashanna \& Katoshevski]{2019_Gupta_etal}
{\sc \au{Gupta, A.~K.}, \au{Chen, S.~B.}, \au{Yu, L.~E.}, \au{Prhashanna, A.} \& \au{Katoshevski, D.}} \yr{2019}  \at{Cfd study on particle grouping under an oscillatory flow in a wavy duct}.  \jt{Separation and Purification Technology}  \bvol{213},  \pg{303--313}.

\bibitem[Halfi {\em et~al.\/}(2020)Halfi, Arad, Brenner \& Katoshevski]{2020_Halfi_etal}
{\sc \au{Halfi, E.}, \au{Arad, A.}, \au{Brenner, A.} \& \au{Katoshevski, D.}} \yr{2020}  \at{Development of an oscillation-based technology for the removal of colloidal particles from water: Cfd modeling and experiments}.  \jt{Engineering Applications of Computational Fluid Mechanics}  \bvol{14}~(1),  \pg{622--641}.

\bibitem[Halfi {\em et~al.\/}(2019)Halfi, Brenner \& Katoshevski]{2019_Halfi_etal}
{\sc \au{Halfi, E.}, \au{Brenner, A.} \& \au{Katoshevski, D.}} \yr{2019}  \at{Separation of colloidal minerals from water by oscillating flows and grouping}.  \jt{Separation and Purification Technology}  \bvol{210},  \pg{981--987}.

\bibitem[Harte {\em et~al.\/}(2023)Harte, Obrist, Caversaccio, Lajoinie \& Wimmer]{2023_Harte_etal}
{\sc \au{Harte, N.}, \au{Obrist, D.}, \au{Caversaccio, M.}, \au{Lajoinie, G. P.~R.} \& \au{Wimmer, W.}} \yr{2023} Transverse flow under oscillating stimulation in helical square ducts with cochlea-like geometrical curvature and torsion,  \arxiv{arXiv: 2303.15603}.

\bibitem[Hassan \& Kawaji(2008)]{2008_Hassan_Kawaji}
{\sc \au{Hassan, S.} \& \au{Kawaji, M.}} \yr{2008}  \at{{The effects of vibrations on particle motion in a viscous fluid cell}}.  \jt{Journal of Applied Mechanics}  \bvol{75}~(3).

\bibitem[Hassan \& Kawaji(2007)]{2007b_Hassan_Kawaji}
{\sc \au{Hassan, S.} \& \au{Kawaji, S.}} \yr{2007}  \at{Vibration-induced particle drift in a fluid cell under microgravity}.  \jt{Microgravity Science and Technology}  \bvol{19}~(3),  \pg{109--112}.

\bibitem[Hassan {\em et~al.\/}(2006)Hassan, Lyubimova, Lyubimov \& Kawaji]{2006b_Hassan_etal}
{\sc \au{Hassan, S.}, \au{Lyubimova, T.~P.}, \au{Lyubimov, D.~V.} \& \au{Kawaji, M.}} \yr{2006}  \at{{Motion of a sphere suspended in a vibrating liquid-filled container}}.  \jt{Journal of Applied Mechanics}  \bvol{73}~(1),  \pg{72--78}.

\bibitem[Jalal(2018)]{2018_Jalal}
{\sc \au{Jalal, J.}} \yr{2018}  \at{Interaction of spherical particles owing to steady streaming induced by ultrasound}. PhD thesis, Swinburne University of Technology Melbourne, Australia.

\bibitem[Katoshevski {\em et~al.\/}(2005)Katoshevski, Dodin \& Ziskind]{2005_Katoshevski_etal}
{\sc \au{Katoshevski, D.}, \au{Dodin, Z.} \& \au{Ziskind, G.}} \yr{2005}  \at{Aerosol clustering in oscillating flows: Mathematical analysis}.  \jt{Atomization and Sprays}  \bvol{15},  \pg{401--412}.

\bibitem[Kempe \& Fröhlich(2012)]{2012_Kempe_Froehlich}
{\sc \au{Kempe, T.} \& \au{Fröhlich, J.}} \yr{2012}  \at{An improved immersed boundary method with direct forcing for the simulation of particle laden flows}.  \jt{J. Comput. Phys.}  \bvol{231}~(9),  \pg{3663--3684}.

\bibitem[Kleischmann {\em et~al.\/}(2021)Kleischmann, Luzzatto-Fegiz, Rommelfanger, Meiburg \& Vowinckel]{2021_Kleischmann_etal}
{\sc \au{Kleischmann, F.}, \au{Luzzatto-Fegiz, P.}, \au{Rommelfanger, N.}, \au{Meiburg, E.} \& \au{Vowinckel, B.}} \yr{2021} Particle-resolved direct numerical simulations of clay particles in the absence of gravity.  \bt{In {\em EGU General Assembly Conference Abstracts\/}}, {\em EGU General Assembly Conference Abstracts\/} ,  \pg{pp. EGU21--4576}.

\bibitem[Klotsa(2009)]{2009_Klotsa_PhD}
{\sc \au{Klotsa, D.}} \yr{2009}  \at{The dymanics of spheres in oscillatory fluid flows.} PhD thesis, University of Nottingham.

\bibitem[Klotsa {\em et~al.\/}(2007)Klotsa, Swift, Bowley \& King]{2007_Klotsa_etal}
{\sc \au{Klotsa, D.}, \au{Swift, M.~R.}, \au{Bowley, R.~M.} \& \au{King, P.~J.}} \yr{2007}  \at{Interaction of spheres in oscillatory fluid flows}.  \jt{Physical Review E}  \bvol{76}~(5),  \pg{056314}.

\bibitem[Klotsa {\em et~al.\/}(2009)Klotsa, Swift, Bowley \& King]{2009_Klotsa_etal}
{\sc \au{Klotsa, D.}, \au{Swift, M.~R.}, \au{Bowley, R.~M.} \& \au{King, P.~J.}} \yr{2009}  \at{Chain formation of spheres in oscillatory fluid flows}.  \jt{Physical Review E}  \bvol{79}~(2),  \pg{021302}.

\bibitem[Kotas {\em et~al.\/}(2007)Kotas, Yoda \& Rogers]{2007_Kotas_etal}
{\sc \au{Kotas, C.W.}, \au{Yoda, M.} \& \au{Rogers, P.H.}} \yr{2007}  \at{Chain formation of spheres in oscillatory fluid flows}.  \jt{Exp Fluids}  \bvol{42},  \pg{111--121}.

\bibitem[Lane(1955)]{1955_Lane}
{\sc \au{Lane, C.}} \yr{1955}  \at{Acoustic streaming in the vicinity of a sphere}.  \jt{The Journal of the Acoustical Society of America}  \bvol{27}~(5),  \pg{1003--1003}.

\bibitem[Li {\em et~al.\/}(2023)Li, Collis, Brumley, Schneiders \& Sader]{2023_Li_etal}
{\sc \au{Li, P.}, \au{Collis, J.~F.}, \au{Brumley, D.~R.}, \au{Schneiders, L.} \& \au{Sader, J.~E.}} \yr{2023}  \at{Structure of the streaming flow generated by a sphere in a fluid undergoing rectilinear oscillation}.  \jt{J. Fluid Mech.}  \bvol{974},  \pg{A37}.

\bibitem[Longuet-Higgins(1998)]{1998_Longuet-Higgins}
{\sc \au{Longuet-Higgins, M.~S.}} \yr{1998}  \at{Viscous streaming from an oscillating spherical bubble}.  \jt{Proc. R. Soc. Lond. A.}  \pg{pp. 25--742}.

\bibitem[Lyubimov {\em et~al.\/}(2001)Lyubimov, Cherepanov, Lyubimova \& Roux]{2001_Lyubimov_etal}
{\sc \au{Lyubimov, D.~V.}, \au{Cherepanov, A.~A.}, \au{Lyubimova, T.~P.} \& \au{Roux, B.}} \yr{2001}  \at{Vibration influence on the dynamics of a two-phase system in weightlessness conditions}.  \jt{J. Phys. IV France}  \bvol{11},  \pg{Pr6--83--Pr6--90}.

\bibitem[L´Espérance {\em et~al.\/}(2005)L´Espérance, Coimbra, Trolinger \& Rangel]{2005_LEsperance_etal}
{\sc \au{L´Espérance, D.}, \au{Coimbra, C.F.M.}, \au{Trolinger, J.D.} \& \au{Rangel, R.H.}} \yr{2005}  \at{Experimental verification of fractional history effects on the viscous dynamics of small spherical particles}.  \jt{Exp. Fluids}  \bvol{38},  \pg{112--116}.

\bibitem[L´Espérance {\em et~al.\/}(2006)L´Espérance, Trolinger, Coimbra \& Rangel]{2006_LEsperance_etal}
{\sc \au{L´Espérance, D.}, \au{Trolinger, J.~D.}, \au{Coimbra, C. F.~M.} \& \au{Rangel, R.~H.}} \yr{2006}  \at{Particle response to low-reynolds-number oscillation of a fluid in microgravity}.  \jt{AIAA Journal}  \bvol{44}~(5),  \pg{1060--1064}.

\bibitem[Mordant \& Pinton(2000)]{2000_Mordant_Pinton}
{\sc \au{Mordant, N.} \& \au{Pinton, J.-F.}} \yr{2000}  \at{Velocity measurement of a settling sphere}.  \jt{Eur. Phys. J. B}  \bvol{18}~(2),  \pg{343--352}.

\bibitem[Mutlu {\em et~al.\/}(2018)Mutlu, Edd \& Toner]{2018_Mutlu_etal}
{\sc \au{Mutlu, B.R.}, \au{Edd, J.F.} \& \au{Toner, M.}} \yr{2018}  \at{Oscillatory inertial focusing in infinite microchannels}.  \jt{Proc. Natl. Acad. Sci.}  \bvol{115}~(30),  \pg{7682--7687}.

\bibitem[Mutlu {\em et~al.\/}(2020)Mutlu, Dubash, Dietsche, Mishra, Ozbey, Keim, Edd, Haber, Maheswaran \& Toner]{2020_Mutlu_etal}
{\sc \au{Mutlu, B.~R.}, \au{Dubash, T.}, \au{Dietsche, C.}, \au{Mishra, A.}, \au{Ozbey, A.}, \au{Keim, K.}, \au{Edd, J.F.}, \au{Haber, D.A.}, \au{Maheswaran, S.} \& \au{Toner, M.}} \yr{2020}  \at{In-flow measurement of cell–cell adhesion using oscillatory inertial microfluidics}.  \jt{Lab Chip}  \bvol{20},  \pg{1612--1620}.

\bibitem[Otto {\em et~al.\/}(2008)Otto, Riegler \& Voth]{2008_Otto_etal}
{\sc \au{Otto, Fl.}, \au{Riegler, E.~K.} \& \au{Voth, G.~A.}} \yr{2008}  \at{Measurements of the steady streaming flow around oscillating spheres using three dimensional particle tracking velocimetry}.  \jt{Physics of Fluids}  \bvol{20}~(9).

\bibitem[Owen {\em et~al.\/}(2023)Owen, Kechagidis, Bazaz, Enjalbert, Essmann, Mallorie, Mirghaderi, Schaaf, Thota, Vernekar, Zhou, Warkiani, Stark \& Krüger]{2023_Owen_etal}
{\sc \au{Owen, B.}, \au{Kechagidis, K.}, \au{Bazaz, S.R.}, \au{Enjalbert, R.}, \au{Essmann, E.}, \au{Mallorie, C.}, \au{Mirghaderi, F.}, \au{Schaaf, C.}, \au{Thota, K.}, \au{Vernekar, R.}, \au{Zhou, Q.}, \au{Warkiani, M.E.}, \au{Stark, H.} \& \au{Krüger, T.}} \yr{2023}  \at{Lattice-boltzmann modelling for inertial particle microfluidics applications - a tutorial review}.  \jt{Advances in Physics: X}  \bvol{8}~(1),  \pg{2246704}.

\bibitem[Riley(1966)]{1966_Riley}
{\sc \au{Riley, N.}} \yr{1966}  \at{On a sphere oscillating in a viscous fluid}.  \jt{Q. J. Mech. Appl. Maths}  \bvol{19}~(4),  \pg{461 -- 472}.

\bibitem[Riley(1967)]{1967_Riley}
{\sc \au{Riley, N.}} \yr{1967}  \at{Oscillatory viscous flows. review and extension}.  \jt{IMA Journal of Applied Mathematics}  \bvol{3}~(4),  \pg{419--434}.

\bibitem[Riley(2001)]{2001_Riley}
{\sc \au{Riley, N.}} \yr{2001}  \at{Steady streaming}.  \jt{Annu. Rev. Fluid Mech.}  \bvol{33}~(1),  \pg{43--65}.

\bibitem[Rott(1964)]{1964_Rott}
{\sc \au{Rott, Nicholas}} \yr{1964}  \at{Theory of time-dependent laminar flows}.  \jt{Theory of Laminar Flows}  \bvol{4},  \pg{421}.

\bibitem[Ruzal-Mendelevich {\em et~al.\/}(2016)Ruzal-Mendelevich, Katoshevski \& Sher]{2016_Ruzal_etal}
{\sc \au{Ruzal-Mendelevich, M.}, \au{Katoshevski, D.} \& \au{Sher, E.}} \yr{2016}  \at{Controlling nanoparticles emission with particle-grouping exhaust-pipe}.  \jt{Fuel}  \bvol{166},  \pg{116--123}.

\bibitem[Saadatmand \& Kawaji(2010)]{2010_Saadatmand_Kawaji}
{\sc \au{Saadatmand, M.} \& \au{Kawaji, M.}} \yr{2010}  \at{Effect of viscosity on vibration-induced motion of a spherical particle suspended in a fluid cell}.  \jt{Microgravity Science and Technology}  \bvol{22}~(3),  \pg{433--440}.

\bibitem[Satish {\em et~al.\/}(2022)Satish, Leontini, Manasseh, Sannasiraj \& Sundar]{2022_Satish_etal}
{\sc \au{Satish, S.}, \au{Leontini, J.~S.}, \au{Manasseh, R.}, \au{Sannasiraj, S.A.} \& \au{Sundar, V.}} \yr{2022} Numerical investigation on the mean flow fields generated by an oscillating sphere.  \bt{In {\em OCEANS 2022 - Chennai\/}},  \pg{pp. 1--5}.

\bibitem[Simic-Stefani {\em et~al.\/}(2005)Simic-Stefani, Hu \& Kawaji]{2006_Simic-Stefani_etal}
{\sc \au{Simic-Stefani, S.}, \au{Hu, H.~H.} \& \au{Kawaji, M.}} \yr{2005}  \at{Numerical and experimental investigation of solid particle motion in a fluid cell under microgravity}.  \jt{Microgravity Science and Technology}  \bvol{16}~(1),  \pg{301--305}.

\bibitem[Sumner {\em et~al.\/}(2021)Sumner, Mestel \& T]{2021_Sumner_etal}
{\sc \au{Sumner, L}, \au{Mestel, J} \& \au{T, Reichenbach}} \yr{2021}  \at{Steady streaming as a method for drug delivery to the inner ear}.  \jt{Sci Rep.}  \bvol{11}~(1),  \pg{191–233}.

\bibitem[Tchen(1947)]{1947_Tchen}
{\sc \au{Tchen, C.M.}} \yr{1947} {\em Mean value and correlation problems connected with the motion of small particles suspended in a turbulent ﬂuid\/}.  \publ{Doctoral dissertation, TU Delft, Martinus Nijhoff, The Hague}.

\bibitem[Ten~Cate {\em et~al.\/}(2002)Ten~Cate, Niewstad, Derksen \& Van~den Akker]{2002_tenCate_etal}
{\sc \au{Ten~Cate, A.}, \au{Niewstad, C.~H.}, \au{Derksen, J.~J.} \& \au{Van~den Akker, H. E.~A.}} \yr{2002}  \at{Particle imaging velocimetry experiments and lattice-boltzmann simulations on a single sphere settling under gravity}.  \jt{Phys. Fluids}  \bvol{14}~(11),  \pg{4012--4025}.

\bibitem[Tho {\em et~al.\/}(2007)Tho, Manasseh \& Ooi]{2007_Tho_etal}
{\sc \au{Tho, P.}, \au{Manasseh, R.ichard} \& \au{Ooi, A.}} \yr{2007}  \at{Cavitation microstreaming patterns in single and multiple bubble systems}.  \jt{J. Fluid Mech.}  \bvol{576},  \pg{191–233}.

\bibitem[Trolinger {\em et~al.\/}(1996{\natexlab{{\em a\/}}})Trolinger, Lal, McIntosh \& Witherow]{1996a_Trolinger_etal}
{\sc \au{Trolinger, J.D.}, \au{Lal, R.B.}, \au{McIntosh, D.} \& \au{Witherow, W.K.}} \yr{1996{\natexlab{{\em a\/}}}}  \at{Holographic particle-image velocimetry in the first international microgravity laboratory aboard the space shuttle discovery}.  \jt{Appl Opt.}  \bvol{35}~(4),  \pg{681--689}.

\bibitem[Trolinger {\em et~al.\/}(1996{\natexlab{{\em b\/}}})Trolinger, Rangel \& Lal]{1996b_Trolinger_etal}
{\sc \au{Trolinger, J.D.}, \au{Rangel, R.H.} \& \au{Lal, R.B.}} \yr{1996{\natexlab{{\em b\/}}}} Particle mechanics and g-jitter observations in the iml-1 spaceflight.  \bt{In {\em 34th Aerospace Sciences Meeting and Exhibit\/}},  \pg{p. 504}.

\bibitem[Uhlmann(2005)]{2005_Uhlmann}
{\sc \au{Uhlmann, M.}} \yr{2005}  \at{An immersed boundary method with direct forcing for the simulation of particulate flows}.  \jt{J. Comput. Phys.}  \bvol{209}~(2),  \pg{448--476}.

\bibitem[Van~Overveld {\em et~al.\/}(2023)Van~Overveld, Clercx \& Duran-Matute]{2023_vanOverveld_etal}
{\sc \au{Van~Overveld, T. J. J.~M.}, \au{Clercx, H. J.~H.} \& \au{Duran-Matute, M.}} \yr{2023}  \at{Pattern formation of spherical particles in an oscillating flow}.  \jt{Phys. Rev. E}  \bvol{108},  \pg{025103}.

\bibitem[Van~Overveld {\em et~al.\/}(2022)Van~Overveld, Shajahan, Breugem, Clercx \& Duran-Matute]{2022_vanOverveld_etal}
{\sc \au{Van~Overveld, T. J. J.~M.}, \au{Shajahan, M.~T.}, \au{Breugem, W.-P.}, \au{Clercx, H. J.~H} \& \au{Duran-Matute, M.}} \yr{2022}  \at{Numerical study of a pair of spheres in an oscillating box filled with viscous fluid}.  \jt{Phys. Rev. Fluids}  \bvol{7}~(1),  \pg{014308}.

\bibitem[Voth {\em et~al.\/}(2002)Voth, Bigger, Buckley, Losert, Brenner, Stone \& Gollub]{2002_Voth_etal}
{\sc \au{Voth, G.~A.}, \au{Bigger, B.}, \au{Buckley, M.~R.}, \au{Losert, W.}, \au{Brenner, M.~P.}, \au{Stone, H.~A.} \& \au{Gollub, J.~P.}} \yr{2002}  \at{Ordered clusters and dynamical states of particles in a vibrated fluid}.  \jt{Phys. Rev. Lett.}  \bvol{88},  \pg{234301}.

\bibitem[Vowinckel {\em et~al.\/}(2021)Vowinckel, Biegert, Meiburg, Aussillous \& Guazzelli]{2021_Vowinckel_etal}
{\sc \au{Vowinckel, B.}, \au{Biegert, E.}, \au{Meiburg, E.}, \au{Aussillous, P.} \& \au{Guazzelli, {\'E}.}} \yr{2021}  \at{Rheology of mobile sediment beds sheared by viscous, pressure-driven flows}.  \jt{J. Fluid Mech.}  \bvol{921}.

\bibitem[Vowinckel {\em et~al.\/}(2017)Vowinckel, Nikora, Kempe \& Fr{\"o}hlich]{2017_Vowinckel_etal}
{\sc \au{Vowinckel, B.}, \au{Nikora, V.}, \au{Kempe, T.} \& \au{Fr{\"o}hlich, J.}} \yr{2017}  \at{Spatially-averaged momentum fluxes and stresses in flows over mobile granular beds: a dns-based study}.  \jt{Journal of Hydraulic Research}  \bvol{55}~(2),  \pg{208--223}.

\bibitem[Vowinckel {\em et~al.\/}(2019)Vowinckel, Withers, Luzzatto-Fegiz \& Meiburg]{2019_Vowinckel_etal}
{\sc \au{Vowinckel, B.}, \au{Withers, J.}, \au{Luzzatto-Fegiz, Paolo} \& \au{Meiburg, E.}} \yr{2019}  \at{Settling of cohesive sediment: particle-resolved simulations}.  \jt{J. Fluid Mech.}  \bvol{858},  \pg{5–44}.

\bibitem[Wunenburger {\em et~al.\/}(2002)Wunenburger, Carrier \& Garrabos]{2002_Wunenburger_etal}
{\sc \au{Wunenburger, R.}, \au{Carrier, V.} \& \au{Garrabos, Y.}} \yr{2002}  \at{{Periodic order induced by horizontal vibrations in a two-dimensional assembly of heavy beads in water}}.  \jt{Physics of Fluids}  \bvol{14}~(7),  \pg{2350--2359}.

\end{thebibliography}

\end{document}